\documentclass[namedreferences]{solarphysics}

\usepackage[urlcolor=blue,colorlinks=true,linkcolor=red,citecolor=magenta,breaklinks=true]{hyperref}
\usepackage[optionalrh,natbib]{spr-sola-addons} 
\usepackage{graphicx}        
\usepackage{color}           
\usepackage{breakurl}        
\usepackage{rotating}        

\graphicspath{{figs/}}

\renewcommand{\vec}[1]{{\mathbfit #1}}

\renewcommand{\deg}{^\circ}
\newcommand{\Rsun}{R_\odot}
\newcommand{\Bsun}{B_\odot}

\newcommand{\fov}{field of view{\ }}

\newcommand{\ie}{{\it i.e.}{\ }}
\newcommand{\eg}{{\it e.g.}{\ }}

\newcommand{\angstrom}{\mbox{\normalfont\AA}}

\newcommand{\adsurl}[1]{\href{http://adsabs.harvard.edu/abs/#1}{ADS}}
\newcommand{\doiurl}[1]{\href{http://dx.doi.org/#1}{DOI}}

\begin{document}

\begin{article}

\begin{opening}

\title{Coronal Photopolarimetry with the LASCO-C2 Coronagraph over 24 Years [1996--2019]}
\subtitle{\sl{Application to the K/F Separation and to the Determination of the Electron Density}}
 

\author[addressref={aff1},corref,email={philippe.lamy@latmos.ipsl.fr}]{\inits{P. }\fnm{Philippe }\lnm{Lamy}} 	
\author[addressref={aff2},email={antoine.llebaria@lam.fr}]{\inits{A. }\fnm{Antoine }\lnm{Llebaria}} 	
\author[addressref={aff2},email={Bricexp@hotmail.com}]{\inits{B. }\fnm{Brice }\lnm{Boclet}} 	
\author[addressref={aff1},email={hugo.gilardy@latmos.ipsl.fr}]{\inits{H. }\fnm{Hugo }\lnm{Gilardy}} 	
\author[addressref={aff2,aff3},email={mburtin@gmail.com}]{\inits{M. }\fnm{Michael }\lnm{Burtin}} 	
\author[addressref={aff1},email={olivier.floyd@latmos.ipsl.fr}]{\inits{O. }\fnm{Olivier }\lnm{Floyd}}	

\address[id=aff1]{Laboratoire Atmosph\`eres, Milieux et Observations Spatiales, CNRS \& UVSQ, 11 Bd d'Alembert, 78280 Guyancourt, France}
\address[id=aff2]{Laboratoire d'Astrophysique de Marseille, CNRS \& Aix-Marseille Universit\'e, 38 rue Fr\'ed\'eric Joliot-Curie, 13388 Marseille cedex 13, France}
\address[id=aff3]{Voxtok, 34980 Saint-Gely-du-Fesc, France}
       
\runningauthor{Lamy {\it et al.}}
\runningtitle{Coronal Photopolarimetry}
 
\begin{abstract}

We present an in-depth characterization of the polarimetric channel of the Large-Angle Spectrometric COronagraph ``LASCO-C2'' onboard the {\it Solar and Heliospheric Observatory} (SoHO).
The polarimetric analysis of the white-light images makes use of polarized sequences composed of three images obtained though three polarizers oriented at $+60\deg$, $0\deg$ and $-60\deg$, complemented by a neighboring unpolarized image, and relies on the formalism of Mueller.
The Mueller matrix characterizing the C2 instrument was obtained through extensive ground-based calibrations of the optical components and global laboratory tests.
Additional critical corrections were derived from in-flight tests relying prominently on roll sequences and on consistency criteria (\eg the ``tangential'' direction of polarization).  
Our final results encompass the characterization of the polarization of the white-light corona, of its polarized radiance, of the two-dimensional electron density, and of the K-corona over two solar cycles.
They are in excellent agreement with measurements obtained at several solar eclipses except for slight discrepancies affecting the innermost part of the C2 \fov.

\end{abstract}
\end{opening}

\section{Introduction}
       
Observing the solar corona in polarized white-light has been actively pursued for decades with different purposes.
\begin{itemize}
\item
Map the polarization and the polarized radiance and compare the measurements with photopolarimetric models of the corona 
(\eg \cite{Baumbach1938}; \cite{Saito1950}; \cite{saito1964polarigraphic}; \cite{arnquist1970coronal}; \cite{nikolsky1977polarization}; \cite{saito1977study}; \cite{durst1982two}; \cite{clette1985observations}; \cite{Koutchmy1996}; \cite{gabryl1999polarization}; \cite{kim2011eclipse}; \cite{Skomorovsky2012}; \cite{vorobiev2017imaging}).
\item
Investigate the polarization of different coronal structures 
(\eg \cite{michard1965contributions}; \cite{pepin1970observations}; \cite{durst1976observations}; \cite{nikolsky1977polarization}; \cite{Jacoub1976}; \cite{koutchmy1977polarimetric}; \cite{tanabe1992optical}; \cite{kim2011eclipse}) and as a function of the solar cycle \citep{badalyan1997white}. 
\item
Verify that the observed polarization and its orientation are consistent with the Thomson theory, at least in the inner corona where the K component dominates 
(\eg \cite{Ney1961}; \cite{Badalyan1993}; \cite{Koutchmy1993}; \cite{Filippov1994}).
\item
Separate the $K$ and $F$ components 
(\eg \cite{Kluber1958intensities}; \cite{Ney1961}; \cite{durst1982two}).
\item
Retrieve the electron density 
(\eg \cite{Allen1947}; \cite{VdH1950electron}; \cite{Kluber1958intensities}; \cite{Ney1961}; \cite{saito1977study}; \cite{durst1982two}; \cite{Sivaraman1984}; \cite{Raju1986}; \cite{hayes2001deriving}; \cite{Skomorovsky2012}).
\item
Reconstruct coronal structures such as coronal mass ejections (CMEs) in three dimensions (\eg \cite{moran2004three}, \cite{Dere2005}).
\item
Search for anomalous polarization which could be diagnostic of relativistic electrons as proposed by \cite{Molodensky1973} and further developed by \cite{Inhester2015thomson} (\eg \cite{koutchmy1971observations}; \cite{kishonkov1975precise}; \cite{Clette1992}; \cite{vorobiev2017imaging}).
\end{itemize}
      
With the invention of the coronagraph by Lyot (\cite{lyot1932etude}), the white-light inner solar corona has been accessible from high altitude ground-based sites, and K-coronameters such as the Mark III, Mark IV, and K-Cor at the {\it Mauna Loa Solar Observatory} (Hawaii) have routinely obtained maps of the polarized brightness $pB$ up to a typical elongation of $\approx$1.5 $\Rsun$ from the center of the solar disk.
Access from the ground to the outer corona beyond this distance is only possible during the rare total eclipses and only for a few minutes.
Measuring the polarization further requires excellent sky conditions, but remains anyway complicated by the intrinsic polarization of the Earth atmosphere. 
During eclipses, scattering of the solar light from outside the central band of totality and of the coronal light in the central band (the aureole)
involves different mechanisms difficult to model, in particular if the solar elevation is low.
Until the widespread use of CCD detectors, photographic plates or films were the standard technique with their intrinsic non-linearity and therefore stringent requirements for appropriate calibrations.
A notable exception was the scanning photometer implemented by \cite{Ney1960}.
In spite of these difficulties, several highly skillful observers have obtained valuable measurements of the coronal polarization, thus laying down the foundations for the understanding of the two components, the Kontinuerlich (K) corona and the Fraunhofer (F) corona, for their characterization and separation, and finally for the determination of the coronal electron density.
The introduction of CCDs has alleviated the shortcomings of the photographic technique, and several high quality observational results have been obtained although the challenge of the terrestrial atmosphere remains. 
With the advent of space coronagraphy, one would have hoped that routine polarization observations of the whole corona were within reach.
A first attempt was realized with the OSO-7 white-light coronagraph \citep{Koomen1975} with small concentric polarizer rings cemented to the vidicon faceplate.
This rudimentary setup together with the poor photometric performances of the vidicon detector precluded any meaningful measurements.
The {\it Solar Maximum Mission} (SMM) Coronagraph/Polarimeter was thoughtfully conceived as a polarimetric instrument \citep{macqueen1980high} and extensively characterized before its launch.
Here again, the vidicon detector limited the performances and very few photopolarimetric results have been reported, essentially radial profiles of the polarized radiance (\eg \cite{Munro1977}; \cite{saito1977study}; in fact, we are not aware of any polarization results.

The Large-Angle Spectrometric COronagraph (LASCO), a set of three coronagraphs \citep{brueckner1995large} onboard the {\it Solar and Heliospheric Observatory} (SoHO) opened an entirely new era in the domain of photometric and polarimetric investigations of the white-light corona.
LASCO-C2, one of the two externally occulted coronagraphs of direct interest to this present article, has been in nearly continuous operation since January 1996, recording the brightest part of the solar corona accessible to LASCO from 2.2 to 6.5 R${}_\odot$. 
Minor interruptions occurred since 1996 for various instrumental and spacecraft reasons except for two major events: i) the accidental loss of SoHO during a roll maneuver on 25 June 1998 resulted in a long data gap until recovery on 22 October 1998, and ii) the failure of the gyroscopes caused another gap from 21 December 1998 to 6 February 1999 when nominal operation resumed. 
During its first year of operation, the attitude of SoHO was set such that its reference axis was aligned along the sky-projected direction of the solar rotational axis resulting in this direction being ``vertical'' (\ie along the y-axis of the CCD detector) with solar north up on the LASCO images.
Starting on 10 July 2013 and following the failure of the motor steering its antenna, SoHO was periodically (every three months) rolled by 180$\deg$ to maximize telemetry transmission to Earth.
On 29 October 2010 and still on-going, the attitude of SoHO was changed to simplify operation, the reference orientation being fixed to the perpendicular to the ecliptic plane causing the projected direction of the solar rotational axis to oscillate between $\pm$7$\deg$ 15' around the ``vertical'' direction on the LASCO images.

Equipped with a CCD camera and extensively calibrated on the ground and in space, LASCO-C2 has the capability of producing highly accurate absolute photometric measurements (\cite{llebaria2006photometric}, \cite{gardes2013photometric}, \cite{Colaninno2015}).
It further offers a good potential for polarization measurements although the selected technique, a set of three polarizers oriented at $+60\deg$, $0\deg$, $-60\deg$ and mounted on a wheel, is far from being ideal for the corona (Section~\ref{Sub:Implementation}).
C2 was thoroughly characterized in the laboratory so as to determine its Mueller matrix and particular attention was paid to the retardation effect introduced by the two folding mirrors.
Several results relying on the polarization measurements of the corona have been published during the first years of operation of LASCO, for instance on the electron density derived from polarized radiance images $pB$, see \cite{lamy1997electronic}, \cite{llebaria1999global}, \cite{lamy2002solar} and \cite{quemerais2002two}.
However, as data accumulated and as we proceeded with new tests and analysis, in particular the three-dimensional reconstruction of the electron density by time-dependent tomography \citep{vibert2016time}, we realized that the instrumental corrections we had introduced so far were insufficient, thus warranting an in-depth investigation of the question.
\cite{moran2006solar} made a parallel effort, but we emphasize that the effects they have studied were already known to us, and therefore corrected for in our results published since 1999.
Furthermore, they have not reported any quantitative results for the polarization of the corona, so that it is difficult to assess the ultimate validity of their proposed corrections.
We will show in our investigation that they are indeed insufficient and that additional instrumental effects must be corrected for.

Our article is organized as follows. 
First, we introduce the question of the polarization of the white-light corona and we describe the technique implemented on LASCO-C2 and the relevant observations.
The next section presents the method of polarization analysis based on the Mueller formalism, the determination of the Mueller matrices and the calibration of the instruments.
A specific section is devoted to the question of the separation of the K and F components since this represents a major application of the polarization measurements.
The practical implementation of the analysis, the initial results and critical tests then reveals the necessity of additional corrections which are dealt with in a subsequent section.
Final results for the polarization, the polarized radiance $pB$, the electron density, and the radiance of the K-corona over 24 years are presented and subsequently compared to results from solar eclipses whenever available.

\section{Measurements of the Polarization of the Solar Corona}
     \label{Sec:Measurements} 

\subsection{Overview}	\label{Sub:Overview}

\begin{figure}[htpb!]
\begin{center}
	\includegraphics[width=0.5\textwidth]{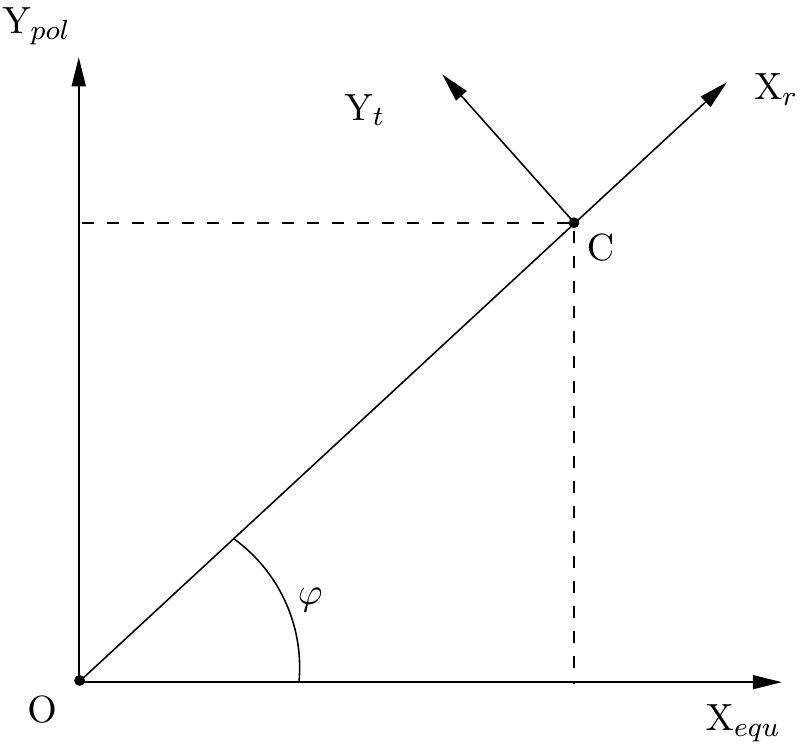}
	\caption{Coordinate system in the plane of the sky with origin O at the center of the Sun and with axis $\vec{OX_{equ}}$ and $\vec{OY_{pol}}$ oriented along the western equatorial and northern polar directions, respectively.}
	\label{PolarSystemCoord}
\end{center}
\end{figure}

The physical processes which are responsible for the polarization of the coronal light, Thomson scattering by the electrons for the K-corona, light scattering by dust particles for the F-corona, result in a state where the polarization is linear and ``tangential''.
The latter property results from the fact that the electric field of the light wave is perpendicular to the scattering plane defined by the observer, the center of the Sun and the scattering element, either electron or dust particle.
For the electrons, this is strictly true when they are at rest or have velocities much less than that of light which is generally the case.
Considering Figure~\ref{PolarSystemCoord}, at any point C of the corona, the polarization is perpendicular to the radius vector going through the center of the Sun, hence along $\vec{CY_{t}}$, that is ``tangential''.

\begin{figure}[htpb!]
	\centering
	\includegraphics[width=0.8\textwidth]{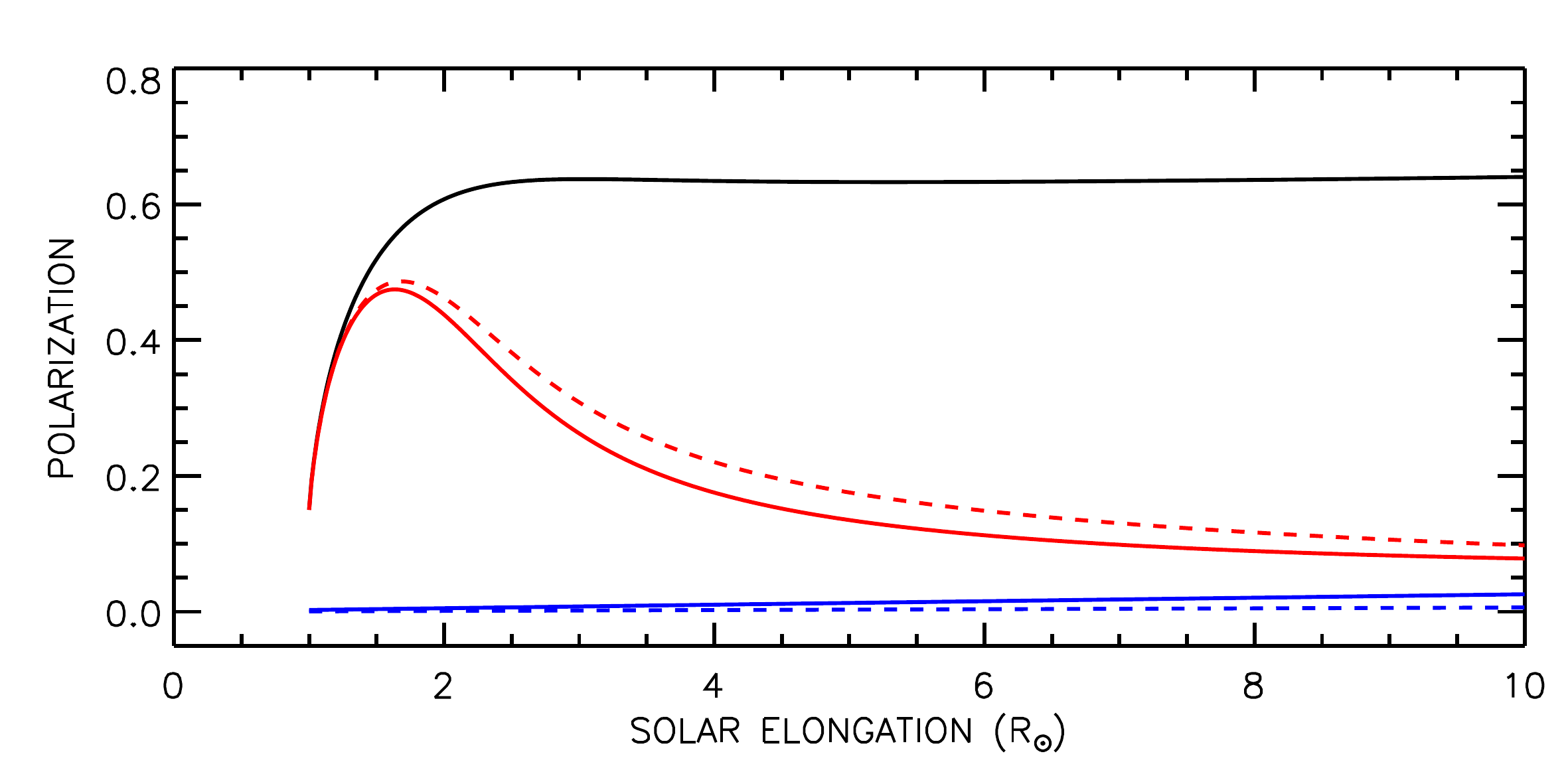}
	\caption{Radial profiles of the polarization of the K (Black line) and F (blue lines) components and of the global corona (red lines) in the equatorial (solid lines) and polar (dashed lines) directions.
	Note that we used a maximum model for the K-corona.}
	\label{FigPkPf}
\end{figure}

The above processes obey different physics so that the radial profiles of the polarization of the K and F coronae are radically different.
Thomson scattering predicts a simple behaviour of the polarization of an electron : it increases with increasing scattering angle to reach a maximum of 1 at $90\deg$ and decreases symmetrically beyond.
The profile of the polarization of an axi-symmetric K-corona $p_K(r)$ is known to have a remarkable property, it rapidly increases as the elongation increases and reaches a nearly constant value of $\approx$0.64 beyond $\approx$2.2 R${}_\odot$.
This is illustrated by the profile displayed in Figure~\ref{FigPkPf} which has been obtained by using the classical model of \cite{baumbach1937strahlung} for the electron density $N_{e}(r)$ of a K-corona of the maximum type and integrating the Thompson scattering along the line-of-sight. 
A deep coronal hole produces essentially the same result:
for instance, using the profile of $N_{e}(r)$ along the north polar direction determined by \cite{fisher1995physical} from eclipse measurements, we found that the polarization of the K-corona remains  between 0.6 and 0.65 in the range 2.5 to 10 R${}_\odot$.
This remarkable property is crucial for the separation of the K and F components as we will see later in Section~\ref{Sec:Separation}.

The F-corona results from diffraction of the solar light by interplanetary dust particles distributed along the line-of-sight and is therefore unpolarized in a first approximation.
The two radial profiles of $p_F$ displayed in Figure~\ref{FigPkPf} for the equatorial and polar directions were constructed by \cite{lamy1986volume} by bridging measurements of the corona as reviewed by \cite{koutchmy1985f} and those of the inner zodiacal light of \cite{Leinert1982}. 
It should be realized that, whereas the Thomson formalism unambiguously describes the polarization of electrons, the situation is far more complex for interplanetary dust particles. 
On the one hand, many poorly known physical parameters come into play (composition, shape and size distribution of the dust grains) and the only general theory, the Mie scattering theory, is strictly valid for spherical particles on the other hand.

The global polarization, that is the quantity $p$ which is accessible to the observer, weights the individual polarization profiles by those of the respective radiance profiles $B_K(r)$ and $B_F(r)$ according to the following equation:
\begin{equation}
p=\frac{p_K\;B_K+p_F\;B_F}{B_K+B_F}
\end{equation}
Consequently, $p$ is controlled by the K-corona in the inner part and by the F-corona in the outer part.

\subsection{Implementation for LASCO-C2 and Observations}	
\label{Sub:Implementation}

Owing to the properties of the polarization of the corona (linear and tangential), the most appropriate and efficient way to determine it would consist in measuring the polarized radiance with two analyzers respectively oriented along the radial and tangential directions at each point (or spatial element) of the corona.
As an historical note, such an axially symmetric analyzer appears to have been used for the first time by M. Waldmeier at the eclipse of 1954.
In its most elaborated form, two exchangeable polarizer wheels with their center aligned with the center of the solar disk, each one divided in 12 sectors with individual polarizers oriented either along the radial (wheel \number1) or tangential (wheel \number2) directions are implemented, and the wheels are rotated during the exposures to homogenize the transmission \citep{koutchmy1971observations}.
The slight error resulting from the finite angular extent of the sectors can be estimated and corrected for.
This however results in a complex mechanical system which could not be considered for the LASCO coronagraph because of limited resources. 
We were compelled to implement the most simple solution, similar to that of the SMM Coronagraph/Polarimeter \citep{macqueen1980high}, which consists in using three identical linear polarizers with orientations at +60${}^o$, 0${}^o$ and -60${}^o$ with respect to the $\vec{OX_{equ}}$ direction and mounted on a wheel. 
This wheel has two additional slots, one with a ``clear'' window for measuring the total radiance $B$ of the corona and the other with a neutral density filter for calibration purpose. 
All polarizers were manufactured by the Meadowlark company (Frederick, Colorado, USA): disks of dichroic Polaroid foil were cut out of the same sheet (Kodak HN22), cemented between polished glass plates using an index of refraction matching that of the cement and mounted in aluminum barrels. 
Attention was paid to the mechanical fixation of the barrels on the polarizer wheel in order to avoid any strain which would have affected the performances of the polarizers.
Whereas the above method based on three polarizers is quite simple to implement, it is far from being satisfactory from the point of view of the quality of the measurements. 
It can be shown that the accuracy on the polarization and its angle is not isotropic and that is it depends upon the position angle of the considered point in the corona \citep{lazarides1992etude}.
It will further be shown that it is affected by additional problems probably related to the nature of the Polaroid foil.

\begin{figure}[htpb!]
	\centering
	\includegraphics[width=\textwidth]{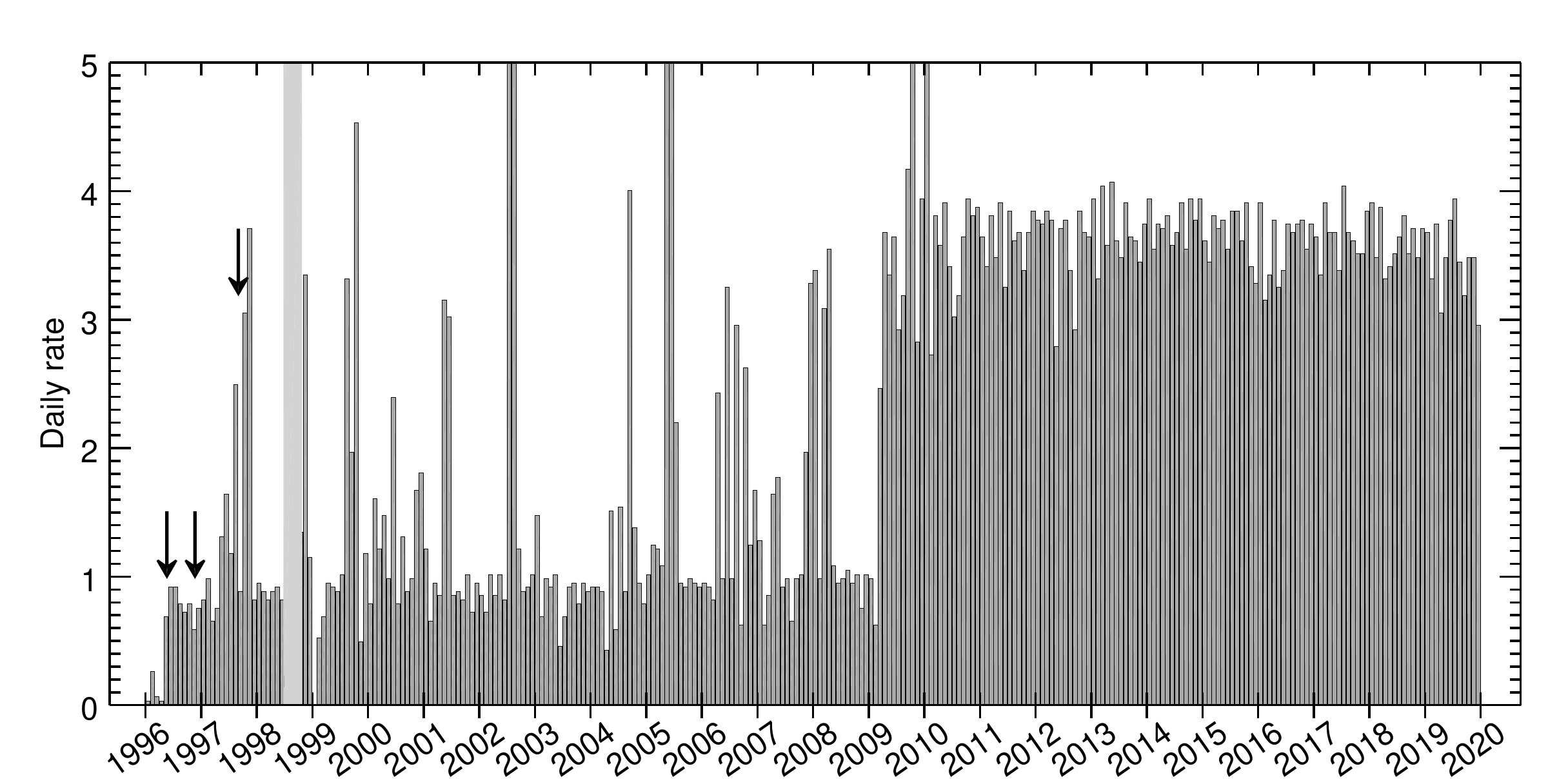}
	\caption{Monthly averaged daily rate of the polarization sequences obtained with the LASCO-C2 coronagraph with the orange filter in the binned format of 512$\times$512 pixels.
		See text for the out-of-scale values of the rate.}
	The arrows indicate the roll sequences of the spacecraft used for calibration. 
	\label{FigImRate}
\end{figure}

A polarization sequence is composed of three polarized images of the corona obtained with the three polarizers and an unpolarized image forming altogether a quadruplet, prominently taken with the orange filter (bandpass of 540-640~nm) in the binned format of 512$\times$512 pixels in order to improve the signal-over-noise ratio of the polarized images (a few sequences were taken in the full format of 1024$\times$1024 pixels).
Figure~\ref{FigImRate} displays the chronogram of the polarization sequences, typically one per day until 2008 inclusive and four per day thereafter when additional telemetry became available following the decommissioning of several SoHO instruments.
Special polarization took place a few times as shown by the peaks in the chronogram.
The most notable were performed during the following time intervals: 6 July to 13 August 2002 (849 sequences), 28 May to 5 June 2005 (530 sequences), and 11 to 18 January 2010 (253 sequences).
Sequences with the blue and red filters were occasionally taken but have not been processed due to the lack of proper calibration.

\section{Analysis of the LASCO Polarized Images}

Among the different methods available to analyze the state of polarization of a given optical system, the formalism of \cite{mueller1943memorandum} conveniently handles measurable physically quantities and is well adapted to the case of partially polarized light. 
The state of polarization is fully determined by the four Stokes parameters I, Q, U, V regrouped in a vector forming a column ($1\times4$) matrix. 
The $4\times4$ Mueller matrix $\vec{M}$ characterizes an optical system by relating the Stockes parameters of the incident beam $\vec{S_{in}}$ to those of the output beam $\vec{S_{out}}$ via:
\begin{equation}
\vec{S_{out}}= \vec{M}\;\vec{S_{in}}
\end{equation}
In the most general case, the sixteen coefficients $m_{ij}$ of the matrix $\vec{M}$  must be determined by sixteen independent measurements. 
We now show that, in the case of the corona, the problem can be solved by determining only three coefficients for each of the three configurations corresponding to the three polarizers, that is a total of nine coefficients.

At a point C of the corona (Figure~\ref{PolarSystemCoord}), let $\vec{CX_{r}}$ and $\vec{CY_{t}}$ be the radial and tangential directions.
As discussed above, the polarization of the coronal light is linear, hence $V=0$, and tangential, \ie along $\vec{CY_{t}}$. 
Let $\vec{S_{c}}=[I_{c},Q_{c},U_{c},0]^{-1}$, $p_{c}$, and $\alpha_{c}$ be respectively the Stokes vector of the coronal light, its polarization and the direction of polarization in the (C,$\vec{X_{r}}$,$\vec{X_{t}}$) reference frame. 
It is well know that:
\begin{equation}
p_{c} = \sqrt{\frac{Q_{c}^{2}+U_{c}^{2}}{I_{c}^{2}}} \\
\end{equation}
\begin{equation}
\tan 2\;\alpha_{c} = \frac{U_{c}}{Q_{c}} .
\end{equation}

Note that we will determine $\alpha_{c}$ and compare it to its theoretical values $\alpha_{c}=90\deg$ as a test of the quality of our measurements. 
For the corona and using the classical notation B$_{r}$ and B$_{t}$ for the radiances in respectively the radial and tangential directions:
\begin{eqnarray}
I_{c} & = & B_{t} + B_{r} \nonumber \\
Q_{c} & = & B_{t} - B_{r} \\
U_{c} & = & 0 \nonumber
\end{eqnarray}

We must however work in a fixed coordinate system and we naturally choose that defined by (O, $\vec{X_{equ}}$, $\vec{Y_{pol}}$). 
By construction, the principal axis of maximum transmittance of the polarizer oriented at $0\deg$, and the direction of the rows of the CCD detector are closely aligned with $\vec{OX_{equ}}$; accordingly the direction of the columns of the CCD corresponds to $\vec{Y_{pol}}$. 
The Stokes vector $\vec{S_{cp}}=[I_{cp},Q_{cp},U_{cp},0]^{-1}$ expressed in the (O, X$_{equ}$, Y$_{pol}$) coordinate system is related to $\vec{S_{c}}$ by a rotation of angle $-\varphi$ around the direction of propagation and we have: 
\begin{eqnarray}
I_{c} & = & I_{cp} \nonumber \\
Q_{c} & = & Q_{cp}\;\cos 2\varphi - U_{cp}\;\sin 2\varphi \\
U_{c} & = & Q_{cp}\;\sin 2\varphi + U_{cp}\;\cos 2\varphi \nonumber
\end{eqnarray}
and therefore:
\begin{equation}
p_{cp} = \sqrt{\frac{Q_{cp}^{2}+U_{cp}^{2}}{I_{cp}^{2}}} \nonumber\\
\end{equation}
\begin{equation}
\tan 2\;(\alpha_{c}+\varphi) = \frac{U_{cp}}{Q_{cp}}
\end{equation}
It can be readily checked that $p_{c}=p_{cp}$ so that the total radiance and its polarization are independent of the coordinate system as they should, whereas this is obviously not the case of the direction of polarization.

Let $\vec{S_{p}}$ be the Stockes vector of the coronal light exiting the coronograph characterized by its  Mueller matrix $\vec{M}$, both expressed in the same fixed coordinate system (O, X$_{equ}$, Y$_{pol}$). 
We thus have:
\begin{equation}
\vec{S_{p}}=\vec{M}\;\vec{S_{cp}}
\end{equation}
and in particular for the total intensity
\begin{equation}
I_{p}=m_{11}\;I_{cp}+m_{12}\;Q_{cp}+m_{13}\;U_{cp}
\end{equation}
When using the three analyzing polarizers successively, we secure three images of the polarized radiance of the corona such that we have at each pixel:
\begin{eqnarray}
\nonumber I_{1} & = & m_{11}(0)\;I_{cp} +m_{12}(0)\;Q_{cp}+m_{13}(0)\;U_{cp} \\ 
I_{2} & = & m_{11}(-60)\;I_{cp} +m_{12}(-60)\;Q_{cp}+m_{13}(-60)\;U_{cp} \\
 \nonumber I_{3} & = & m_{11}(+60)\;I_{cp} +m_{12}(+60)\;Q_{cp}+m_{13}(+60)\;U_{cp}
\end{eqnarray}
where $0$, $-60$, and $+60$ correspond to the orientation of the three polarizers.
Let us introduce the so-called IPMV (Intensity Polarization Modification Vector) matrix:
\begin{equation}
\vec{\chi}=\left( \begin{array}{ccc}
m_{11}(0) & m_{12}(0) &  m_{13}(0)\\
m_{11}(-60) & m_{12}(-60) & m_{13}(-60) \\
m_{11}(+60) & m_{12}(+60) & m_{13}(+60)
\end{array} \right)
\end{equation}
The problem simplifies to inverting $\vec{\chi}$ so as to determine the Stockes vector $\vec{S_{cp}}$ via:
\begin{equation}
\left( \begin{array}{c}
I_{cp} \\ Q_{cp} \\ U_{cp}
\end{array} \right) =
\chi^{-1}\left( \begin{array}{c}
I_{1} \\ I_{2} \\ I_{3}
\end{array} \right)
\end{equation}
In the case of a perfect optical system, we would have:
\begin{eqnarray}
I_{cp} & = &  2\;(I_{1}+I_{2}+I_{3})/3 \nonumber \\
Q_{cp} & = &  2\;(I_{1}-I_{2}-I_{3})/3 \\
U_{cp} & = &  2\;(I_{2}-I_{3})/\sqrt{3} \nonumber
\end{eqnarray}
Up to now, it is implicitly assumed that all quantities are expressed in absolute unit of radiance and therefore the Mueller and the IPMV matrices must have absolute, dimensionless coefficients which correctly relate radiances. 
Their determination requires that both the input and output Stockes vectors be expressed in absolute unit of radiance such as $W\,m^{-2}\,st^{-1}\,\mu{}m^{-1}$, a very challenging calibration task further complicated by the vignetting inherent to externally occulted coronographs (see further detail below). 
We explain in the next section that it is in practice possible to work with relative coefficients and perform at the end, a global calibration of the total radiance.

\section{Determination of the Mueller Matrix}
\label{MuellerMatrix}

One key advantage of the Mueller formalism is that the global matrix of an instrument is equal to the product of its individual components. 
Therefore, two methods to determine the Mueller matrix of an instrument are possible, either by isolating each optical component for which the matrix is known or globally for the whole instrument.

\subsection{Component calibration}
\label{ComponentCalib}

A linear polarizer whose principal axis of transmittance makes an angle $\theta$ with respect to the $\vec{CX_{equ}}$ direction is characterized by a Mueller matrix:
\begin{equation}
{\scriptsize
\vec{M_{P}(\theta)}=\frac{k_{1}}{2}\;
\left[ \begin{array}{cccc}
1+\varepsilon & (1-\varepsilon)\cos 2\theta &  (1-\varepsilon)\sin 2\theta & 0 \\
(1-\varepsilon)\;\cos 2\theta & (1-\sqrt{\varepsilon})^{2}\;\cos^{2} 2\theta+2\sqrt{\varepsilon} & (1-\sqrt{\varepsilon})^{2}\;\cos 2\theta\;\sin 2\theta & 0 \\
(1-\varepsilon)\;\sin 2\theta & (1-\sqrt{\varepsilon})^{2}\;\cos 2\theta\;\sin 2\theta & (1-\sqrt{\varepsilon})^{2}\;\sin^{2} 2\theta+2\sqrt{\varepsilon} & 0 \\
0 & 0 & 0 & 2\sqrt{\varepsilon}
\end{array} \right]}
\end{equation}
\noindent where $\varepsilon$ is the ratio of the principal transmittances of the polarizer k$_{1}$ and k$_{2}$: $\varepsilon=k_{1}/k_{2}$. 
The three Mueller matrices characterizing the three linear polarizers in a given spectral band are therefore given by setting $\theta$ equal to the three angles $0\deg$, $-60\deg$, and $+60\deg$.

LASCO-C2 further incorporates two identical folding mirrors. 
Although they have been coated to minimize their polarization, they both work at an unfavorable incidence angle of $45\deg$ and therefore their Mueller matrices must be introduced. 
In the chosen reference frame and owing to their geometry, the two mirrors have identical matrices:
\begin{equation}
\vec{M_{m}}=\frac{1}{2}\;
\left[ \begin{array}{cccc}
R_{p}+R_{s} & R_{p}-R_{s} & 0 & 0 \\
R_{p}-R_{s} & R_{p}+R_{s} & 0 & 0 \\
0 & 0 & 2\sqrt{R_{p}.R_{s}}\;\cos\delta & 2\sqrt{R_{p}.R_{s}}\;\sin\delta \\
0 & 0 & -2\sqrt{R_{p}.R_{s}}\;\sin\delta & 2\sqrt{R_{p}.R_{s}}\;\cos\delta
\end{array} \right]
\end{equation}
\noindent where R$_{p}$ and R$_{s}$ are the reflectances of the components respectively parallel and perpendicular to the incidence plane, and where $\delta$ is the phase difference between these two components. 
So in the case of LASCO-C2,
\begin{equation}
\vec{M_{C2}}=\vec{M_{P}(\theta)}\;\vec{M_{m}}\;\vec{M_{m}}
\end{equation}

\begin{figure}[htpb!]
	\centering
	\includegraphics{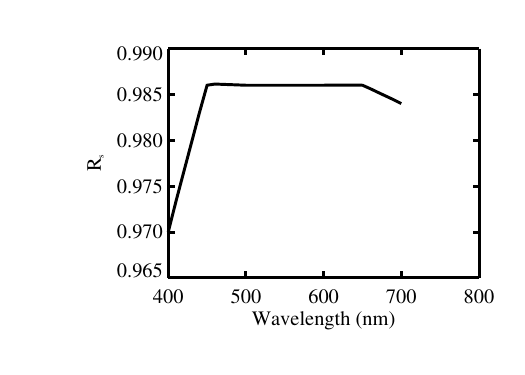}
	\includegraphics{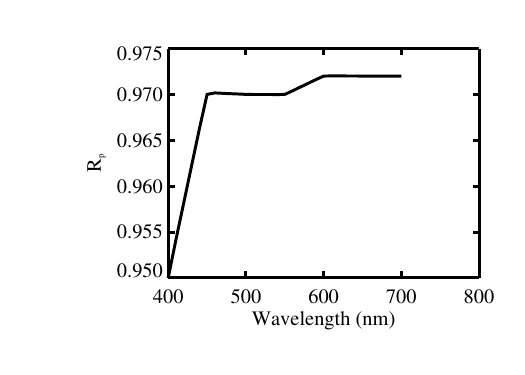}
	\includegraphics{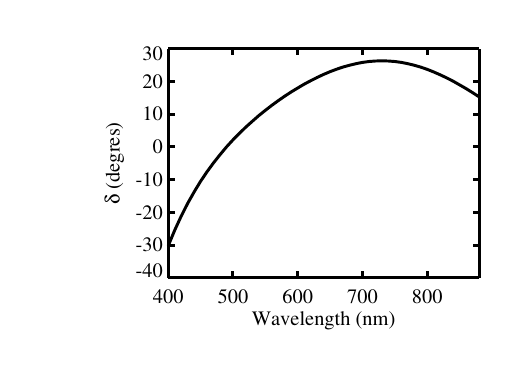}
	\caption{Spectral variations of the reflectances $R_{s}$ and $R_{p}$ (upper panels) and of the phase difference $\delta$ (lower panel) of a LASCO-C2 folding mirror at an incidence angle of $45\deg$.}
	\label{MirrorDelta}
\end{figure}

Ellipsometry measurements of a spare mirror were performed by CMO-LETI (Grenoble, France) and the spectral variations of $R_{p}$, $R_{s}$, and $\delta$ at an incidence angle of $45\deg$ are displayed in Figure~\ref{MirrorDelta}.

The field of view of LASCO-C2 implies that light rays deviate by at most $\pm2\deg$ from the nominal incident angle. 
Additional measurements have therefore been performed at $43\deg$ and $47\deg$ but the resulting differences are negligibly small. 
This is fortunate, otherwise it would have been necessary to consider a Mueller matrix for each pixel of the CCD detector, or at least, for groups of pixels.

\begin{figure}[htpb!]
	\centering
	\includegraphics{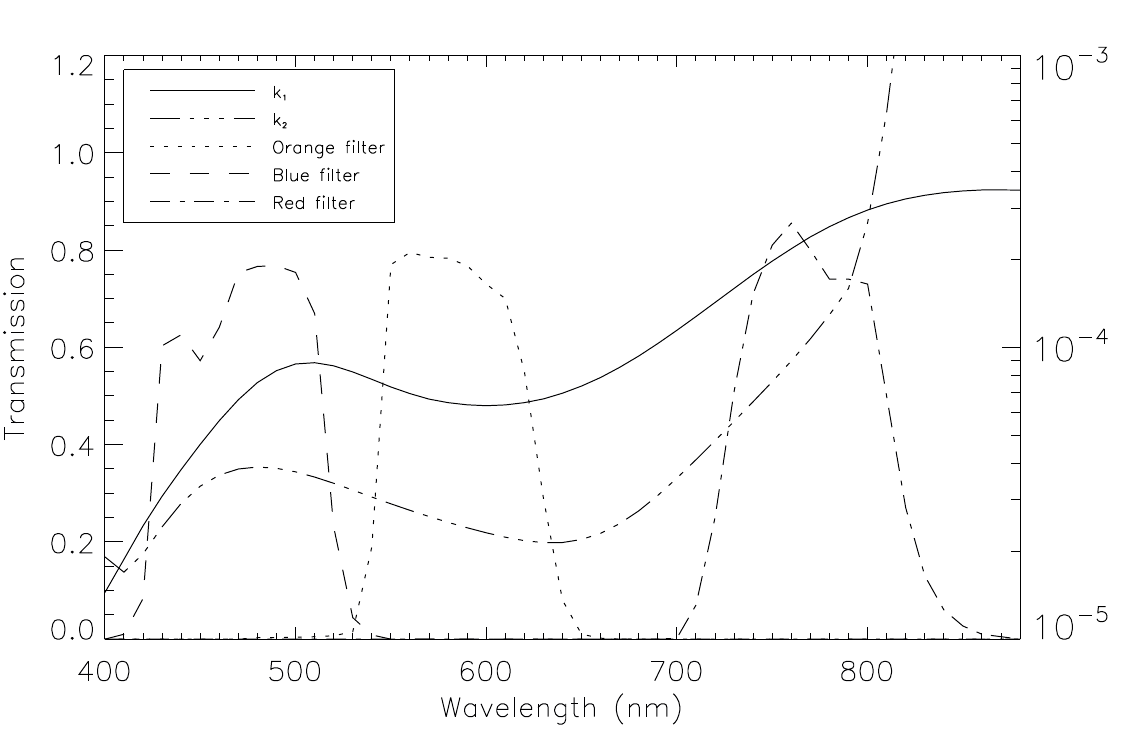}
	\caption{The spectral principal transmittance of the LASCO-C2 polarizers $k_{1}$ (left scale) and $k_{2}$ (right scale) together with the spectral transmittance of the blue, orange and red filters (left scale).}
	\label{Transmissions}
\end{figure}

Figure~\ref{Transmissions} displays the spectral variations of the principal transmittances $k_{1}$ and $k_{2}$ together with the transmission profiles of the LASCO "blue", "orange" and "red" filters.
These filters are sufficiently broad that the spectral variations of the properties of the polarizers and of the mirrors must be taken into account.
Therefore the Mueller coefficients were averaged by considering the transmissions of the optics $T_{0}(\lambda)$, of the filters $T_{f}(\lambda)$, the quantum efficiency of the CCD $\eta(\lambda)$, and the spectrum of the coronal light which is nearly similar to that of the Sun $B_{\odot}(\lambda)$:
\begin{equation}
\overline{m_{ij}}=\frac{\int_{\lambda_{1}}^{\lambda_{2}}\;m_{ij}\;T_{0}(\lambda)\;T_{f}(\lambda)\;\eta(\lambda)\;B_{\odot}(\lambda)\;d\lambda}{\int_{\lambda_{1}}^{\lambda_{2}}\;T_{0}(\lambda)\;T_{f}(\lambda)\;\eta(\lambda)\;B_{\odot}(\lambda)\;d\lambda}
\end{equation}
where the units of $B_{\odot}(\lambda)$ must involve photons (\eg $photon\,sec^{-1}\,cm^{-1}\,st^{-1}$).
Note that we neglect the slight reddening of the F-corona.
Table~\ref{table:mcoeff} displays the first three coefficients of the Mueller matrix for the three filters and the three polarizers of LASCO-C2.

\begin{figure}[htpb!]
	\centering
	\includegraphics[width=\textwidth]{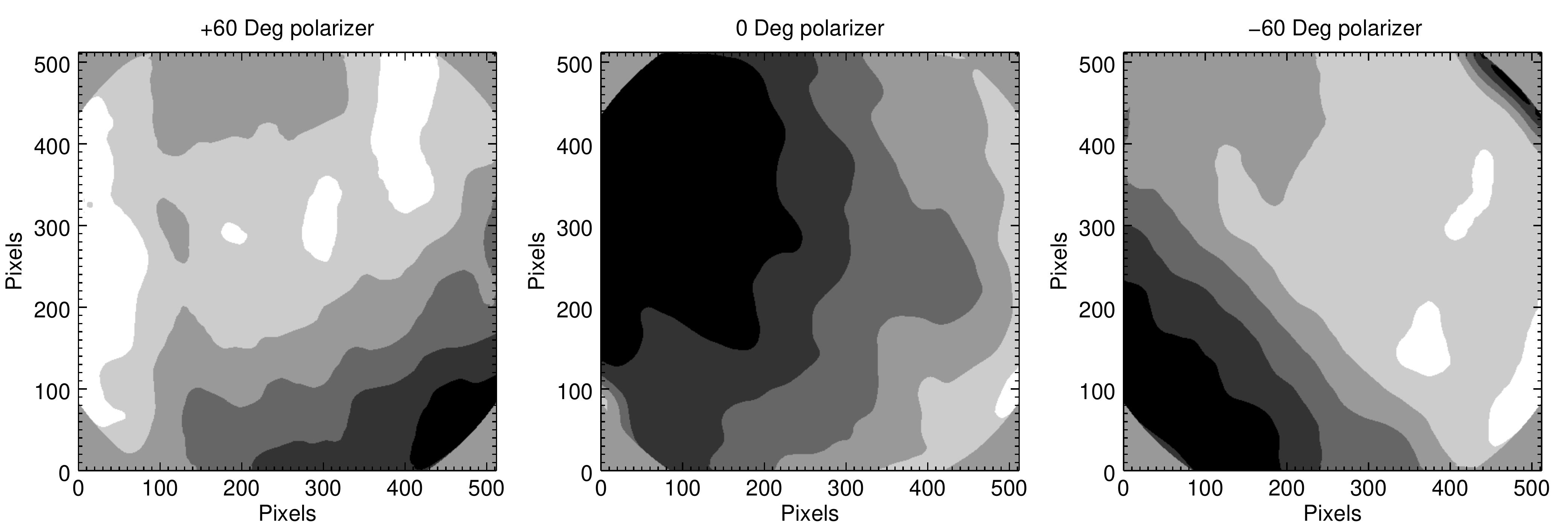}
	\caption{Images of the principal transmittance $k_{1}$ of the three LASCO-C2 flight polarizers in the orange bandpass. 
	The dynamics of the grey levels extends from 1.0 (white) to 1.08 (black).}
	\label{ImageK1C2}
\end{figure}

The principal transmittances of the polarizers may not be uniform over their area.
Therefore, each polarizer was calibrated in the laboratory by illuminating it with uniform light so as to obtain images of $k_{1}$ and $k_{2}$.    
Figure~\ref{ImageK1C2} displays the results for the $k_{1}$ transmittance of the three LASCO-C2 flight polarizers in the orange bandpass.
Here again, we avoided introducing a Mueller matrix for each pixel by directly correcting the polarized images themselves for the non-uniformity of the k$_{1}$ principal transmittance (that of $k_{2}$ are disregarded since $\epsilon$ is much less than 1).
The $k_{1}$ images were normalized by imposing that the mean value over the image $\overline{k_{1}}$ be equal to the value $k_{1}(\lambda)$ averaged over the bandpass of a given filter $T_{f}(\lambda)$ via:

\begin{equation}
\overline{k_{1}}=\frac{\int_{\lambda_{1}}^{\lambda_{2}}\;k_{1}(\lambda)\;T_{f}(\lambda)\;d\lambda}{\int_{\lambda_{1}}^{\lambda_{2}}\;T_{f}(\lambda)\;d\lambda}
\end{equation}

\noindent where $k_{1}(\lambda)$ and $T_{f}(\lambda$) are given in Figure~\ref{Transmissions} and where the integral extends over the bandpass.

\begin{table}
\caption{The first three coefficients of the Mueller matrix for the three filters and the three polarizers of LASCO-C2.}
\begin{tabular}{ccccc} 
\hline
Filter & Polarizer & $\mbox{m}_{11}$ & $\mbox{m}_{12}$ & $\mbox{m}_{13}$\\[3pt]
\hline
Blue & $0\deg$ & 0.244 & 0.244 & 0. \\
Blue & $-60\deg$ & 0.250 & -0.128 & -0.212 \\
Blue & $+60\deg$ & 0.250 & -0.128 & 0.212 \\
Orange & $0\deg$ & 0.233 & 0.233 & 0. \\
Orange & -$60\deg$ & 0.236 & -0.120 & -0.170 \\
Orange & $+60\deg$ & 0.236 & -0.120 & 0.170 \\
Red & $0\deg$ & 0.387 & 0.386 & 0. \\
Red & $-60\deg$ & 0.390 & -0.196 & -0.216  \\
Red & $+60\deg$ & 0.390 & -0.196 & 0.216 \\
\hline
\end{tabular}
\label{table:mcoeff}
\end{table}

\subsection{Global calibration}
\label{GlobalCalib}

The general principle of determining the global Mueller matrix of an imaging optical system consists in illuminating it with uniform light beams of different, known states of  polarization. 
In our case, only three different states of linear polarization are required since we are interested in only the first three coefficients $m_{11}$, $m_{12}$, $m_{13}$.
We used an external reference polarizer successively oriented at angles of $0\deg$, $45\deg$, and $90\deg$ with respect to the $\vec{OX_{equ}}$ direction and illuminated by a ``double opal'' light source of radiance $I_{i}$. 
The corresponding Stockes vectors are:

\begin{eqnarray}
\vec{S_{i}} & = & \frac{1}{2}\;
\left( \begin{array}{c} k_{1}^{r} - k_{2}^{r} \\ k_{1}^{r} - k_{2}^{r} \\ 0 \\ 0 \end{array}
\right)\;I_{i}\,\,\textrm{ for the } \textrm{$0\deg$  orientation} \\
\vec{S_{i}} & = & \frac{1}{2}\;
\left( \begin{array}{c} k_{1}^{r} - k_{2}^{r} \\ 0 \\ k_{2}^{r} - k_{1}^{r} \\ 0 \end{array}
\right)\;I_{i}\,\,\textrm{ for the } \textrm{$45\deg$  orientation} \\
\vec{S_{i}} & = & \frac{1}{2}\;
\left( \begin{array}{c} k_{1}^{r} - k_{2}^{r} \\ k_{2}^{r} - k_{1}^{r} \\ 0 \\ 0 \end{array}
\right)\;I_{i}\,\,\textrm{ for the } \textrm{$90\deg$ orientation}
\end{eqnarray}

\noindent where $k_{1}^{r}$ and $k_{2}^{r}$ are the principal transmittances of the reference polarizer.

Let $I_{A}$, $I_{B}$, and $I_{C}$ be the signals recorded by a given pixel of the CCD detector for the $0\deg$, $45\deg$ and $90\deg$ orientations. 
Solving the system of the three linear equations, we obtain:
\begin{eqnarray}
m_{11}&=&\frac{I_{A}+I_{C}}{k_{1}^{r}\;I_{i}\;(1+\epsilon^{r})} \\
m_{12}&=&\frac{I_{A}-I_{C}}{k_{1}^{r}\;I_{i}\;(1+\epsilon^{r})} \\
m_{13}& =&\frac{I_{A}+I_{C}-2\;I_{B}}{k_{1}^{r}\;I_{i}\;(1+\epsilon^{r})}
\end{eqnarray}
where $\varepsilon^{r}=k_{2}^{r}/k_{1}^{r}$.
In addition, a fourth measurement was performed without the reference polarizer yielding:
\begin{equation}
I_{D}=m_{11}\;I_{i}
\end{equation}
and offering a check of consistency.

For an imaging systems such as the LASCO-C2 coronograph, images $I_{A}$, $I_{B}$, $I_{C}$, and $I_{D}$ are obtained and in turn, maps of the three Mueller coefficients. 
In practice, there are serious difficulties with this calibration method.
\begin{itemize}
\item The determination of the absolute values of the $m_{ij}$ requires a rigorous absolute calibration of the instrument so as to accurately relate the recorded images to the radiance $I_{i}$.
\item The color temperature of the light source (a quartz-iodine lamp with $T_{color}$ $\approx$ 2000K) is substantially different from that of the Sun.
\item Non-uniformities, roughly axially symmetric are present in all images and were traced to a stray reflection by the mirror-polished front face of the external occulter onto the second opal of the light box.
\end{itemize}

To circumvent these problems so as to allow a comparison with the components calibration, we introduced the ratios $m_{12}/m_{11}$ and $m_{13}/m_{11}$. 
It should be underlined that these ratios are indeed relevant quantities as they directly enter the expressions of the polarization and its angle. 
We calculated average values $<I_{A}>$, $<I_{B}>$, and $<I_{C}>$ by taking the mean of the corresponding pixel values, avoiding the stray reflections from the occulter. 
Table \ref{table:mratio} presents these results for the component and global calibrations in the case of the orange filter.
The agreement between the two determinations ranges from excellent ($\sim4\%$), to fair ($\sim30\%$), to poor ($\sim46\%$) without any clear trend. 
In view of the inherent difficulties with the global calibration, we decided to use the Mueller matrix resulting from the component calibration for the pipeline processing of the LASCO-C2 polarized images.

\begin{table}
\caption{Comparison of the ratios of the first three coefficients of the Mueller matrix of the LASCO-C2 coronograph with the orange filter resulting from the component and global calibrations.}
\begin{tabular}{rcccc}
\hline
Polarizer & \multicolumn{2}{c}{$m_{12}/m_{11}$} & \multicolumn{2}{c}{$m_{13}/m_{11}$} \\
\hline
 & CC & GC & CC & GC \\[3pt]
\hline
$0\deg$ & 1 & 0.96 & 0 & 0.15 \\
$-60\deg$ & -0.51 & -0.65 & -0.72 & -0.55 \\
$+60\deg$ & -0.51 & -0.35 & +0.72 & +0.75 \\
\hline
Note : & \multicolumn{4}{l}{CC = Component calibration} \\
 & \multicolumn{4}{l}{GC = Global calibration} \\
\hline
\end{tabular}
\label{table:mratio}
\end{table}

\subsection{Laboratory test}
\label{LabTest}

\begin{figure}[htbp!]
	\centering
	\includegraphics[width=0.5\textwidth]{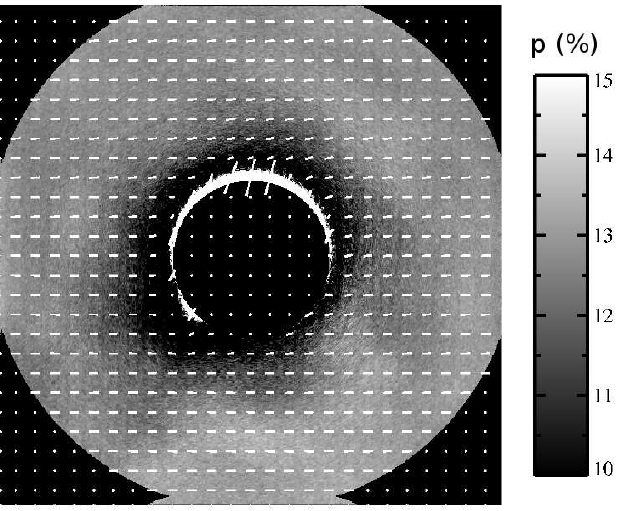}
	\includegraphics{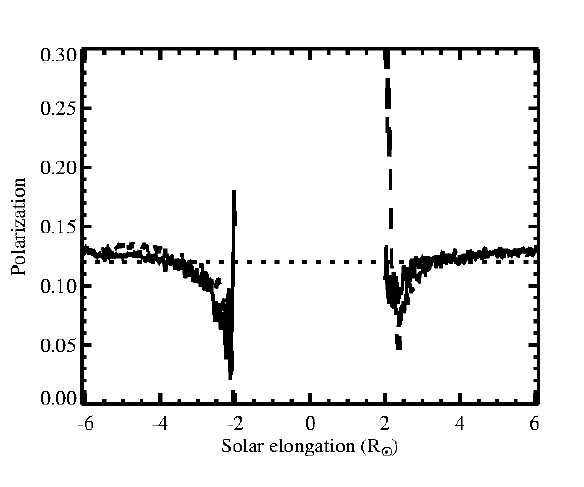}
	\caption{Results of the laboratory test  of LASCO-C2 (orange filter) with a stack of plates producing a linear polarization of 0.12 oriented at $0\deg$.
	Left panel: polarization map with the direction of polarization indicated by white bars.
	Right panel: horizontal and vertical profiles through the center of the  polarization map (note that the pixel scale is converted to solar radius).}
	\label{Polar12}
\end{figure}

A test of the performance of the polarization analysis was performed during the campaign of final verification at the Naval Research Laboratory (Washington, USA).
A uniformly illuminated stack of plates producing a linear polarization of 0.12 was placed in front of the coronagraphs and triplets of polarized images were obtained and processed as described in the above section. 
Figure~\ref{Polar12} illustrates the results obtained in the orange bandpass with the external polarizing device oriented along the $\vec{OX_{equ}}$ direction ($0\deg$). 
The central part suffers from unpolarized light reflected back by the front face of the external occulter, an effect already noted in the session of global calibration. 
Otherwise, the direction of polarization is well recovered with a dispersion of $\sim6.5\deg$ (FWHM) as well as the polarization itself which ranges from 0.12 to 0.13 in the outer circular region not affected by the stray reflection.

\subsection{Calibration of the total radiance}
\label{CalibRadiance}

Our procedure leaves the total intensity $I_{p}=I_{cp}$ uncalibrated. 
This limitation is circumvented by introducing the routine unpolarized images $I_{0}$ systematically taken before or after the three polarized images. 
Those images $I_{0}$ are calibrated in unit of mean solar radiance $\overline{\Bsun}$ as described in \cite{llebaria2006photometric} following a procedure which involves thousands measurements of stars present in the C2 field of view.

\begin{figure}[htbp!]
	\centering
	\includegraphics[width=\textwidth]{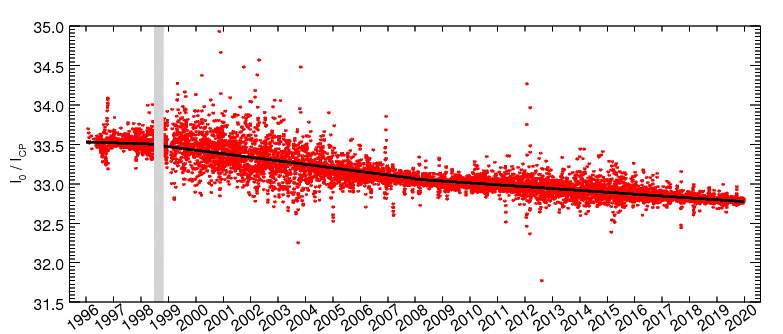}
	\caption{Temporal variations of the ratio $I_{0}/I_{cp}$ of the intensity of the``clear'' unpolarized image to that resulting from the polarization analysis.}
	\label{C2CalCoef}
\end{figure}

For each quadruplet $I_{0}$, $I_{1}$, $I_{2}$, and $I_{3}$, we calculate the mean value of the ratio $I_{0}/I_{cp}$ where $I_{cp}$ results from the polarization analysis of $I_{1}$, $I_{2}$, and $I_{3}$. 
Figure~\ref{C2CalCoef} displays the temporal variations of this ratio and reveals a continuous decrease which reflects the global degradation of the transmittance of the three polarizers which amounts 
to a modest 1.8\% over 20 years.
Combining the calibrations of $I_{0}$ and of $I_{0}/I_{cp}$ allows calibrating the polarized radiance $pB$ of the corona as observed by LASCO-C2 in units of $\overline{\Bsun}$.

\begin{figure}[htbp!]
	\centering
	\includegraphics[width=\textwidth]{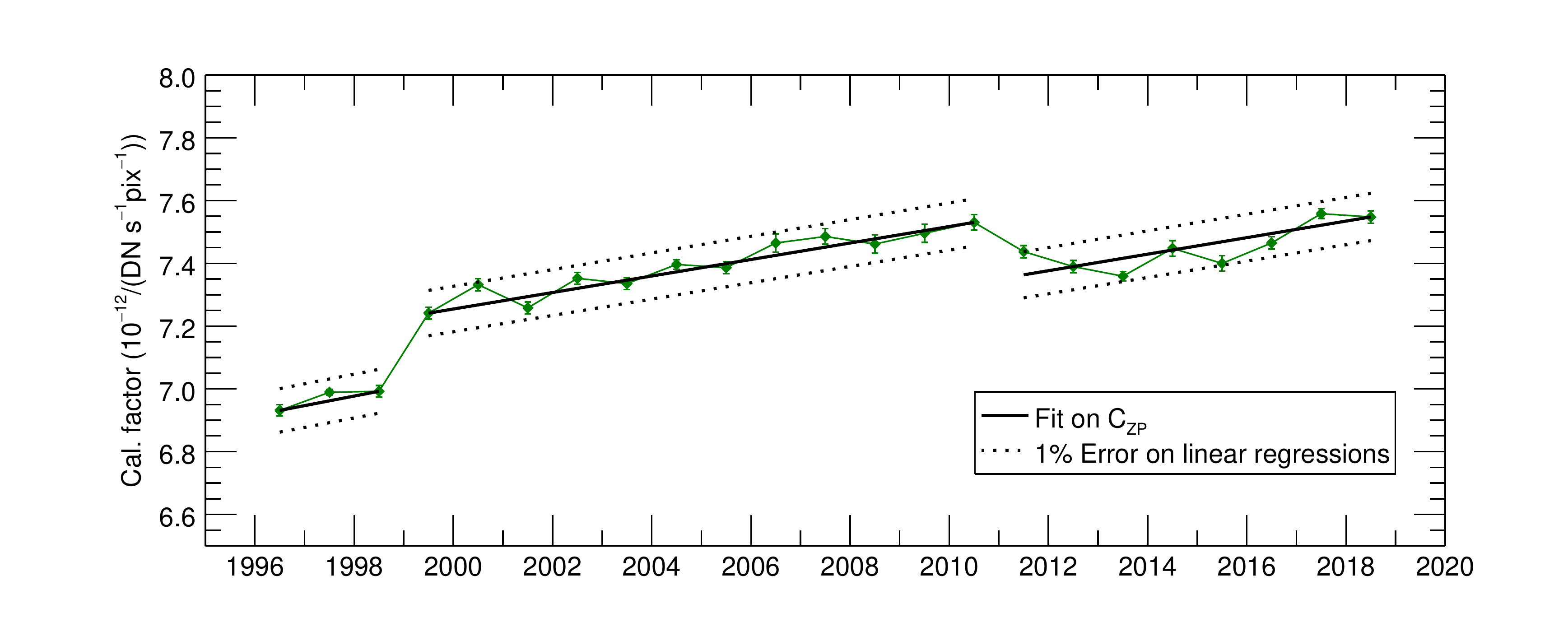}
	\caption{Temporal variation of the LASCO-C2 calibration factor for the orange filter in units of $10^{-12}\overline{\Bsun}/(DN\,sec^{-1}\,pix^{-1})$ deduced from our photometric measurements of stars.}
	\label{Cal_Factor}
\end{figure}

The LASCO-C2 calibration factor was first published by \cite{llebaria2006photometric} and  extended by \cite{gardes2013photometric}.
\cite{Colaninno2015} have later presented an independent determination confirming our results.
In the framework of this present study, we ran our specific calibration processing over 24 years of LASCO-C2 data to produce an homogeneous determination of the temporal variation of its calibration factor as displayed in Figure~\ref{Cal_Factor}.
The linear fit suggests three regimes separated by two small jumps in opposite directions.
The first one in 1999, undoubtedly a consequence of several months of ``hibernation'' when SoHO lost its pointing, corresponds to a decrease of sensitivity of 3.5\%.
The second one in 2010 indicates a surprising increase of sensitivity of $\approx$1.3\% which probably has its origin in the electronics of the instrument.
Inside each of the three regimes, the deviations of the measurements from the linear fits do not exceed 1\%.
The continuous decline of the sensitivity in the interval [1999 -- 2011] offers a good assessment of the evolution of C2; it amounts to a mere 0.3\% per year, a quite remarkable performance.

\section{Separation of the K and F Coronae}
	\label{Sec:Separation}

A real coronagraph suffers from instrumental stray light so that the observed radiance $B$ amounts to:
\begin{equation}
\label{EquB}
B = B_K+B_F+B_{S}
\end{equation}
The stray light $B_{S}$ mostly results from light diffracted by the various occulters, apertures and stops, and is therefore axially symmetric (except for a narrow sector corresponding to the pylon holding the occulters) and unpolarized ($p_{S}=0$) to first order. 
The C2 images display additional, faint stray structures such as arcs which do not seriously affect the K/F analysis. 

The observed polarized radiance $pB$ is therefore given by:
\begin{equation}
\label{EqupB}
pB = p_K\;B_K+p_F\;B_F
\end{equation}
showing that, in its most general form, the problem of separating $B_K$ and $B_F$ using Equations \ref{EquB} and \ref{EqupB} is intractable.
Fortunately and as well known, the respective radial variations of the four terms $p_K$, $p_F$, $B_K$, and $B_F$ very much help and allow to solve the problem.

Observations and simulations using models such as introduced in Section \ref{Sub:Overview} show that for $r\lesssim 8\;\Rsun$, the inequality:
\begin{equation}
\label{Inequ}
p_F\;B_F\ll p_K\;B_K
\end{equation}
is satisfied. 
Further making the classical assumption $p_F=0$ allows to strictly write:
\begin{equation}
pB=p_K\;B_K
\end{equation}
At this point, two routes are possible to obtain $B_K$.

The first route consists in assuming a model of $p_K$ such as given in Figure~\ref{FigPkPf}, and calculating $B_K$ according to
\begin{equation}
\label{B_K_ratio}
B_K=pB/p_K
\end{equation}
This is justified by the robust "asymptotic" behaviour of $p_K(r)$ beyond $\approx$ 2.2$\Rsun$, which is almost independent of the coronal electron density profiles (Figure~\ref{FigPkPf}).

The second route consists in inverting the integral equation: 
\begin{equation}
\label{pB_Int}
pB=\int_{los}{\sigma_{1,e}\;N_{e}\;dl}
\end{equation}
so as to retrieve the electron density $N_{e}$ and then calculating $B_K$ via the integral:
\begin{equation}
B_K=\int_{los}{\sigma_{2,e}\;N_{e}\;dl}
\end{equation}
where $\sigma_{1,e}$ and $\sigma_{2,e}$ are the relevant electron cross-sections for Thomson scattering, and ``$los$'' stands for line of sight.
Such a method or close variant versions have been implemented in the past, for instance by \cite{Kluber1958intensities}, \cite{Munro1977}, \cite{saito1977study}, and \cite{durst1982two}, but limited to a few radial directions, most often equatorial and polar.
\cite{quemerais2002two} have developed a full two-dimensional inversion that they have applied to LACO-C2 images and they have shown that the two routes produce consistent results for $B_K$. 
However additional assumptions must be introduced, notably the symmetry, either spherical or cylindrical, of the electron density.

Ultimately, a ``mixed'' route may even be considered where, instead of using a model of $p_K$, it is calculated from $N_e$ which itself comes from the inversion of the $pB$ integral as given by Equation \ref{pB_Int} \citep{lamy1997electronic}.
Then $B_K$ is obtained from Equation \ref{B_K_ratio}.

The elongation at which the inequality \ref{Inequ} no longer holds very much depends upon the relative behaviours of $p_K$, $B_k$, $p_F$, and $B_F$ as a function of solar elongation, but also latitude (in a broad stroke, equatorial or polar regions) and the level of solar activity. 
Simulations with models such as given in Figure~\ref{FigPkPf} show that $B_K$ is correctly retrieved up to $\sim7\Rsun$ in the most unfavorable situations. 
This insures that the above procedures always apply to the C2 images.

The question of deriving $B_F$ and $B_S$, that is separating these two unpolarized components is beyond the scope of the present article and will be dealt in a separate article. 

\section{Implementation, First Results and Critical Tests}
	\label{Implementation}

\subsection{Implementation}
The original data stream coming from the spacecraft represents the lowest level data, known as Level-0. 
Once received at the Naval Research Laboratory, this Level-0 data is processed into FITS files of individual images with documented headers and forms the Level-0.5 data set.
No corrections are applied at this stage.
Level-0.5 images are then distributed to the participating institutes for processing, calibration, and analysis.
However, the process experienced a considerable slown-down in 2015 to a point of accumulating a delay of one year.
As a consequence, we decided in October 2015 to definitively use the ``quick look data'' produced by the Goddard Space Flight Center instead of the Level-0.5 data.
Strictly speaking, these two data sets are identical except for slightly less missing telemetry blocks in the Level-0.5 data.

The LASCO team at the Laboratoire d'Astrophysique de Marseille (formerly Laboratoire d'Astronomie Spatiale) has developed a two-stage procedure which corrects for all instrumental effects and process the raw data to calibrated physical images of the corona.
The in-flight performances of C2 are continuously monitored so as to update these corrections as well as the absolute calibration.

First, a preprocessing is applied to all images and performs the following tasks.
\begin{itemize}
\item Bias correction.
The bias level of the CCD detector evolves with time; it is continuously monitored using specific blind zones, and systematically subtracted from the images.
\item Exposure time equalization.
Small random errors in the exposure times are corrected using a method developed by \citet{llebaria2001highly} in which relative and absolute correction factors are determined.
This method works extremely well for the routine (unpolarized) images because of their high cadence but less so for the less frequent polarized images, especially during the first 14 years when only daily polarization sequences were taken. 
The Naval Research Laboratory later developed an alternative method with similar performances \citep{morrill2006calibration}.
\item Missing block correction.
Telemetry losses result in blocks of $32 \times 32$ pixels sometime missing in the images.
Different solutions are implemented to restore the missing signal depending upon the location of these blocks \citep{pagot2014automated}.
\item Cosmic rays correction.
The impacts of cosmic rays (and stars as well) are eliminated from the images using the procedure of opening by morphological reconstruction developed by \citet{pagot2014automated}.
\end{itemize}

The polarized images further undergo the following processes.
\begin{itemize}	
	\item   
	Rebinning to 512$\times$512 pixels. This practically applies to the few 1024$\times$1024 pixels images to bring them to a common format for polarization analysis. 
	\item
	Correction for the transmission of the polarizers using images of the $k_{1}$ coefficient.
	\item
	Polarimetric analysis based on the Mueller procedure.
	It is applied to each triplet of polarized images and returns images of the total radiance, the polarization and the angle of polarization.	
	\item Vignetting correction.
This instrumental effect is removed from the radiance images using a geometric model of the 2-dimensional vignetting function of C2 \citep{llebaria2004lessons}.
	\item
	Absolute calibration of the total radiance derived from the photometric measurements of thousand of observations of stars present in the C2 field of view \citep{llebaria2006photometric,gardes2013photometric} and updated in Figure~\ref{Cal_Factor}.	
	\item
	K/F separation.
\end{itemize}

\subsection{First Results}
\begin{figure}[htbp!]
	\centering
	\includegraphics{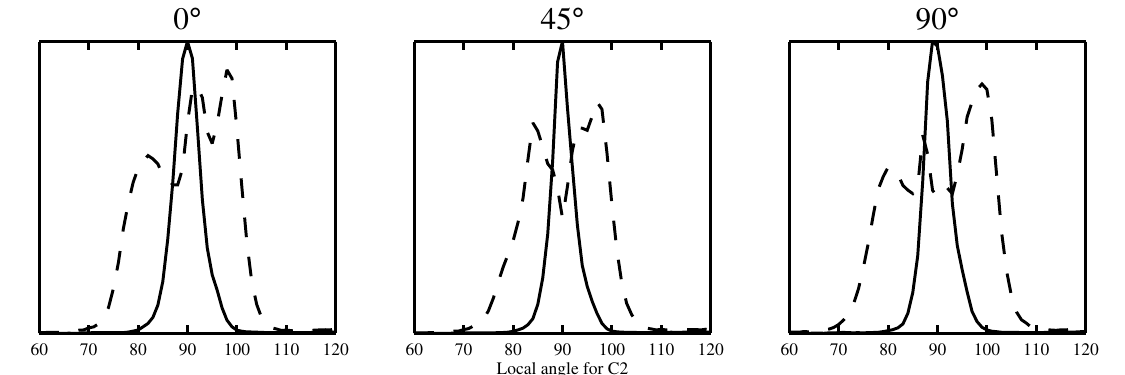}
	\caption{Histograms of the local angle of polarization calculated from images obtained during the September 1997 roll sequence.
	The three panels correspond to the three roll angles, $0\deg$, $45\deg$ and $90\deg$ as indicated.
	Dashed lines: initial results. Solid lines: results after correcting for the global transmittance of the polarizers. 
	The histograms have been arbitrarily scaled for better legibility.}
	\label{FigHistC2}
\end{figure}

The polarimetric analysis and K/F separation were first systematically performed on the whole set of polarized images acquired over almost ten years of LASCO operation. 
As part of our program of validation, we performed several tests which soon revealed various anomalies. 
A first problem was noted with the distributions of the local angle of polarization: whereas correctly centered at $90\deg$, it was broader than expected (Figure~\ref{FigHistC2}).

A second problem affected the K/F separation best seen on synoptic maps which revealed that K-corona structures, prominently streamers, were conspicuously visible in the F-corona especially during the periods of high activity.

\begin{figure}[htbp!]
	\centering
	\includegraphics[width=\textwidth]{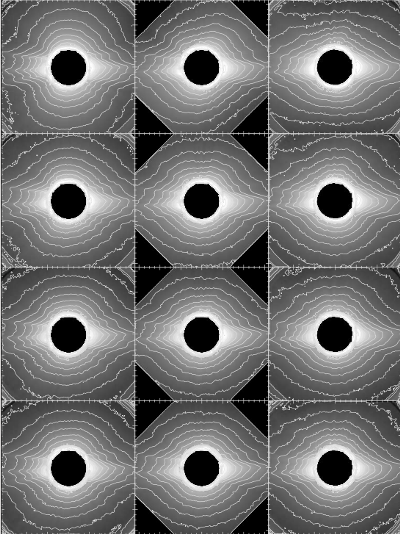}
	\caption{LASCO-C2 images of the radiance $B_K$ of the K-corona calculated from polarization sequences secured during the SoHO roll sequence of September 1997. 
	The three columns correspond to roll angles of $0\deg$ (left), $45\deg$ (middle), and $90\deg$ (right). 
	Upper row : basic polarization analysis with no corrections. 
	Second row : the correction for the global transmission of the $0\deg$ polarizer has been introduced.
	Third row : the correction $S(x,y)$ has been further introduced.
	Lower row: the optimized procedure for exposure time equalization has been introduced.}
	\label{FigBkC2}
\end{figure}

The most crucial test was offered by the shape of the K-corona derived from the observations secured during roll sequences: the SoHO spacecraft was rotated around its axis pointed to the Sun and allowed to dwell at specified roll angles (see detail in Table \ref{table:rollsequence}). 
The most useful one was that of September 1997 since it included a sequence at a roll angle of $45\deg$, markedly different from the  orientations of the polarizers.
Although this sequence extended over $\approx20$ hours, the large scale corona was not expected to change much at a time close to the minimum of activity.
Inspecting the C2 images (upper row in Figure~\ref{FigBkC2}), one can remark that, whereas the streamer belt remains approximately consistent, this is not the case of the general shape of the corona, especially outside the equatorial region. 

At this stage, we were facing the situation reminiscent of that experienced by \cite{leinert1981} when they analyzed the polarization measurements coming from their photometers aboard the HELIOS spacecraft: puzzled by their results they had to introduce extensive corrections in order to retrieve meaningful results, illustrating the difficulties inherent to polarization analysis with limited capabilities.
It took us several years of effort to thoroughly circumvent the problems, to perform systematic tests, and finally to derive proper corrections described in the next section.

\begin{table}
\caption{Journal of the images taken during different roll sequences.}
\label{table:rollsequence}
\begin{tabular}{c c c c c c}
\hline
Date & Time & Telescope & Roll angle  & Polarizer & Exp. Time (sec) \\[3pt]
\hline
1996 May 21  & 14:42:18 UT & C2 & $82\deg$   &  None      & 6.093 \\
1996 May 21  & 14:47:41 UT & C2 & $82\deg$   &  $+60\deg$  & 22.68 \\
1996 May 21  & 14:53:04 UT & C2 & $82\deg$   &  $0\deg$    & 25.09 \\
1996 May 21  & 14:58:27 UT & C2 & $82\deg$   &  $-60\deg$  & 17.37  \\
1996 May 21  & 23:42:02 UT & C2 & $0\deg$    &  None      & 25.46 \\
1996 May 21  & 20:09:22 UT & C2 & $0\deg$    &  $+60\deg$  & 17.37  \\
1996 May 21  & 20:14:44 UT & C2 & $0\deg$   &  $0\deg$     & 25.09 \\
1996 May 21  & 20:20:08 UT & C2 & $0\deg$    &  $-60\deg$  & 25.09 \\
1996 Nov. 21 & 21:20:10 UT & C2 & $89\deg$   &  None      & 25.39 \\
1996 Nov. 21 & 21:22:42 UT & C2 & $89\deg$   &  $+60\deg$  & 100.09 \\
1996 Nov. 21 & 21:26:29 UT & C2 & $89\deg$   &  $0\deg$    & 100.09 \\
1996 Nov. 21 & 21:30:17 UT & C2 & $89\deg$   &  $-60\deg$  & 100.19 \\
1996 Nov. 22 & 09:45:10 UT & C2 & $0\deg$    &  None      & 25.09 \\
1996 Nov. 22 & 09:47:43 UT & C2 & $0\deg$    &  $+60\deg$  & 100.09 \\
1996 Nov. 22 & 09:51:29 UT & C2 & $0\deg$    &  $0\deg$    & 100.09 \\
1996 Nov. 22 & 09:55:16 UT & C2 & $0\deg$    &  $-60\deg$  & 100.09 \\
1997 Sep. 02 & 22:18:05 UT & C2 & $0\deg$    &  None      & 25.09 \\
1997 Sep. 02 & 22:20:36 UT & C2 & $0\deg$    &  $+60\deg$  & 100.09 \\
1997 Sep. 02 & 22:24:22 UT & C2 & $0\deg$    &  $0\deg$    & 100.09 \\
1997 Sep. 02 & 22:28:07 UT & C2 & $0\deg$    &  $-60\deg$  & 100.09 \\
1997 Sep. 03 & 09:46:27 UT & C2 & $45\deg$   &  None      & 25.09 \\
1997 Sep. 03 & 09:48:57 UT & C2 & $45\deg$   &  $+60\deg$  & 100.09 \\
1997 Sep. 03 & 09:52:44 UT & C2 & $45\deg$   &  $0\deg$    & 100.09 \\
1997 Sep. 03 & 09:56:30 UT & C2 & $45\deg$   &  $-60\deg$  & 100.09 \\
1997 Sep. 03 & 17:51:41 UT & C2 & $90\deg$   &  None      & 25.09 \\
1997 Sep. 03 & 17:54:12 UT & C2 & $90\deg$   &  $+60\deg$  & 100.09 \\
1997 Sep. 03 & 17:58:36 UT & C2 & $90\deg$   &  $0\deg$    & 100.09 \\
1997 Sep. 03 & 18:03:21 UT & C2 & $90\deg$   &  $-60\deg$  & 100.09 \\
\hline
\end{tabular}
\end{table}

\section{Improvements of the Polarization Analysis}
\label{Improvement}

\subsection{Adjustment of the global transmission of the polarizers}
Slightly different transmissions of the polarizers were first suspected as a possible cause of the above problems, a route independently explored by \cite{moran2006solar} whose derived correction factors for the C3 polarizers using two different methods.
We introduced a different method, namely the minimization of the width of the histograms of the local angle of polarization, and considered not a single image as done by \cite{moran2006solar}, but the whole set of the seven polarization sequences obtained during the three roll maneuvers (Table \ref{table:rollsequence}).
We applied a standard computational technique which searches the optimal values of the transmissions that simultaneously minimizes all histogram widths.
It turned out that only one polarizer needed an adjustment, namely the $0\deg$ one with a factor of 0.98, and this turned out to be extremely efficient in reducing the width of the distributions to $\approx6\deg$ as illustrated in Figure~\ref{FigHistC2}.

We note that \cite{moran2006solar} did not implement any correction for the transmission of the C2 polarizers contrary to those of C3, but concentrated their attention to the two folding mirrors.
As already emphasized in Section~\ref{ComponentCalib}, the polarizing properties of these mirrors were known before launch and as far as we are concerned, were already introduced in our Mueller formalism. 
Our approach of correcting for the transmission of the C2 polarizers is further justified by two arguments: i) it would be rather surprising that the C3 polarizers alone needed corrections and not those of C2, and ii) the hard, low-polarization coating of the mirrors is certainly less prone to degradation that the Polaroid foils.
As a matter of curiosity, we confronted the two approaches of tuning i) the properties of the mirrors, and ii) the transmissions of the polarizers, to minimize the widths of the histograms, but this time over seven years of observation.
Indeed, the first approach is capable of yielding results almost as good as the second one, but with values of the mirror parameters substantially different from those obtained by \cite{moran2006solar}.
In fact, we introduced their values in our calculations and obtained results far worse than those displayed in Figure~\ref{FigHistC2}.
This clearly demonstrates that the problem resulted from the polarizers, namely the $0\deg$, and not from the mirrors as proposed by \cite{moran2006solar}.

Figure~\ref{FigBkC2} dramatically illustrates the improvement of the shape of the K-corona resulting from our corrections when comparing the first and second rows. 
The north-south  distortions have almost disappeared, and the three $B_K$ images obtained at the three different roll angles in September 1997 are close to identical.
There does however remains some discrepancies, for instance in the north-east quadrant and we explain in the next section how they were corrected.

\subsection{Global correction}
At this stage, it was difficult to trace the remaining discrepancies to a specific problem or problems with the optical components.
Our approach was consequently to derive a global correction function $S(x,y)$ to be directly applied to the $pB$ images produced by the polarization analysis.
Here again, we took advantage of the roll sequence performed in September 1997 and the technical details of the derivation of $S(x,y)$ are presented in Appendix I.
However, our procedure left its mean level $\gamma = \overline{S(x,y)}$ undetermined and this shortcoming was solved by imposing the condition that the K/F separation led to the smoothest possible F-corona.
This was performed on images obtained during the period of high activity of solar cycle 23 characterized by intense streamers when the K/F separation is most sensitive to the correct determination of the K-corona.
We looked for the most regular F-corona profiles as function of $\gamma$ by inspecting circular profiles extracted at $4\Rsun$. 
The third row in Figures~\ref{FigBkC2} was obtained with the optimum values of $\gamma$.
The most noticeable improvements brought by the introduction of the $S(x,y)$ correction is best seen on image at the roll angles of $0\deg$ and $90\deg$.

\subsection{Ultimate Improvement of the polarization procedure}
\begin{figure}[htbp!]
	\centering
	\includegraphics[width=\textwidth]{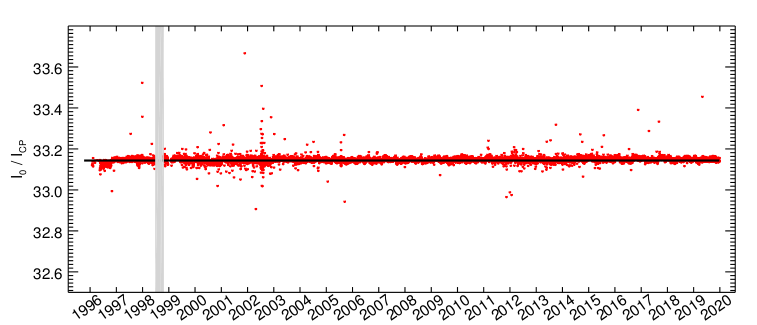}
	\caption{Temporal variations of the ratio $I_{0}/I_{cp}$ of the radiance of the unpolarized image to that resulting from the final polarization analysis with fine tuning of the exposure time equalization.}
	\label{C2CalCoefOpt}
\end{figure}

The preprocessing of the level-0.5 LASCO images as described in Section \ref{Implementation} includes a step of exposure time equalization required to correct for the errors in exposure times.
They result from random time delays in the reception and transmission of the information between the three processors involved in opening and closing the shutte\citep{morrill2006calibration}.
These errors were first detected when flickers were noticed in the movies constructed from image series; they typically remain within 2 to 3\% but sometimes reach up to $\pm50$\% between successive images.
Building homogeneous temporal sequences for scientific purposes requires a relative accuracy of 0.1\% in the short term (a few days) and better than 1\% on the long term. 
The problem is even more acute for the polarization analysis as its accuracy is critically dependent upon the rigorous timing of the three exposures of the three polarized images.
Llebaria and Thernisien (2001) developed a method relying on the short-time stability of the corona; it is based on an image-to-image regression and a long-term correction to circumvent the drifts induced by the minute but unavoidable residual inaccuracies. 
This method has been successfully applied to the routine unpolarized images since their high cadence guaranties the condition of stability of the corona, but less so for the low cadence polarized images. 
In fact, we consider that the dispersion in the $I_{0}/I_{cp}$ ratio (Figure~\ref{C2CalCoef}) stems in part from the imperfections in exposure time equalization.
The behaviour of this ratio may be interpreted in terms of a long trend evolution due to the slow irreversible degradation of the transmittance of the three polarizers and high frequency fluctuations due to the above imperfections (an additional trend may be noted correlated to the solar cycles; it most likely results from the condition of stability being less satisfied during solar maxima compared to minima).
We therefore needed to fine tune the corrections for the errors in exposure times. 
Such errors have in fact the same impact as a change in the transmittance of the polarizers since they both result in an incorrect unbalance between the three polarized images.
For practical simplicity, we combined these two sources of errors in a single treatment and used the most sensitive test on the local polarization angle for optimization: for each polarization sequence, its distribution must be centered at $90\deg$ and be as narrow as possible.
The $0\deg$ polarizer was taken as a reference and the ratios of the transmittances ``$+60\deg$/$0\deg$'' and ``$-60\deg$/$0\deg$'' were explored in the range 0.91 to 1.09.
To speed up the process, a multi-resolution approach was implemented, starting with a step of 0.03 and reducing the range of exploration by a factor of 2 at each successive iteration.
At the end of the process when an estimated accuracy of 0.1\% was reached, the absolute values of the transmittance of the polarizers were determined by comparing the sum of the three corrected polarized images with the associated unpolarized image of the sequence.
Therefore, the whole process simultaneously corrects for the long-term evolution of the polarizers and the errors in exposure time to the ultimate accuracy allowed by the underlying assumptions spelled above.
Consequently, the $I_{0}/I_{cp}$ ratio becomes time-independent and assumes a constant value of 33.18 with an rms deviation of only $\pm$0.001 as illustrated in Figure~\ref{C2CalCoefOpt}; note the achieved improvement by comparison with Figure~\ref{C2CalCoef}.

\begin{figure}[htbp!]
	\centering
	\includegraphics[width=\textwidth]{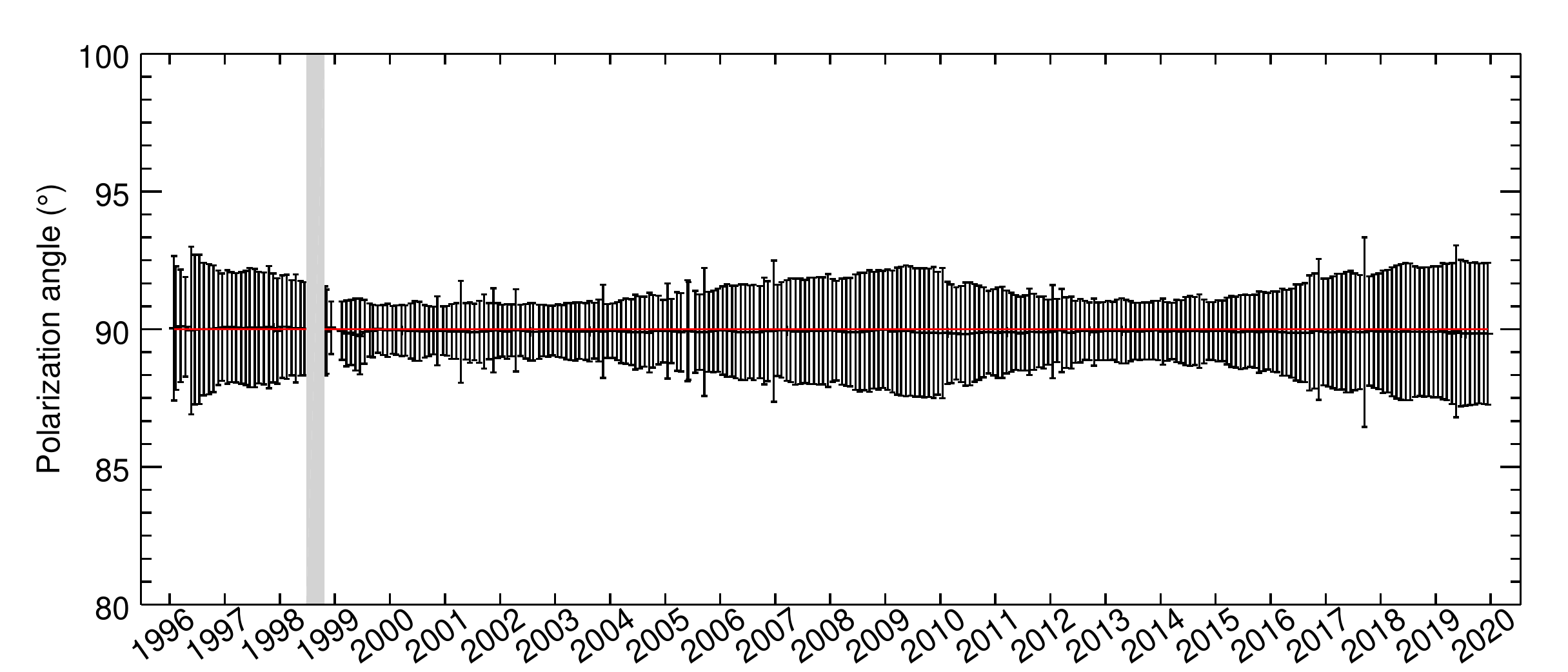}
	\caption{Temporal evolution of the local angle of polarization over 24 years. 
	The error bars represent the standard deviations of the monthly values.}
	\label{FigEvolHistC2}
\end{figure}

\begin{figure}[htbp!]
	\centering
	\includegraphics[width=\textwidth]{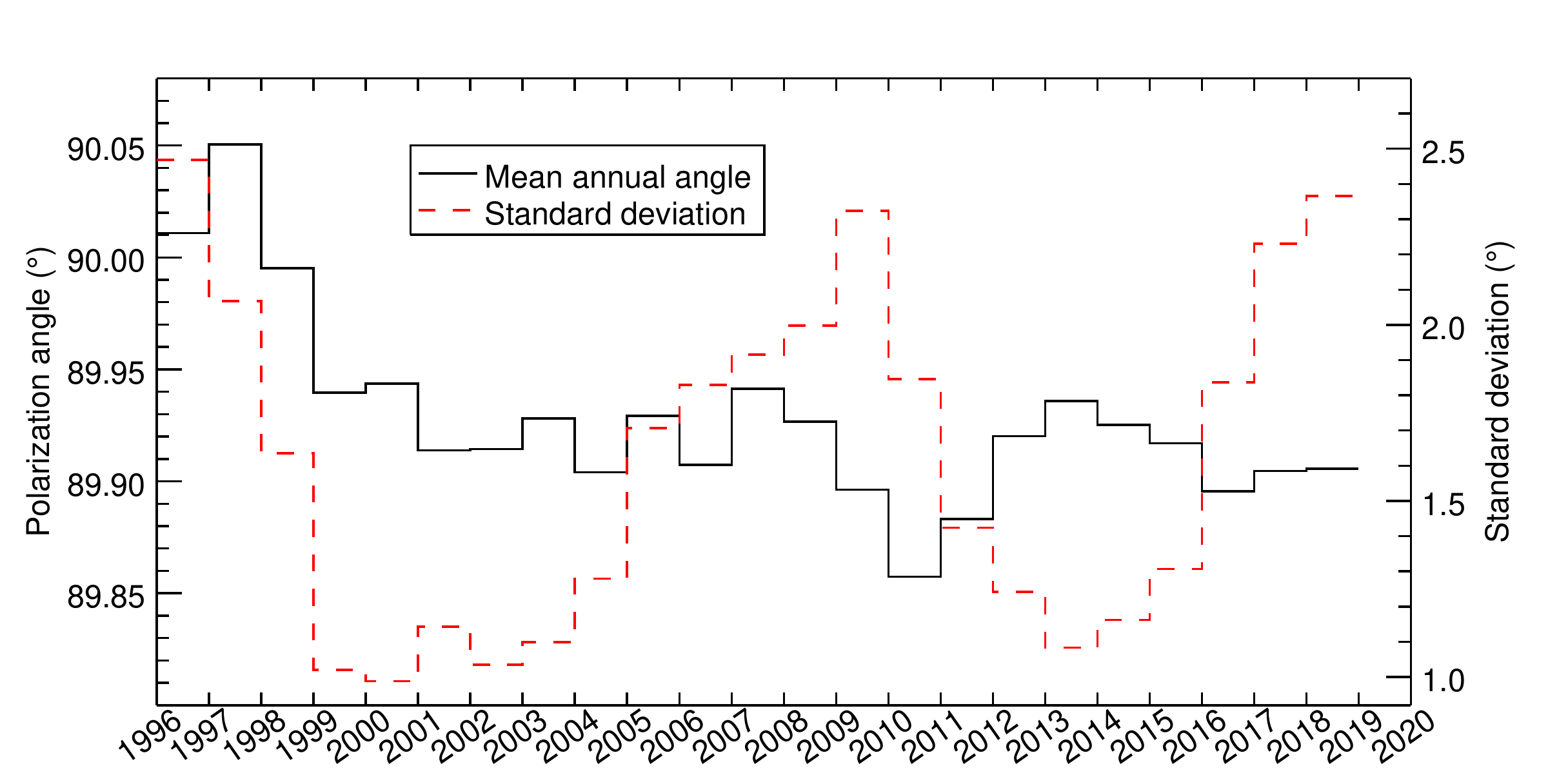}
	\caption{Temporal evolution of the annually averaged value of the local angle of polarization (left scale) and of its standard deviations (right scale) over 24 years.}
	\label{FigStatParamPol}
\end{figure}

Figure~\ref{FigEvolHistC2} displays the monthly averaged values of the local angle of polarization $\overline{\alpha_c}$ and their standard deviations throughout the 24 years of LASCO-C2 observation.
A complementary view is offered by Figure~\ref{FigStatParamPol} where the annually averaged values of the local angle of polarization and their standard deviations are displayed over 24 years.
The remaining deviation of $\overline{\alpha_c}$ from the theoretical value of 90$\deg$ is typically 0.2$\deg$ with slightly larger values during the first few years of operation.
It is interesting to note in the above two figures that the standard deviation varies in opposition with the solar cycle.
Phases of high activity are characterized by increased number of streamers whose large polarization improves the signal-over-noise ratio, thus resulting in lower values of the standard deviation.

\section{Final Results for the Photopolarimetric Properties of the Corona}

Over 20500 sequences of polarization have been accumulated by LASCO-C2 at the end of 2018 and it is quite challenging to present such a large amount of data in a synthetic form.
For this section, we selected different presentations which hopefully give an overview of the two-dimensional photopolarimetric properties of the corona and the derived science products over two solar cycles.

\begin{itemize}
	\item Maps at three phases of solar activity during each of the two solar cycles SC 23 and SC 24 as well as the corresponding profiles along the equatorial and polar directions.
	\item Multi-annual synoptic maps at two elongations 2.7 and 5.5$\Rsun$.
	\item Monthly averaged temporal profiles extracted from the above synoptic maps at 2.7$\Rsun$ (all quantities) and 5.5$\Rsun$ (only for the polarization and the polarized radiance). 
In both cases, the quantity of interest is averaged over all latitudes (hence labeled ``global''), over the northern and southern hemispheres, and in two sectors 30$^{\circ}$ wide centered along the equatorial and polar directions.
For the temporal profiles of the ``global'' quantities, we superimposed the temporal variation of the total photospheric magnetic flux (TMF) as the proxy of solar activity which was found by \cite{barlyaeva2015mid} to best match the integrated radiance of the K-corona.
The TMF was calculated from the Wilcox Solar Observatory photospheric field maps by Y.-M.~Wang according to a method described by \cite{wang2003fluctuating}.
\end{itemize}

\subsection{Polarization of the corona}
\label{Polar}

\begin{figure}[htpb!]
	\centering
	\includegraphics[width=1.\textwidth]{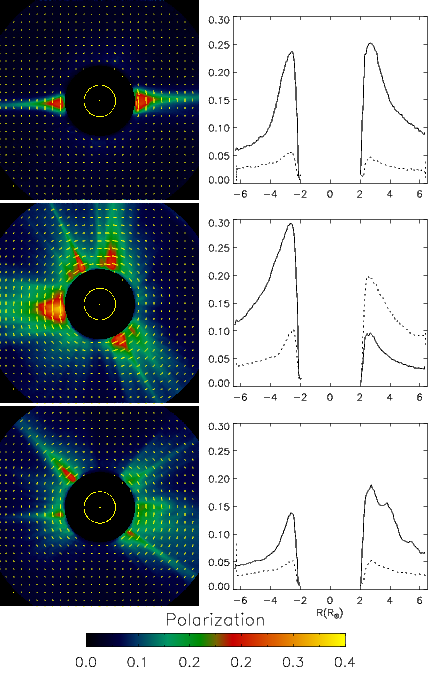}
	\caption{Polarization maps and radial profiles of the corona at three phases of activity of SC 23: 
	low (16 May 1996, upper panels), high (14 Nov. 1999, middle panels), and declining (15 May 2004, lower panels).
	The direction of polarization is indicated by yellow bars whose length is scaled to the polarization.
	The yellow circles represent the solar disk and solar north is up.
	The profiles are extracted along the equatorial (solid lines) and polar (dotted lines) directions.}
	\label{FigPolarVecC2sc23}
\end{figure}

\begin{figure}[htpb!]
	\centering
	\includegraphics[width=1.\textwidth]{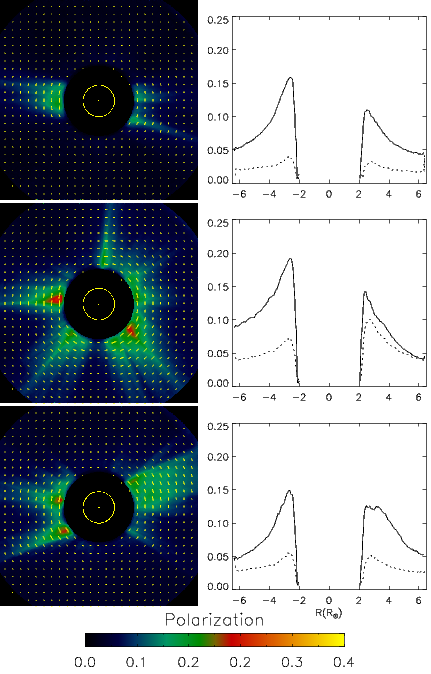}
	\caption{Same as Figure~\ref{FigPolarVecC2sc23} for three phases of activity of SC 24: 
	low (16 Apr. 2009, upper panels), high (16 Apr. 2014, middle panels), and declining (1 Sep. 2016, lower panels).}
	\label{FigPolarVecC2sc24}
\end{figure}

Figures~\ref{FigPolarVecC2sc23} and \ref{FigPolarVecC2sc24} display the maps of the polarization at three phases of solar activity during solar cycles 23 and 24 as well as the corresponding profiles along the equatorial and polar directions.
The six maps are shown with the same color scale and the profiles with the same scale so as to emphasize the striking difference of the general polarization level between the two solar cycles, consistent with the difference in their strength, SC24  being weaker than SC23.
The radial variation is remarkably consistent with the model presented in Figure~\ref{FigPkPf} with a monotonic decrease with increasing elongation.
At the inner edge of the C2 \fov ($\approx$2.2 R${}_\odot$), the brightest streamers culminate at a peak polarization of $\approx$0.4, here again consistent with Figure~\ref{FigPkPf} which illustrates the case of a corona of the maximum type.
Coronal holes are characterized by very low polarizations, in the range 0.025--0.05 during the minimum of solar cycle 22/23 and even lower, in the range 0.03--0.02 during the anomalous minimum of solar cycle 23/24.

\begin{figure}[htpb!]
\centering
\includegraphics[width=0.83\textheight, angle=90]{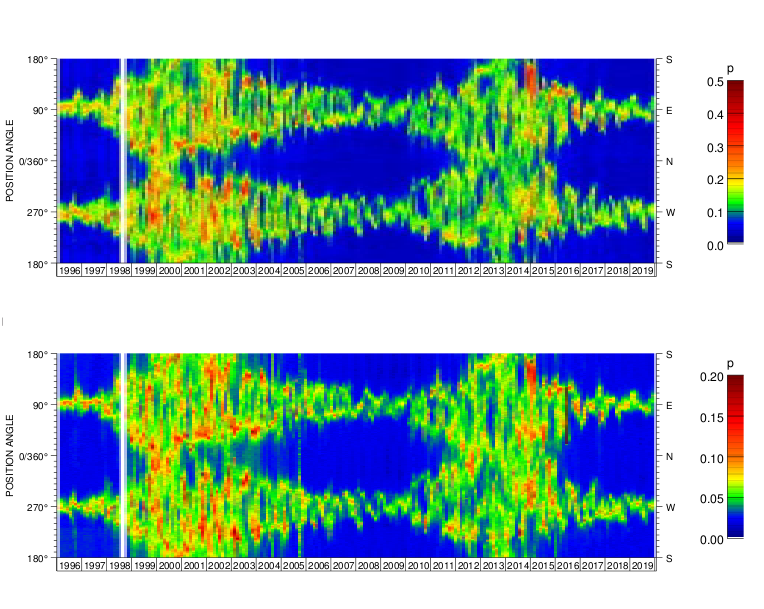}
\caption{Multi-annual synoptic maps of the coronal polarization over 24 years [1996--2019] at two elongations 2.7$\Rsun$ (left panel) and 5.5$\Rsun$ (right panel).
Note the different color scales for the two maps.
The white band corresponds to missing data when SoHO lost its pointing.}
\label{pSyno}
\end{figure}

The multi-annual synoptic maps of the coronal polarization over 24 years [1996--2019] shown in Figure~\ref{pSyno}) conspicuously illustrate its spatial and temporal evolutions.
The global pattern closely follows that of the streamer belt: a rapid broadening as solar activity increases with peak values being recorded during the maxima, followed by its progressive narrowing as the activity declines. 
These distributions strikingly confirm the difference in the general polarization level between the two solar cycles, consistent with the difference in their strength, as already noted above.
A noteworthy peculiarity is the patches of high polarizations present from late 2014 to beginning of 2015 in the south-east region. 
They are connected to the anomalous surge of the radiance of the corona discovered by \cite{Lamy2017}.

\begin{figure}
\centering
\includegraphics[width=\textwidth]{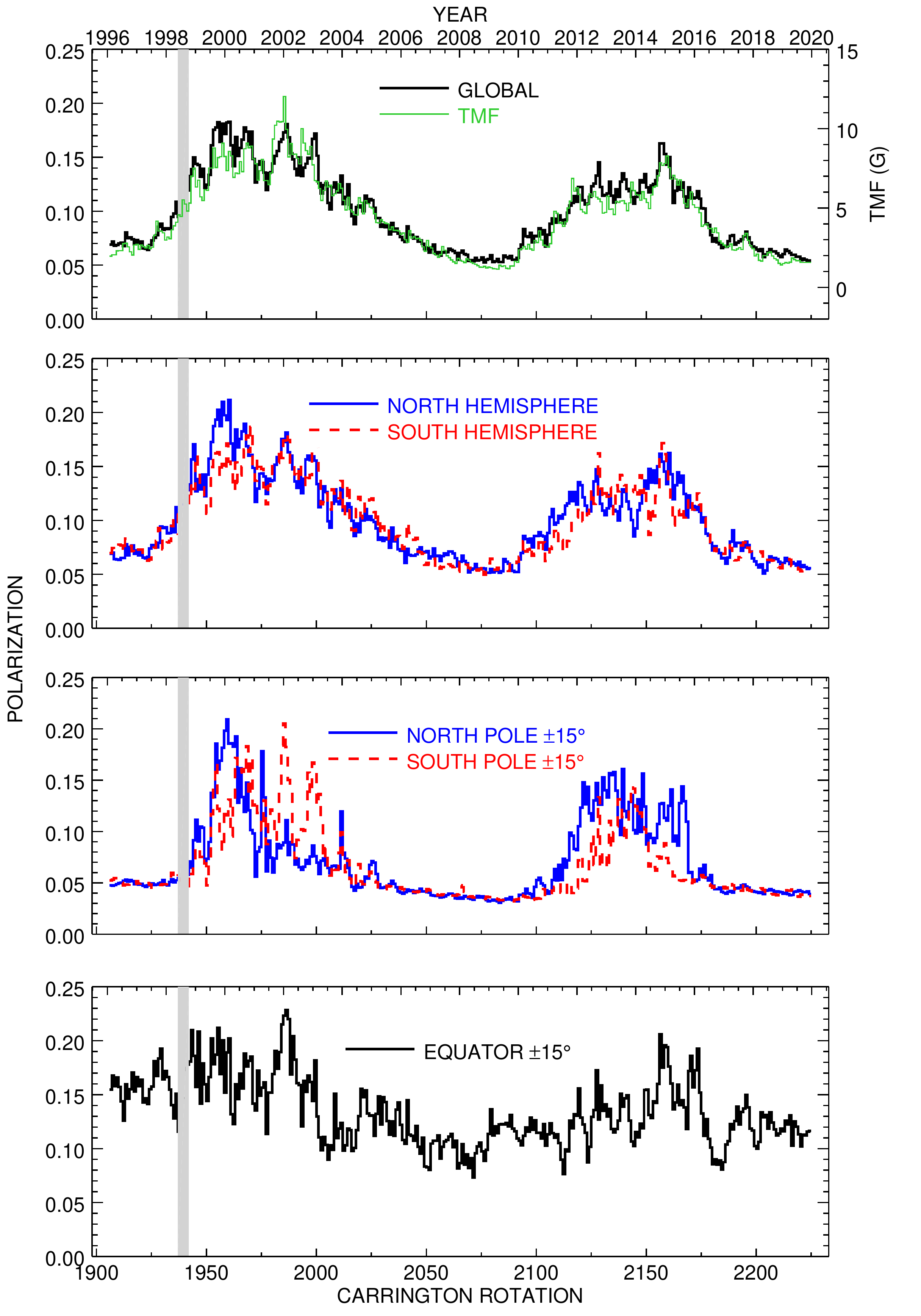}
\caption{Monthly averaged profiles extracted from the multi-annual synoptic map of the coronal polarization at 2.7$\Rsun$, globally, in the two hemispheres, and in polar and equatorial sectors.
The gray bands correspond to missing data when SOHO lost its pointing.}			
\label{PActiv27}
\end{figure}

\begin{figure}
\centering
\includegraphics[width=\textwidth]{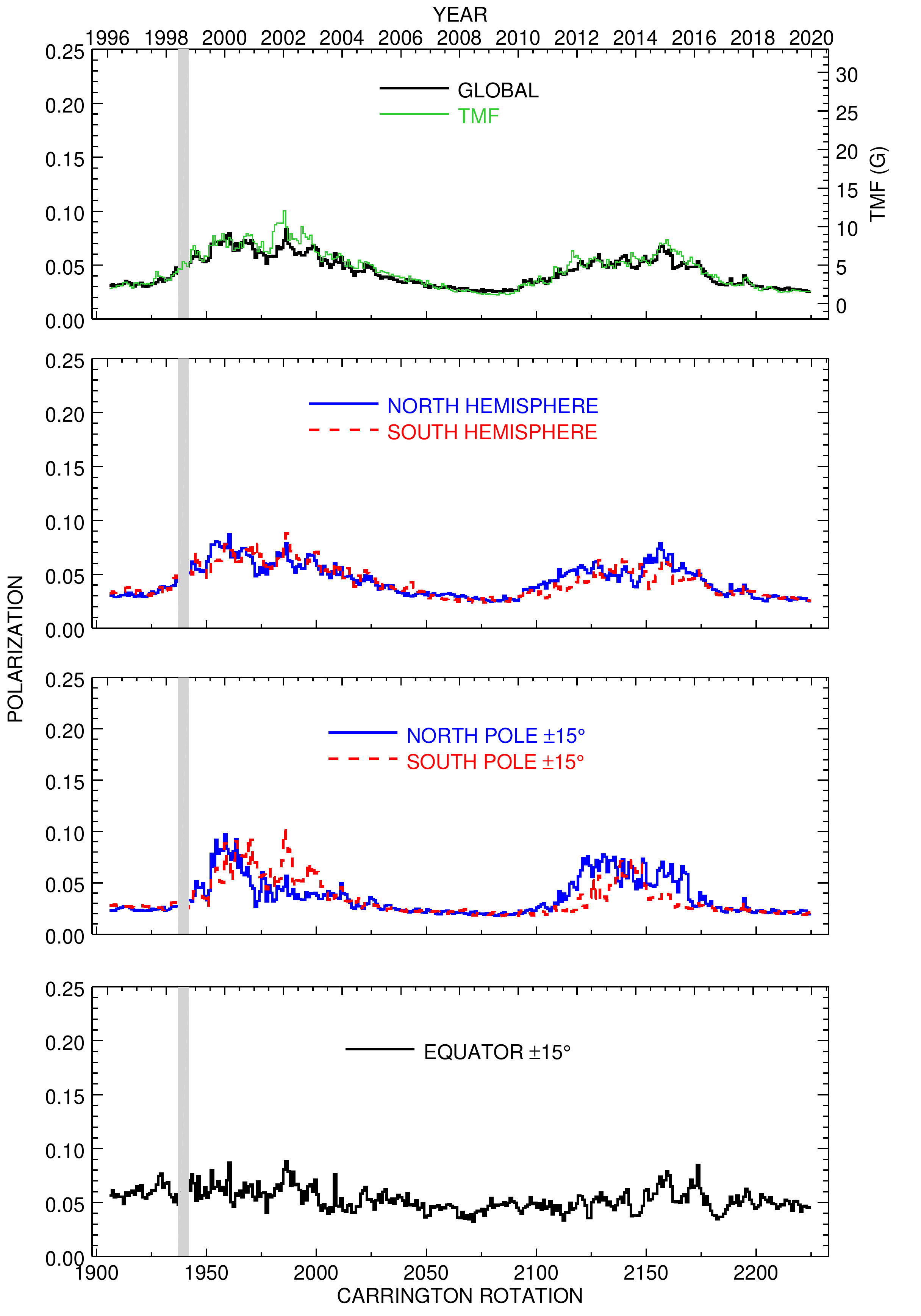}
\caption{Same as Figure~\ref{PActiv27} at 5.5$\Rsun$.}
 \label{PActiv55}
\end{figure}

A more detailed quantitative view of the coronal polarization is offered by the monthly averaged temporal profiles displayed in Figure~\ref{PActiv27} and Figure~\ref{PActiv55}).
Note the excellent agreement of the temporal variations of the globally integrated polarizations with that of the TMF, not only in the case of the relative strengths of the two maxima but also on detailed fluctuations within the maxima.

Because the polarization, the polarized radiance, the electron density, and the radiance of the K-corona share many properties, they will be further discussed altogether in Subsection~\ref{properties} below.

\subsection{Polarized radiance of the corona}   

\begin{figure}[htpb!]
\centering
\includegraphics[width=\textwidth]{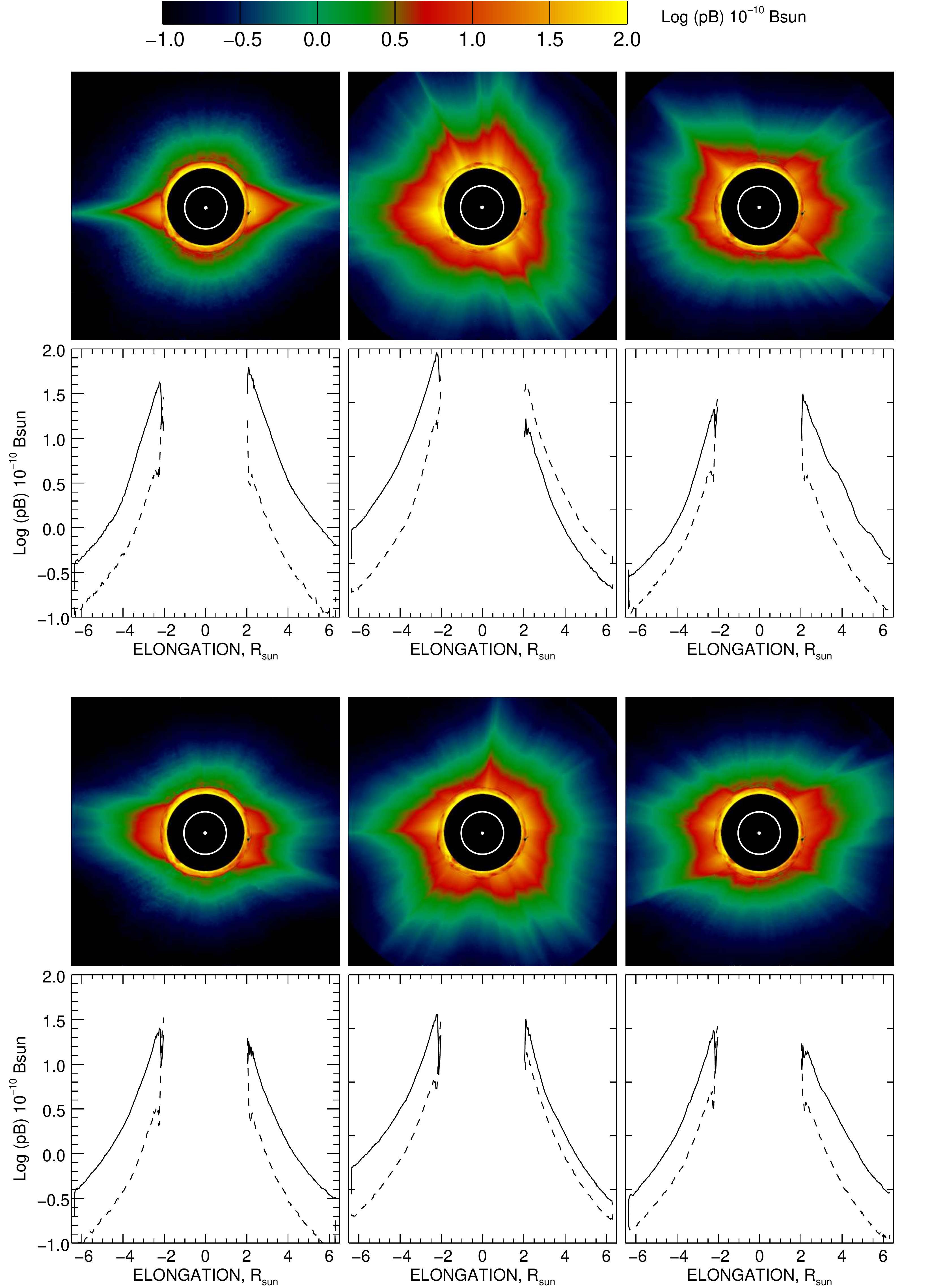}
\caption{Images and radial profiles of the polarized radiance of the corona at three phases of activity of SC 23 (upper two rows) and SC 24 (lower two rows).
The dates are identical to those of Figures~\ref{FigPolarVecC2sc23} and \ref{FigPolarVecC2sc24}: 16 May 1996, 14 Nov. 1999, and 15 May 2004 for SC 23 and 16 Apr. 2009, 16 Apr. 2014, and 1 Sep. 2016 for SC 24. 
The yellow circles represent the solar disk and solar north is up.
The profiles are extracted along the equatorial (solid lines) and polar (dotted lines) directions.}
	\label{FigpBs}
\end{figure}

\begin{figure}[htpb!]
\centering
\includegraphics[width=0.83\textheight, angle=90]{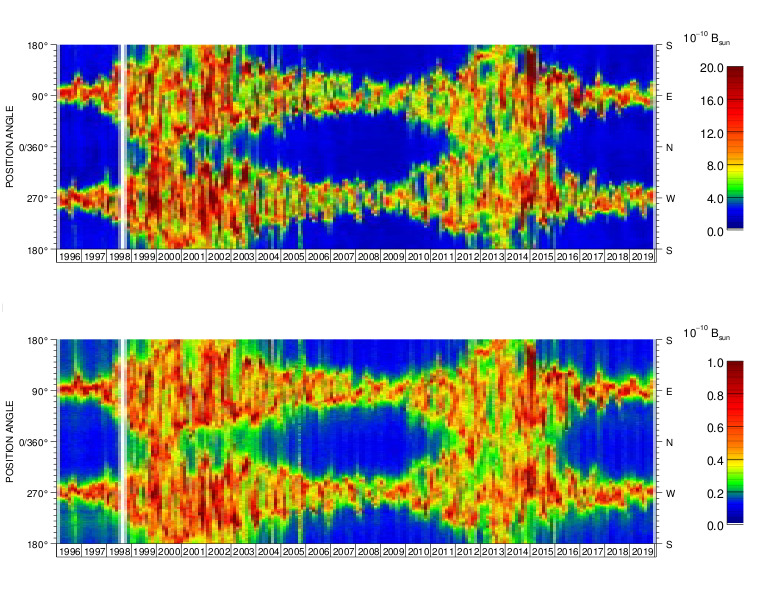}
\caption{Multi-annual synoptic maps of the polarized radiance of the corona over 24 years [1996--2019] at two elongations 2.7$\Rsun$ (upper panel) and 5.5$\Rsun$ (lower panel).}
\label{pBSyno}
\end{figure}

\begin{figure}
      \centering
      \includegraphics[width=\textwidth]{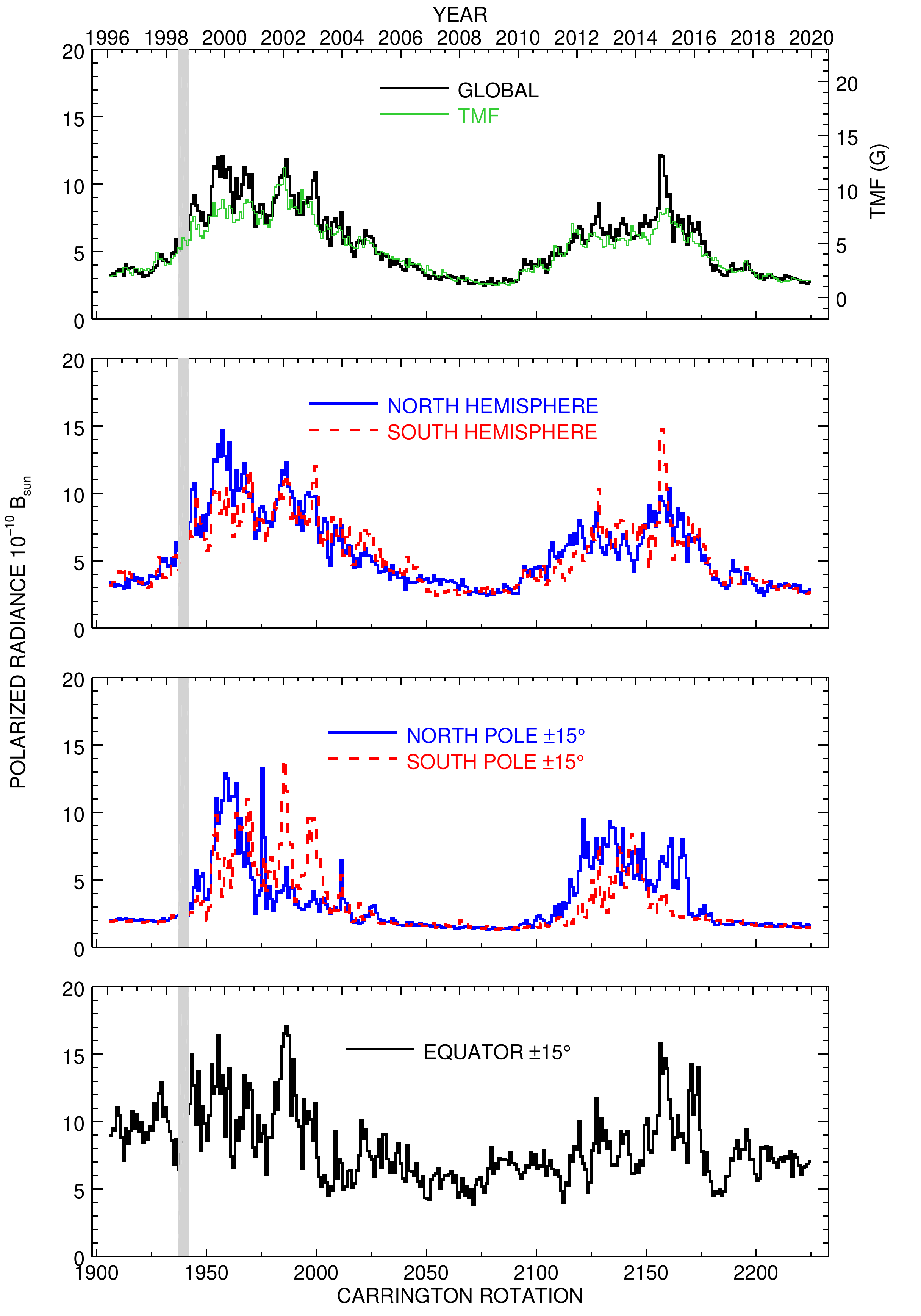}
      \caption{Monthly averaged profiles extracted from the multi-annual synoptic map of the polarized radiance at 2.7$\Rsun$, globally, in the two hemispheres, and in polar and equatorial sectors.}
      \label{pBActiv27}
\end{figure}

\begin{figure}
      \centering
      \includegraphics[width=\textwidth]{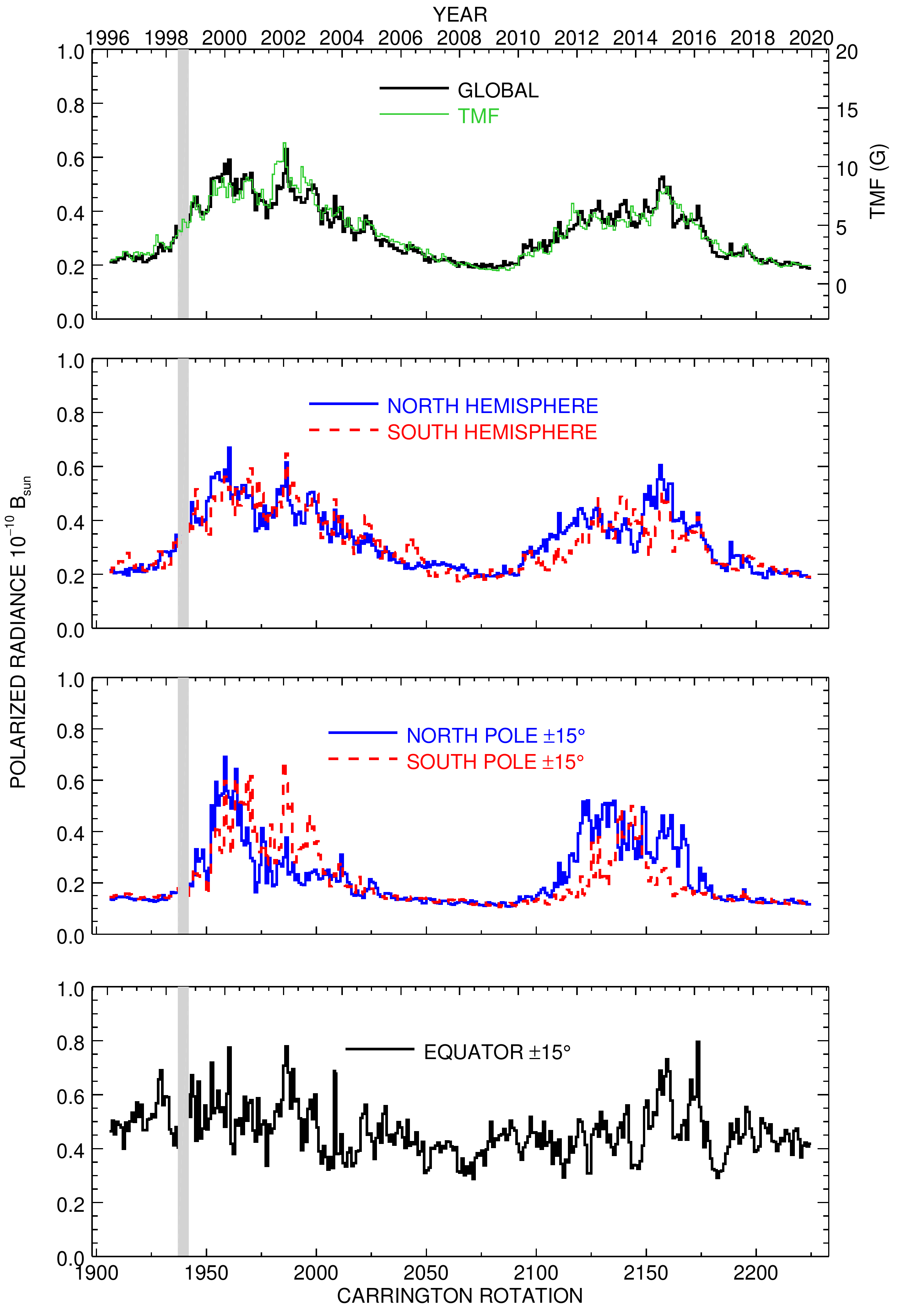}
      \caption{Same as Figure~\ref{pBActiv27} at 5.5$\Rsun$.}
      \label{pBActiv55}
\end{figure}

The presentation of the results for the polarized radiance $pB$ of the corona is similar to that of the polarization, except for a more compact format in the case of the images and profiles as illustrated in Figure~\ref{FigpBs}.  
The dates are identical to those of Figures~\ref{FigPolarVecC2sc23} and \ref{FigPolarVecC2sc24}. 
Figure~\ref{pBSyno} displays the two multi-annual synoptic maps of $pB$ over 24 years [1996--2019] and Figures~\ref{pBActiv27} and \ref{pBActiv55}) display the temporal profiles extracted from these maps.

\subsection{Coronal electron density } 
\label{Ne}

The two-dimensional (2D) distributions of the coronal electron density $N_e$ were obtained by applying the method developed by \cite{quemerais2002two} for the 2D inversion of $pB$ images and previous examples can be found in \cite{lamy2002solar}, \cite{Lamy2014comparing}, and \cite{Lamy2017}.
We display a set of figures similar to those presented above for the polarized radiance except that we limit the monthly averaged profiles to the case of an elongation of 2.7$\Rsun$ for brevity.

\begin{figure}
\centering
\includegraphics[width=\textwidth]{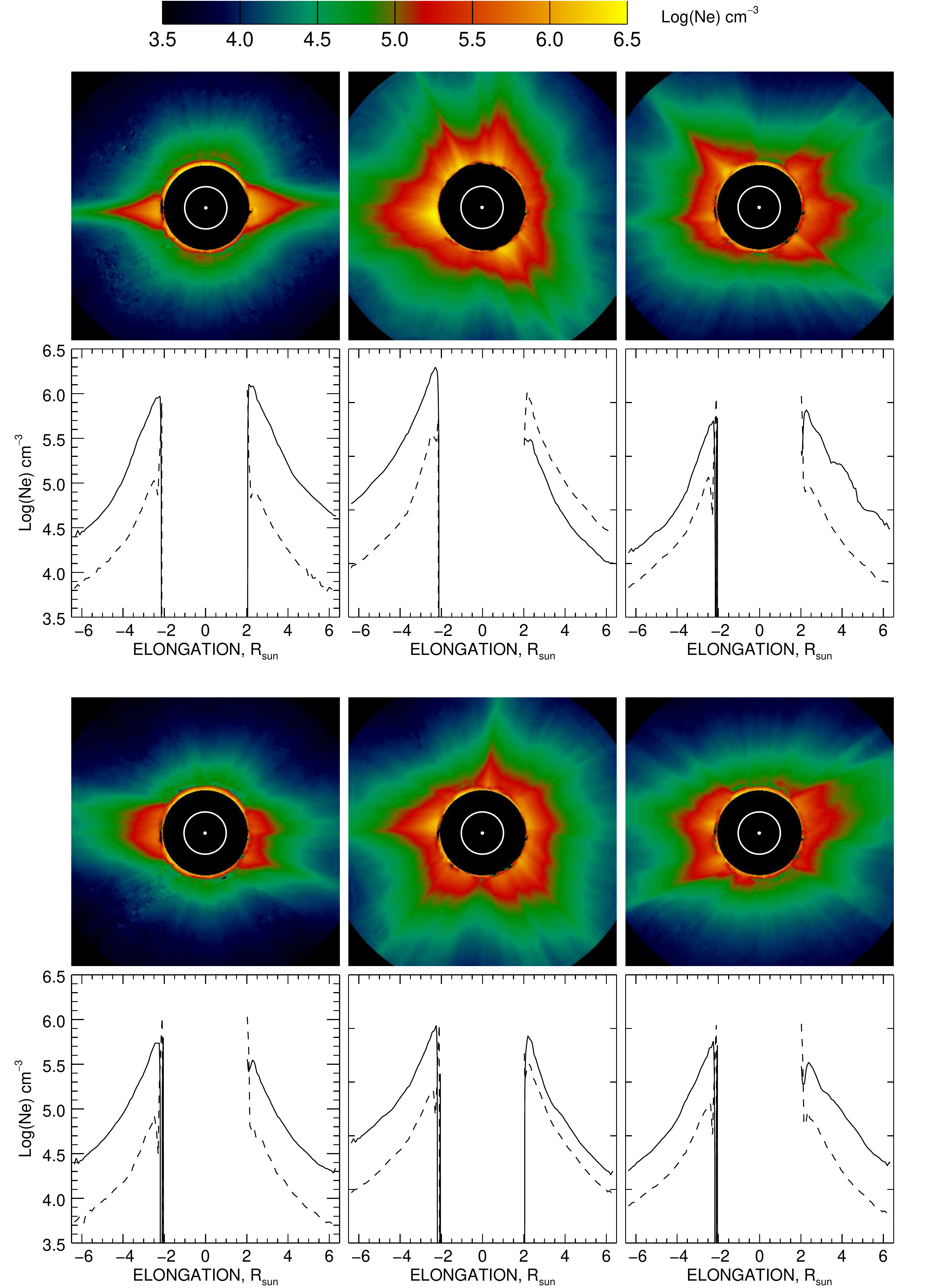}
\caption{Images and radial profiles of the coronal electron density at three phases of activity of SC 23 (upper two rows) and of SC 24 (lower two rows).
The dates are identical to those of Figures~\ref{FigPolarVecC2sc23} and  \ref{FigPolarVecC2sc24}): 16 May 1996, 14 Nov. 1999, and 15 May 2004 for SC23 and 16 Apr. 2009, 16 Apr. 2014, and 1 Sep. 2016 for SC24. 
The yellow circles represent the solar disk and solar north is up.
The profiles are extracted along the equatorial (solid lines) and polar (dotted lines) directions.}
\label{NeSample}
\end{figure}

\begin{figure}[htpb!]
\centering
\includegraphics[width=0.83\textheight, angle=90]{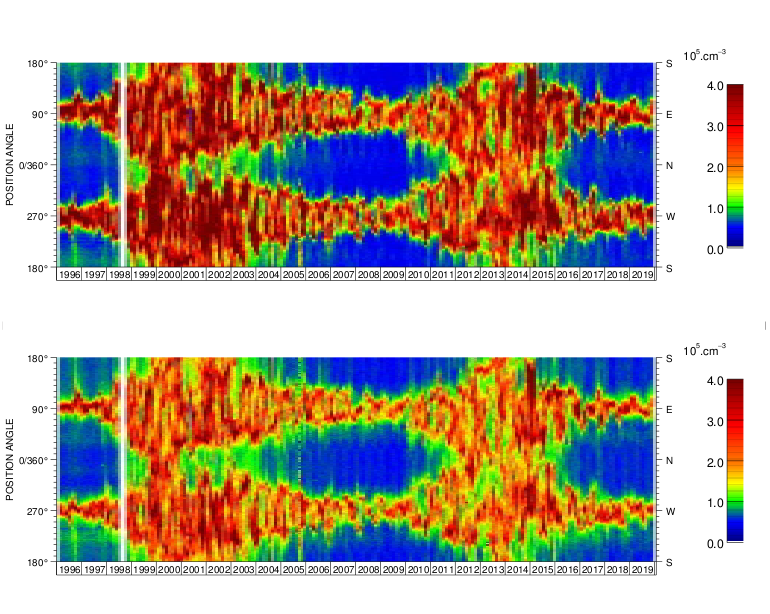}
\caption{Multi-annual synoptic maps of the coronal electron density over 24 years [1996--2019] at two elongations 2.7$\Rsun$ (upper panel) and 5.5$\Rsun$ (lower panel).}
\label{NeSyno}
\end{figure}

\begin{figure}
\centering
\includegraphics[width=\textwidth]{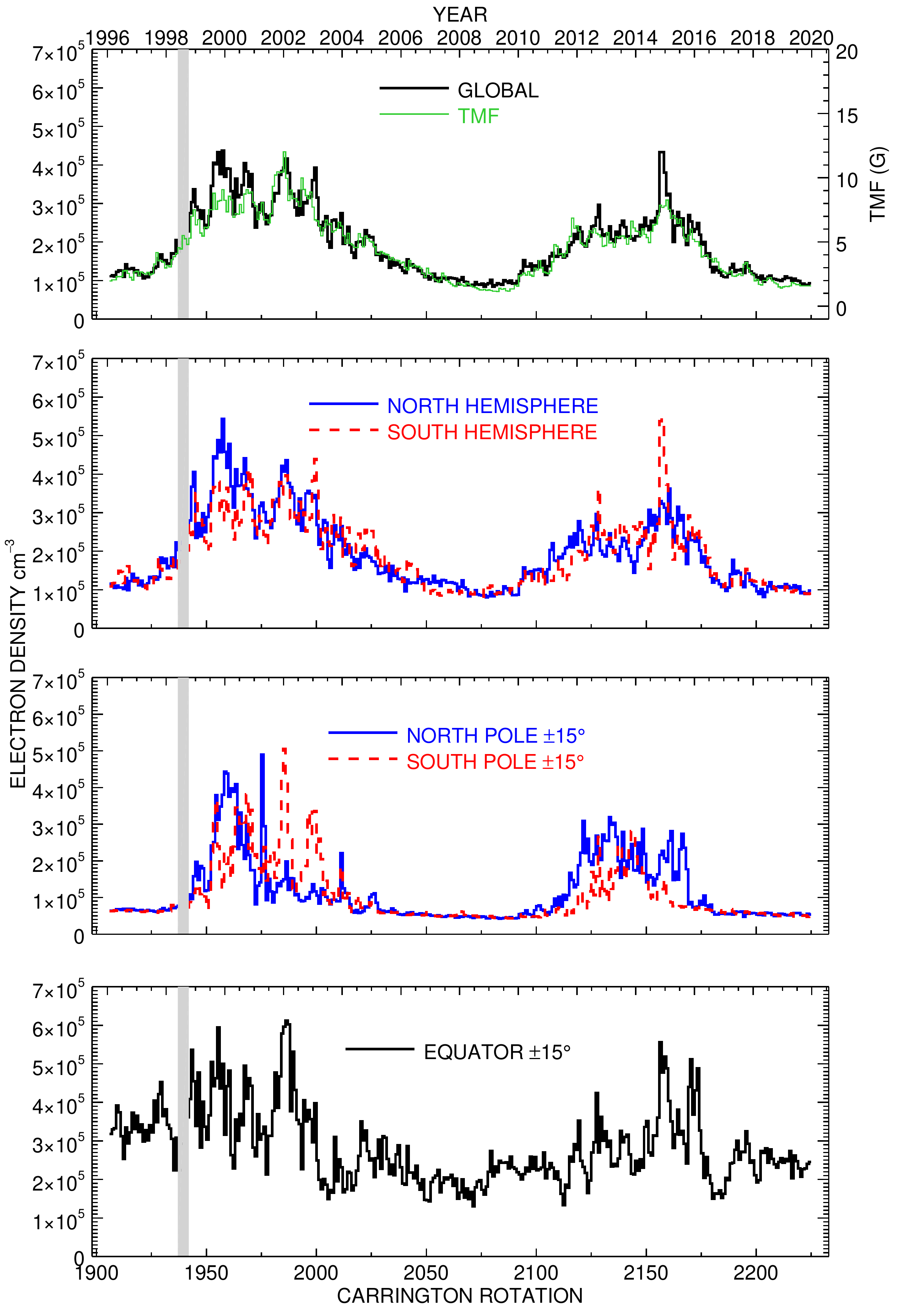}
\caption{Monthly averaged profiles extracted from the pluri-annual synoptic map of the electron density at 2.7$\Rsun$, globally, in the two hemispheres, and in polar and equatorial sectors.}
\label{NeActiv27}
\end{figure}

\begin{itemize}
	\item Maps of $Ne$ at three phases of solar activity during SC 23 and SC 24 as well as the corresponding profiles along the equatorial and polar directions (Figure~\ref{NeSample});
	\item Multiannual synoptic maps of $Ne$ at 2.7 and 5.5$\Rsun$ (Figure~\ref{NeSyno});
	\item Monthly averaged profiles at 2.7$\Rsun$ (Figure~\ref{NeActiv27}).
\end{itemize}

Three-dimensional inversion by time-dependent solar rotational tomography \citep{vibert2016time} of the whole set of $pB$ images is in progress and the resulting $Ne$ ``cubes'' will be released in the near future.

\subsection{K-Corona } 
\label{Kcorona}

\begin{figure}
\centering
\includegraphics[width=\textwidth]{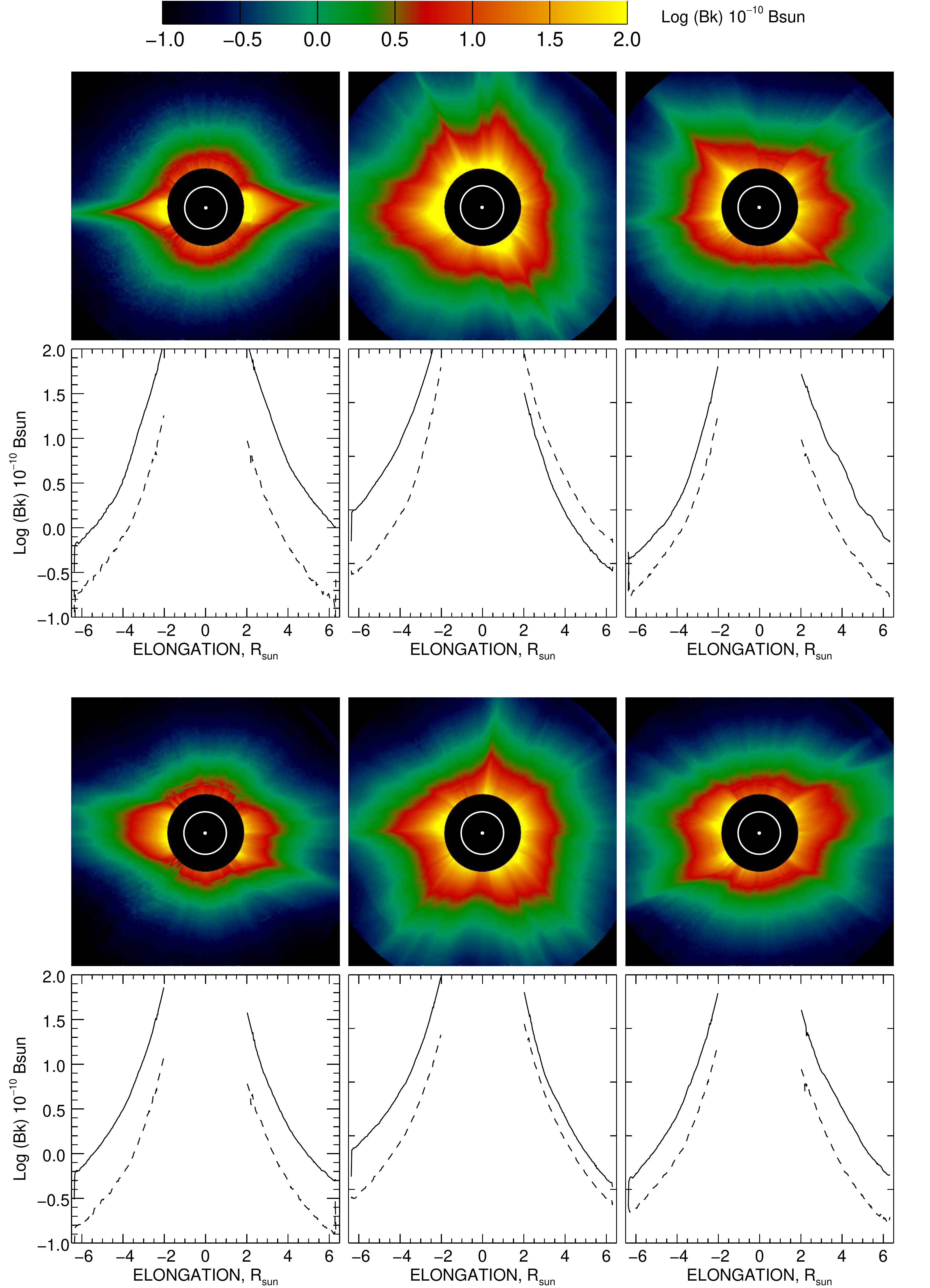}
\caption{Images and radial profiles of the K-corona at three phases of activity of SC 23 (upper two rows) and SC 24 (lower two rows).
The dates are identical to those of Figures~\ref{FigPolarVecC2sc23} and \ref{FigPolarVecC2sc24}: 16 May 1996, 14 Nov. 1999, and 15 May 2004 for SC 23 and 16 Apr. 2009, 16 Apr. 2014, and 1 Sep. 2016 for SC 24. 
The yellow circles represent the solar disk and solar north is up.
The profiles are extracted along the equatorial (solid lines) and polar (dotted lines) directions.}
\label{KSample}
\end{figure}

\begin{figure}[htpb!]
\centering
\includegraphics[width=0.83\textheight, angle=90]{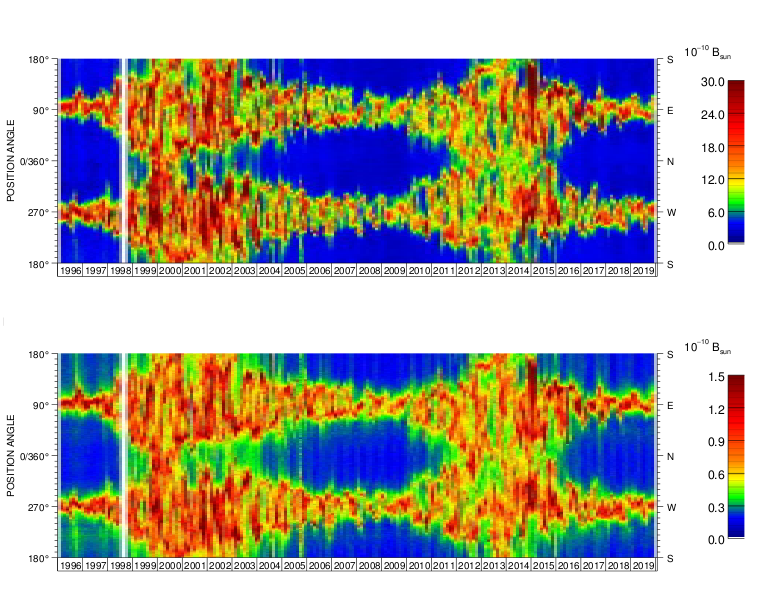}
\caption{Multi-annual synoptic maps of the radiance of the K-corona over 24 years [1996--2019] at two elongations 2.7$\Rsun$ (upper panel) and 5.5$\Rsun$ (lower panel).}
\label{KSyno}
\end{figure}

\begin{figure}
\centering
\includegraphics[width=\textwidth]{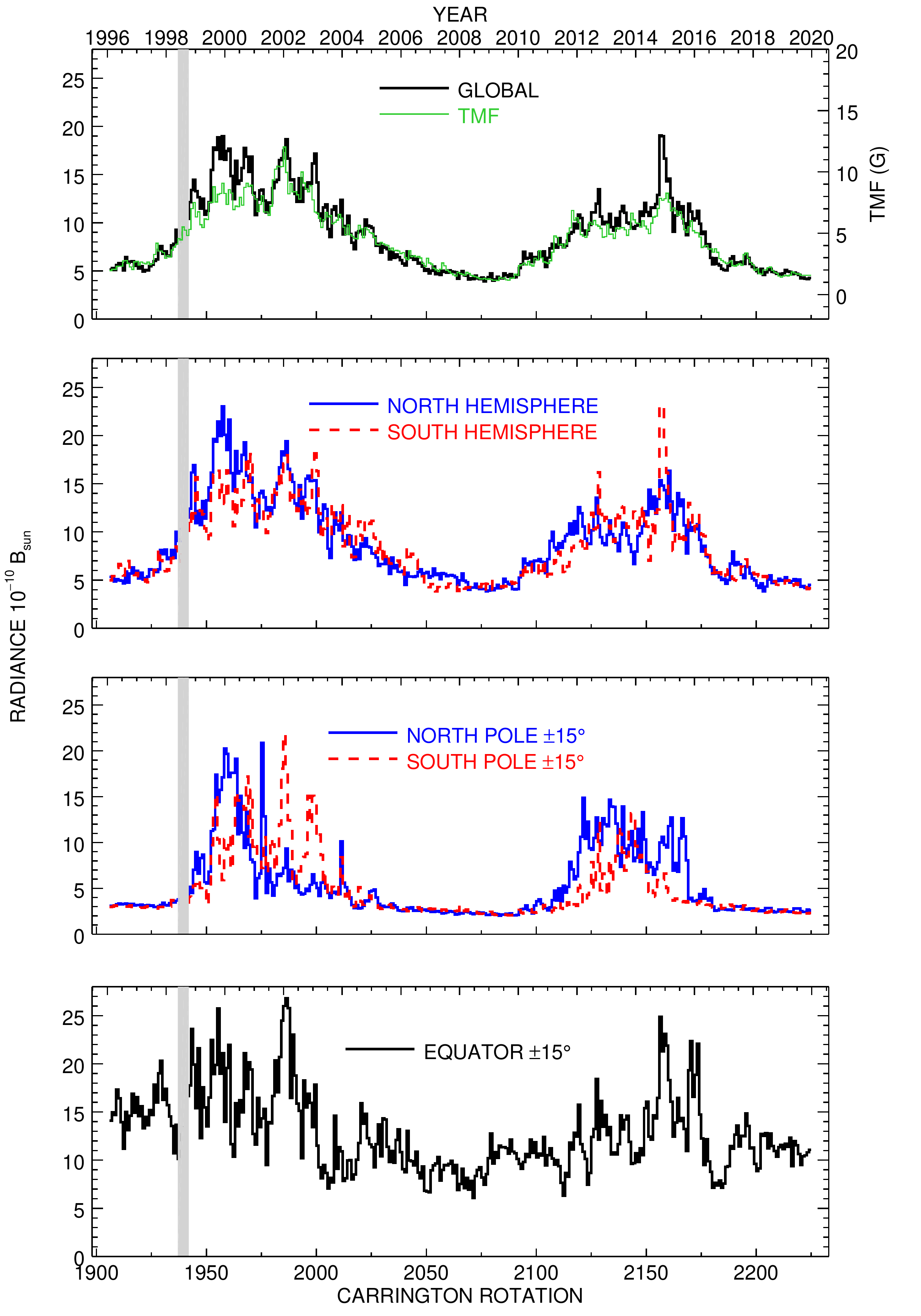}
\caption{Monthly averaged profiles extracted from the multi-annual synoptic map of the radiance of the K-corona at 2.7$\Rsun$, globally, in the two hemispheres, and in polar and equatorial sectors.}
\label{KActiv27}
\end{figure}

Images of the K-corona were calculated according to the method described in Section~\ref{Sec:Separation}, \ie using a model of $p_K$.
We display a set of figures similar to those presented above for the electron density:
\begin{itemize}
	\item Maps of the K-corona at three phases of activity of SC 23 and SC 24 as well as the corresponding profiles along the equatorial and polar directions (Figure~\ref{KSample});
	\item Multi-annual synoptic maps at two elongations 2.7 and 5.5$\Rsun$ (Figure~\ref{KSyno});
		\item Monthly averaged profiles at 2.7$\Rsun$ (Figure~\ref{KActiv27}).
\end{itemize}


\subsection{Common properties of the polarization, the polarized radiance, the electron density, and the radiance of the K-corona} 
\label{properties}

The multi-annual synoptic maps and the extracted monthly averaged profiles conspicuously show that the polarization, the polarized radiance, the electron density, and the radiance of the K-corona all follow the same spatial and temporal patterns both controlled by the solar activity cycle.
This is unsurprising for the last three quantities since they are directly related but less expected for the polarization.
However, this is readily understood on the basis of the paramount contribution of the streamers to the polarization, the evolution of the streamers in both number and brightness being intimately linked to solar activity.

The temporal evolution of the K-corona was extensively studied by \cite{barlyaeva2015mid} over a time interval of 18.5 years [1996.0--2014.5], slightly less than the 24 years considered here but their main conclusions hold and can be straightforwardly extended to the polarization, the polarized radiance, and the electron density.

The ``global'' quantities closely track the total photospheric magnetic flux at an incredible level of detail: main peaks during the two maxima but also minute fluctuations agree extremely well. 
As already alluded in Subsection~\ref{Polar}, the corona experienced a strong surge from late 2014 to beginning of 2015 in the south-east region \citep{Lamy2017}.
It can conspicuously be seen as a patch of high values in the synoptic maps and has a sudden peak in the temporal profiles for all quantities including polarization.
As explained by \cite{Lamy2017}, a specific configuration of the magnetic field that resulted from the interplay of various factors generally prevailing at the onset of the declining phase of solar cycles and which was particularly efficient in the case of SC 24 caused the electrons to be trapped forming a high density, stable bulge.

Differences in the temporal evolutions can already be suspected when comparing the results for the northern and southern hemispheres but are strongly amplified when considering the 30$^{\circ}$ wide polar sectors.
Activity is seen in all four quantities over most of SC 23 with however totally uncorrelated variations between the two sectors.
They also prevailed during SC 24 but even more striking, the activity lasted during approximately four years in the northern sector and only three years in the southern one, with a phase lag of the rising branch of roughly one year.
The profiles in the equatorial sector are at odds with the above ones being almost uncorrelated with the solar cycle except for a weak minimum during the minimum of Solar Cycle 23/24. 
This was explained by \cite{barlyaeva2015mid} as a consequence of the quasi-continuous presence of streamers in the equatorial band irrespective of the phase of solar activity.

\subsection{Periodicities in the equatorial polarization and polarized radiance.}   

\begin{figure}
      \centering
      \includegraphics[width=\textwidth]{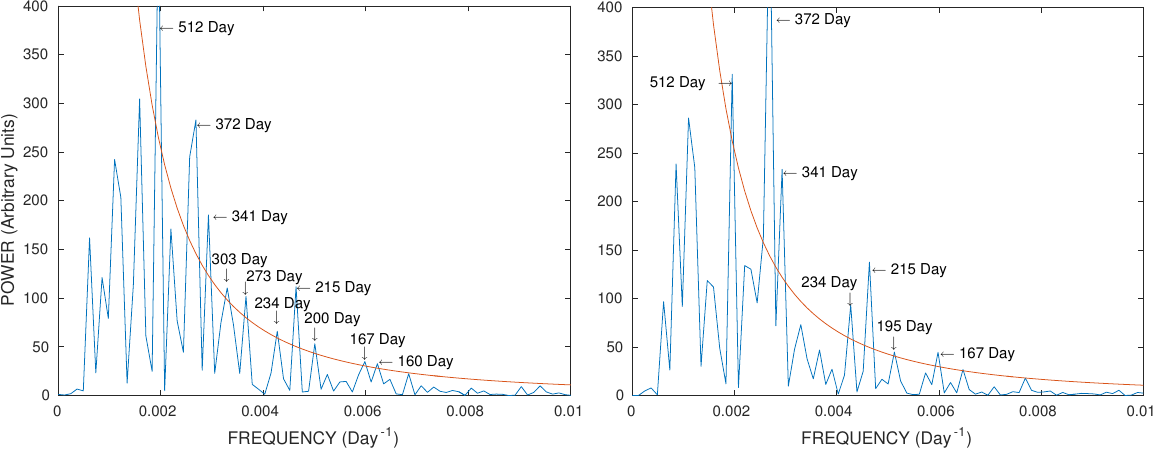}
      \caption{Periodograms of the polarization at 2.7$\Rsun$ (right panel) and the polarized radiance at 5.5$\Rsun$ (left panel) calculated in the equatorial band (equatorial direction $\pm$15$^{\circ}$).
The periods exceeding the 95\% significance levels against the red noise backgrounds (red curves) are individually labeled.}
      \label{C2_activ_fourier}
\end{figure}

Figures~\ref{PActiv27}, \ref{PActiv55}, \ref{pBActiv27}, and \ref{pBActiv55} indicate the presence of an oscillatory pattern in the temporal variation of both $p$ and $pB$ in the equatorial band (equatorial direction $\pm$15$^{\circ}$).
A frequency (periodogram) analysis was performed by first applying a standard Ã¢â‚¬Å“de-trendingÃ¢â‚¬Â (subtraction of a 25-month running average to remove the large scale temporal variations) and then proceeding with a discrete Fourier transform to generate power spectra.
The reality of the periods was ascertained by implemented the test of statistical significance at the 95\% level against the red noise background. 
We limit the presented results to two cases, the polarization at 2.7$\Rsun$ and the polarized radiance at 5.5$\Rsun$, and Figure~\ref{C2_activ_fourier} displays the corresponding power spectra in the frequency domain of the mid-term quasi-periodicities.  
A forest of seven periods spanning the range 160--512 days are statistically significant for both $p$ and $pB$.
None of them match well-known periods such as the 154-day Rieger periodicity \citep{rieger1984154} found in the temporal distribution of flares, sunspots, sunspot areas, and the radio flux F10.7 or the the 1.3-year (475-day) periodicity detected in sunspot and sunspot time series \citep{krivova2002}.
Comparing with the periodicities found in the radiance of the K-corona by \cite{barlyaeva2015mid}, we note that i) the 372-day periodicity in both $p$ and $pB$ in the equatorial band nearly match the $\approx$ 1-year periodicity found in the K-corona over most of S C23 and ii) the two periods of 215 and 234-day of the former lie in the range of $\approx$ 7--8 months found over the ascending and maximum phases of SC 24 of the latter.
Finally, a time-frequency (wavelet) analysis indicated that likewise the case of the radiance of the K-corona, the periodicities in the equatorial polarization and polarized radiance are prominently observed during the maxima of solar activity and are nearly absent during the minima.
 
Whereas the periodicities in both $p$ and $pB$ are most conspicuous in the equatorial band, they are also present in the global corona as visually perceived on their temporal variations (top panels of Figures~\ref{PActiv27}, and \ref{pBActiv27}).
They are collectively known as intermediate or mid-term quasi-periodicities and are often found in the physical features and quantities that reflect solar activity \citep{bazilevskaya2014combined}.
They share the same properties of variable periodicity, intermittency, and largest amplitudes during the maximum phase of solar cycles.
They are thought to be related to the dynamics of the deep layers of the Sun \citep{rieger1984154} and intrinsic to the solar dynamo mechanism (see discussion in \cite{barlyaeva2015mid}).

\begin{figure}
     \centering
     \includegraphics[width=\textwidth]{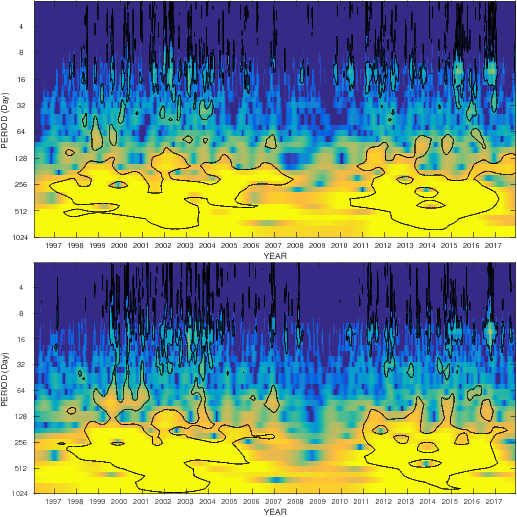}
     \caption{Wavelet spectra of the polarization at 2.7$\Rsun$ (lower panel) and the polarized radiance at 5.5$\Rsun$ (upper panel).
		  The black contours indicate the 95\% significance level.}
     \label{C2_activ_wavelet}
\end{figure}

\subsection{F-Corona} 
\label{Fcorona}

\begin{figure}[htpb!]
\begin{center}
\includegraphics[width=0.85\textwidth]{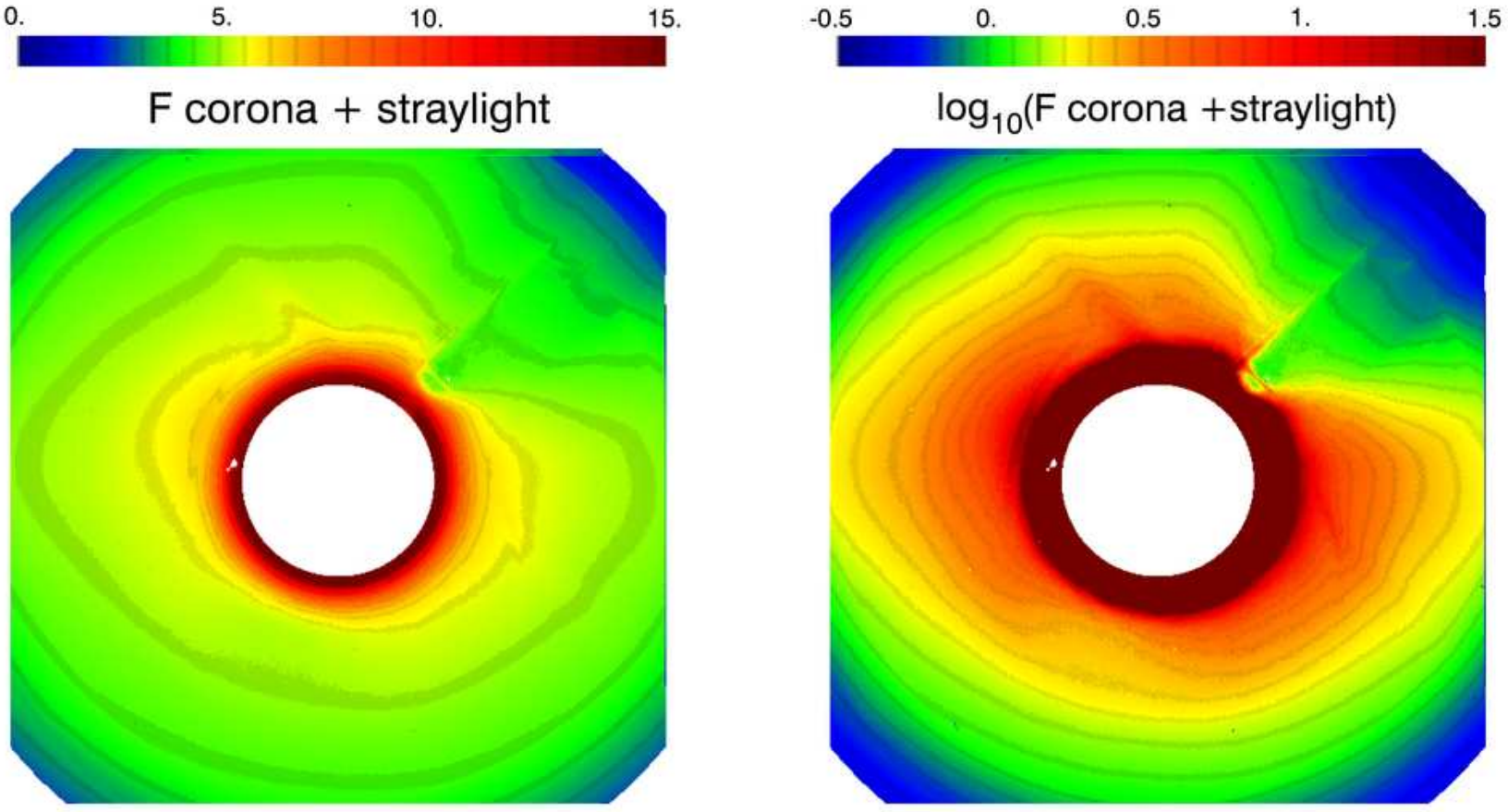}
\caption{Two displays of and image resulting from the separation process where the vignetted F-corona and the stray light are still mixed.
The left display uses a linear scale to best view the diffraction fringe surrounding the occulter.
The right display uses a logarithmic scale to best view the F-corona distorted by several stray light structures as well as by the vignetting from the pylon holding the occulter).}
\label{Fcor_strlight_4}
\end{center}
\end{figure}

As stated in Section~\ref{Sec:Separation}, the K/F separation process allows us retrieving $B_K$ but leaves the two unpolarized components of the observed radiance as a sum $B_F$ + $B_S$ of the F-corona and the stray light. 
Figure~\ref{Fcor_strlight_4} displays an example of such an image which reveals that the stray light pattern is not only composed of the diffraction fringe surrounding the occulter, but also of several structures which distort the expected smooth ``elliptic'' shape of the isophotes of the F-corona.
The separation of these two components to correctly retrieve the F-corona turned out to be an extremely complex task whose outcome will be presented in a forthcoming article.

\section{Uncertainty Estimates and Comparison with Eclipse Results}

The question of the estimation of uncertainties in polarized observations of the solar corona was recently addressed by \cite{frazin2012intercomparison} in the framework of the inter-comparison of the LASCO-C2, SECCHI-COR1, SECCHI-COR2, and Mark IV data where they considered both the polarized radiance $pB$ and the total radiance $B$. 
The error analysis for the C2 polarized data is particularly complex in view of the many sources of error and of the implemented optimization procedure.
Even a Monte-Carlo error propagation, assuming that the uncertainties on the involved parameters could be rigorously quantified, would be difficult to implement in the case of such a procedure.

Let us consider the various situations and consider first the polarization $p$.
As discussed in Section \ref{Improvement}, errors on the relative global transmission between polarizers and on the individual exposure times of the triplet forming a polarization sequence have similar effects which are corrected for by the optimization procedure based on imposing a tangential direction of polarization and minimal dispersion around this direction. 
As illustrated in Figure~\ref{FigStatParamPol}, the standard deviation of the dispersion amounts to typically $2\deg$ which translates to a formal error on the polarization $< 0.1 \%$.
The absolute value of the global transmissions introduces a larger uncertainty which can be assessed by the laboratory calibration (Figure~\ref{Polar12}) and estimated at a relative level of $\approx$8 \% (for instance, $\pm$0.013 for $p$=0.15 and $\pm$0.025 for $p$=0.30).
Local inhomogeneities in the principal transmittance k$_{1}$ of the polarizers are small (Figure~\ref{ImageK1C2}) and are calibrated at a 1 \% level insuring that their influence is negligible.
However, they are based on a single calibration and aging effects can certainly not be excluded.
As discussed in Section \ref{Improvement}, the two folding mirrors were found to have less effects than the polarizers, with the additional argument that their hard, low-polarization coating is certainly less prone to degradation that the Polaroid foils. 

Turning our attention to the polarized radiance $pB$, it benefits from an additional correction through the global correction function $S(x,y)$, although this function may also evolve with time. 
The absolute calibration of $pB$ implements a two-step procedure, first with the ratio $I_{0}/I_{cp}$ whose final value is determined with an accuracy of $\pm$0.001 (Figure~\ref{C2CalCoefOpt}) and second, with the calibration of the unpolarized $B$ images based on photometric
measurements of thousands of observations of stars present in the C2 field of view resulting in an uncertainty of 1 \%.

Finally, the separation of the three components K, F, and stray light S (see Equation 27) to retrieve the radiance of the K-corona requires several assumptions, namely that the F-corona and the stray light are unpolarized and that $p_K$ obeys a prescribed model. 
These assumptions are obviously all sources of uncertainties affecting the determination of the $B_K$ maps which are difficult to assess.
The problem is however alleviated by the fact that our subsequent derivation of the electron density is prominently based on $pB$ data thus by-passing the above uncertainties.

One of the purpose of the aforementioned intercomparison work of \cite{frazin2012intercomparison} was to assess the correctness of the uncertainties derived for the different coronagraphs.
They concluded that the agreement between all of the instrumental $pB$ values were within the quoted uncertainties in bright streamers, but much less so for the coronal holes, except when comparing Mark IV and C2 data.
However, the overlap between the useful fields of view of Mark IV and C2 is very narrow and corresponds to regions where the quality of the data is compromised by the uncorrected sky polarization and the detector bit error for Mark IV and by the diffraction fringe for C2. 
We follow below the relevant intercomparison approach but avoid using the Mark IV data and favor eclipse observations as they allow a much more comfortable overlap with C2.
The prerequisite is the availability of high quality data and this seriously limits the possibilities.
For this present analysis, we could only locate five data sets obtained at four different eclipses, in particular thanks to unpublished data made available to us.

\subsection{Eclipse on 26 February 1998}

\begin{figure}[htpb!]
\begin{center}
\label{}
\includegraphics[width=0.6\textwidth]{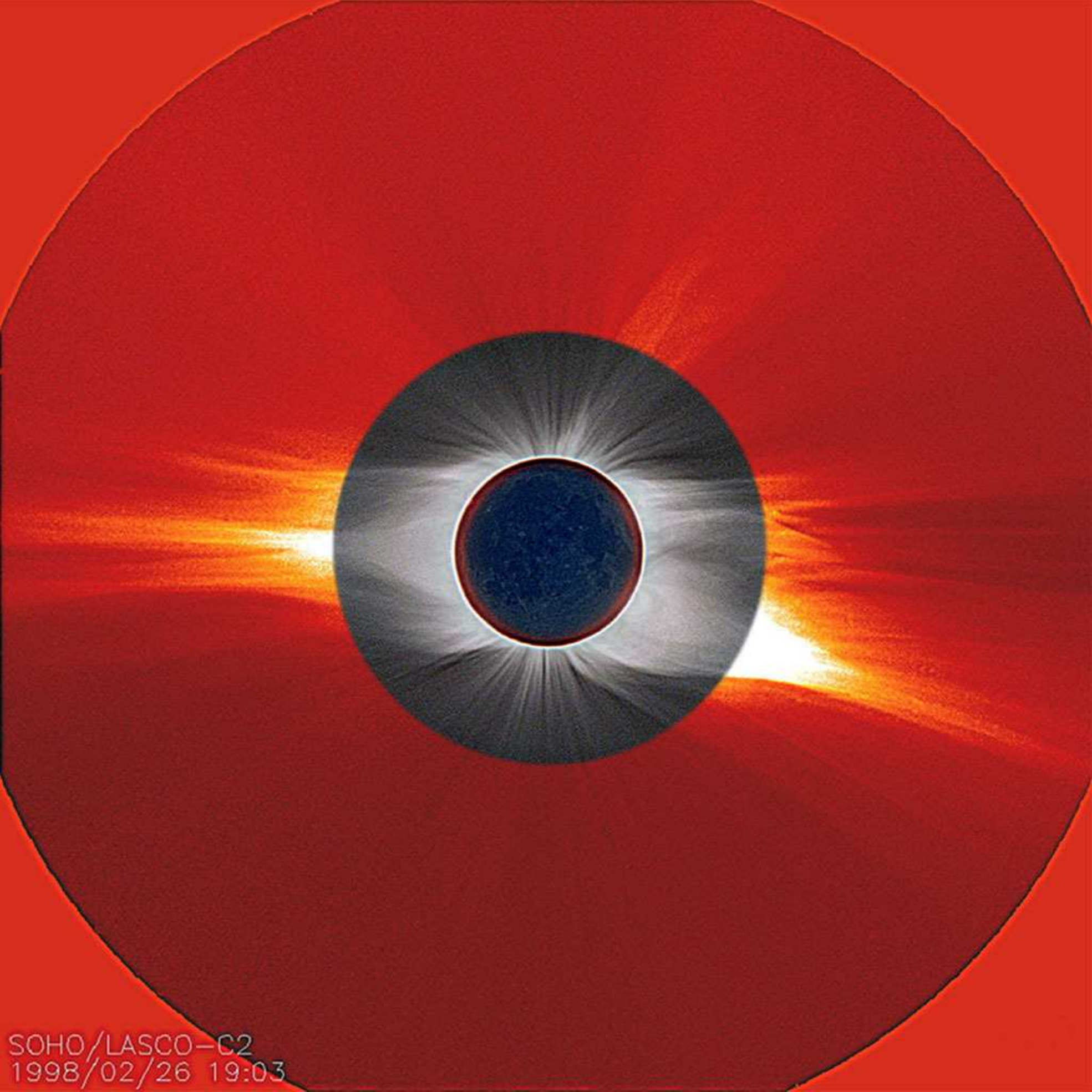} 
\caption{Composite of two images of the eclipse on 26 February 1998, a highly processed image obtained by C. Viladrich of the inner corona and a LASCO-C2 image of the outer corona (courtesy S. Koutchmy).
Solar north is up.}
\label{FigEcl1998}
\end{center}
\end{figure}

Figure~\ref{FigEcl1998} displays a qualitative composite constructed by S. Koutchmy of two images, a highly processed one obtained by C. Viladrich of the inner corona and a LASCO-C2 image of the outer corona, mostly intended to reveal the continuity of the coronal structures.
This eclipse was observed by a team of the High Altitude Observatory (HAO) with their {\it Polarimetric Imager for Solar Eclipse 98} (POISE98) at Westpoint, Cura\c{c}ao.
Their 1000 mm, f/13 telescope was equipped with an uncooled CCD camera of 2034$\times$2034 pixels offering 16-bit digitalization.
The pixel \fov was 3.095 arcsec and the total \fov was 6.56 R${}_\odot$.
Polarization analysis was achieved by a liquid crystal variable retarder operating at 620 nm with a bandpass of 10 nm and polarized images were obtained at four retardance settings and with different exposure times.
Absolute calibration was performed with a standard HAO opal.
Further detail can be found in \cite{lites2000dynamics} who presented two $pB$ profiles of the corona in the southern region.
The images themselves of $pB$ and $B$ used in the present analysis (and from which we derived a $p$ image) were provided to us by D. Elmore.
This HAO observation was bracketed by two LASCO-C2 polarization sequences each taken within a couple of hours. 
The resulting $p$ and $pB$ images were averaged for the purpose of the comparison with the HAO images.

\begin{figure}[htpb!]
\begin{center}
  \includegraphics[width=1.\textwidth]{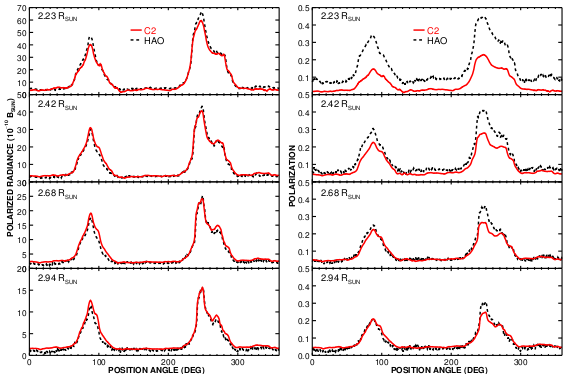} 
\caption{Circular profiles extracted at four elongations from HAO and C2 images of the polarized radiance $pB$ (left column) and from images of the polarization $p$ (right column).}
\label{FigCircProfHAOC2}
  \end{center}
\end{figure}

The high resolution HAO images were processed so as to match the lower spatial scale and the orientation of the C2 images. 
Circular profiles were extracted at four heliocentric distances in the overlap region and Figure~\ref{FigCircProfHAOC2} displays the polarized radiance and the polarization as functions of position angle $PA$ measured from solar North toward East (counter-clockwise direction).
The agreement of the $pB$ data is impressive with just a slight mismatch of the two peak radiances in the streamer belt at 2.23 R${}_\odot$.
The HAO and C2 polarization profiles follow the same pattern, but with a scaling factor which decreases with increasing heliocentric distance so that the agreement becomes very satisfactory at 2.9 R${}_\odot$ except for the peak in the western streamer belt 
( $PA$ $\approx$ 250$\deg$ ).

\begin{figure}[htpb!]
\begin{minipage}{1.\textwidth}
\begin{center}
  \includegraphics[width=0.47\textwidth]{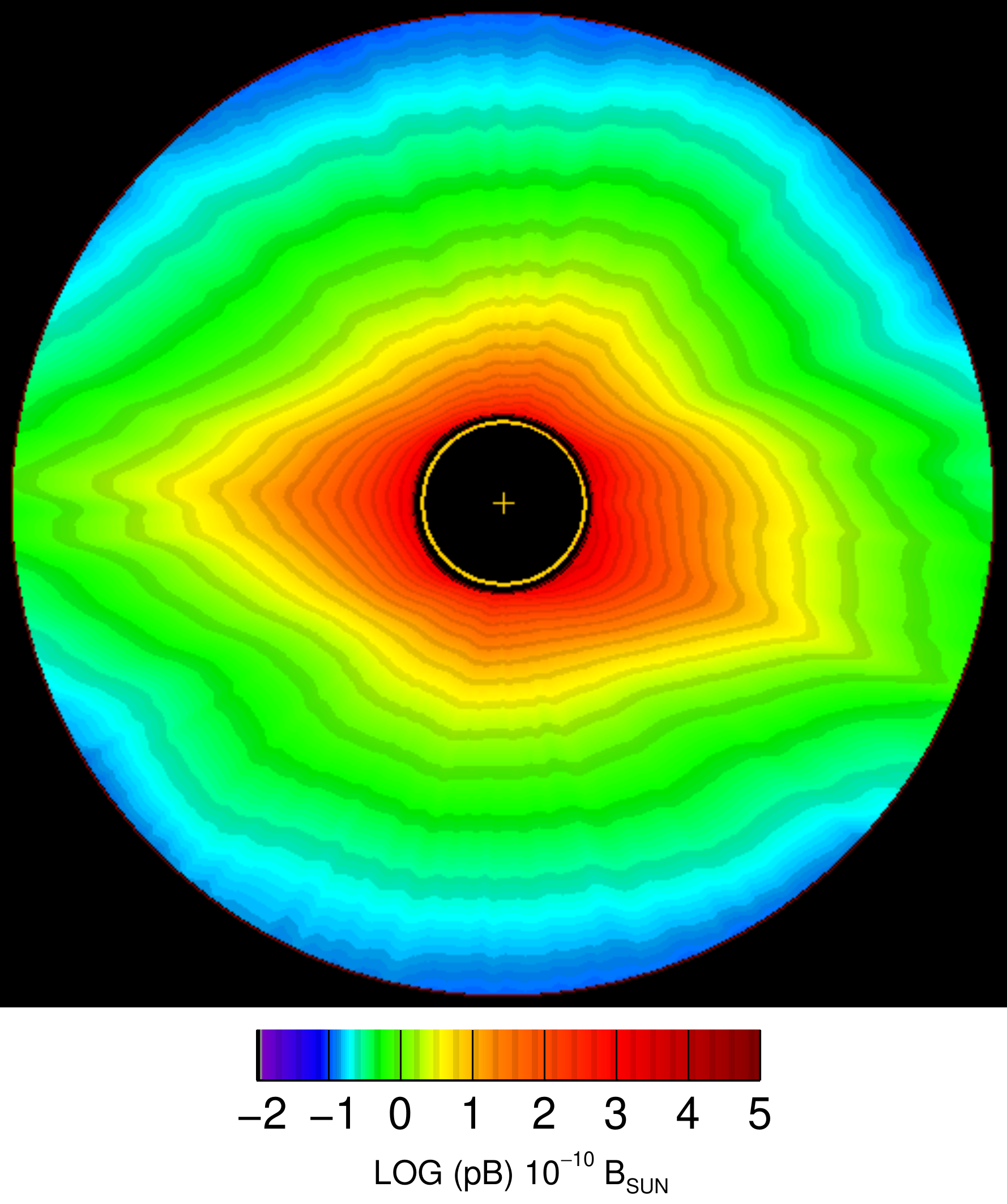} 
\hspace{0.01\textwidth}
  \includegraphics[width=0.47\textwidth]{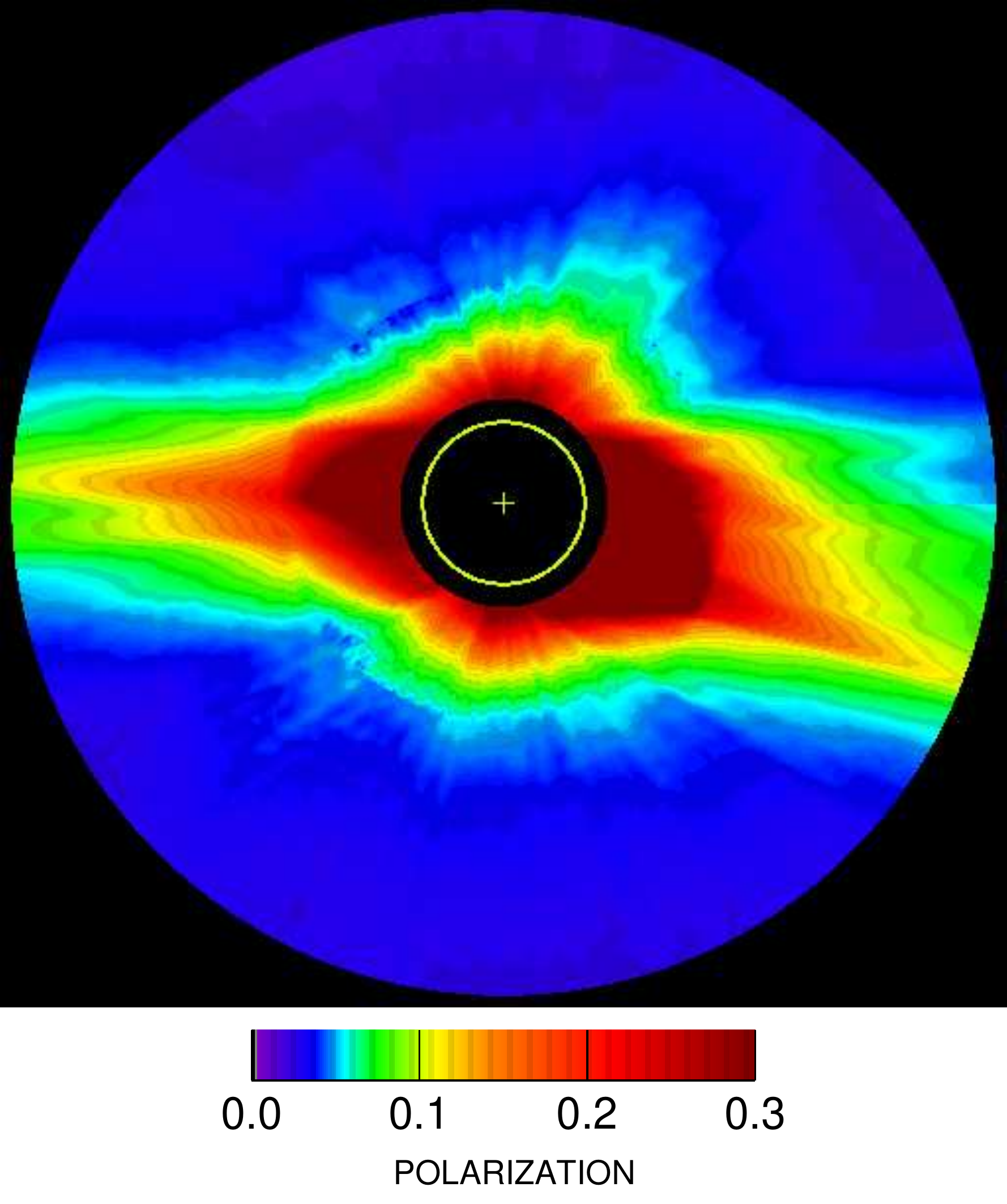}
\hspace{0.01\textwidth}  
\caption{Composites of the HAO and C2 images of the polarized radiance $pB$ (left panel) and the polarization $p$ (right panel).
The yellow circles represent the solar disk with a cross at its center.
Solar north is up.}
\label{FigComposite}
  \end{center}
\end{minipage}
\end{figure}

\begin{figure}[htpb!]
\begin{center}
  \includegraphics[width=1.\textwidth]{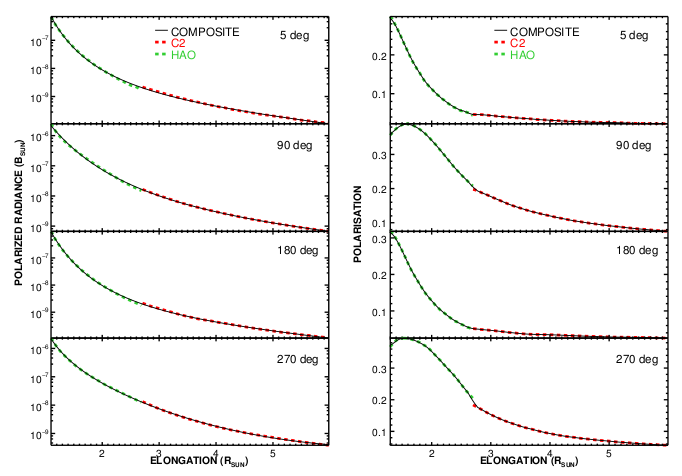} 
\hspace{0.01\textwidth}
\caption{Radial profiles along four directions of the composites of the HAO and C2 images of the polarized radiance $pB$ (left column) and the polarization $p$ (right column).
The profiles are along the north ($PA$ = $5\deg$), east ($PA$ = $90\deg$), south ($PA$ = $180\deg$), and west ($PA$ = $270\deg$) directions.
The separate profiles of the HAO and C2 images are over-plotted for comparison.}
\label{FigRadprofComposite}
  \end{center}
\end{figure}

In view of these excellent results, we decided to build two composites merging the HAO data up to 2.7 R${}_\odot$ and the C2 data beyond as displayed in Figure~\ref{FigComposite}. 
The $pB$ composite was slightly smoothed to iron out minute imperfections at the junction of the two images but with no incidence at all on the photometry as clearly demonstrated by the four radial profiles shown in Figure~\ref{FigRadprofComposite}.
Although less perfect, the polarization composite confirms the excellent continuity between the HAO and C2 data, also illustrated by their radial profiles (Figure~\ref{FigRadprofComposite}).

\begin{figure}[htpb!]
\begin{center}
  \includegraphics[width=1.\textwidth]{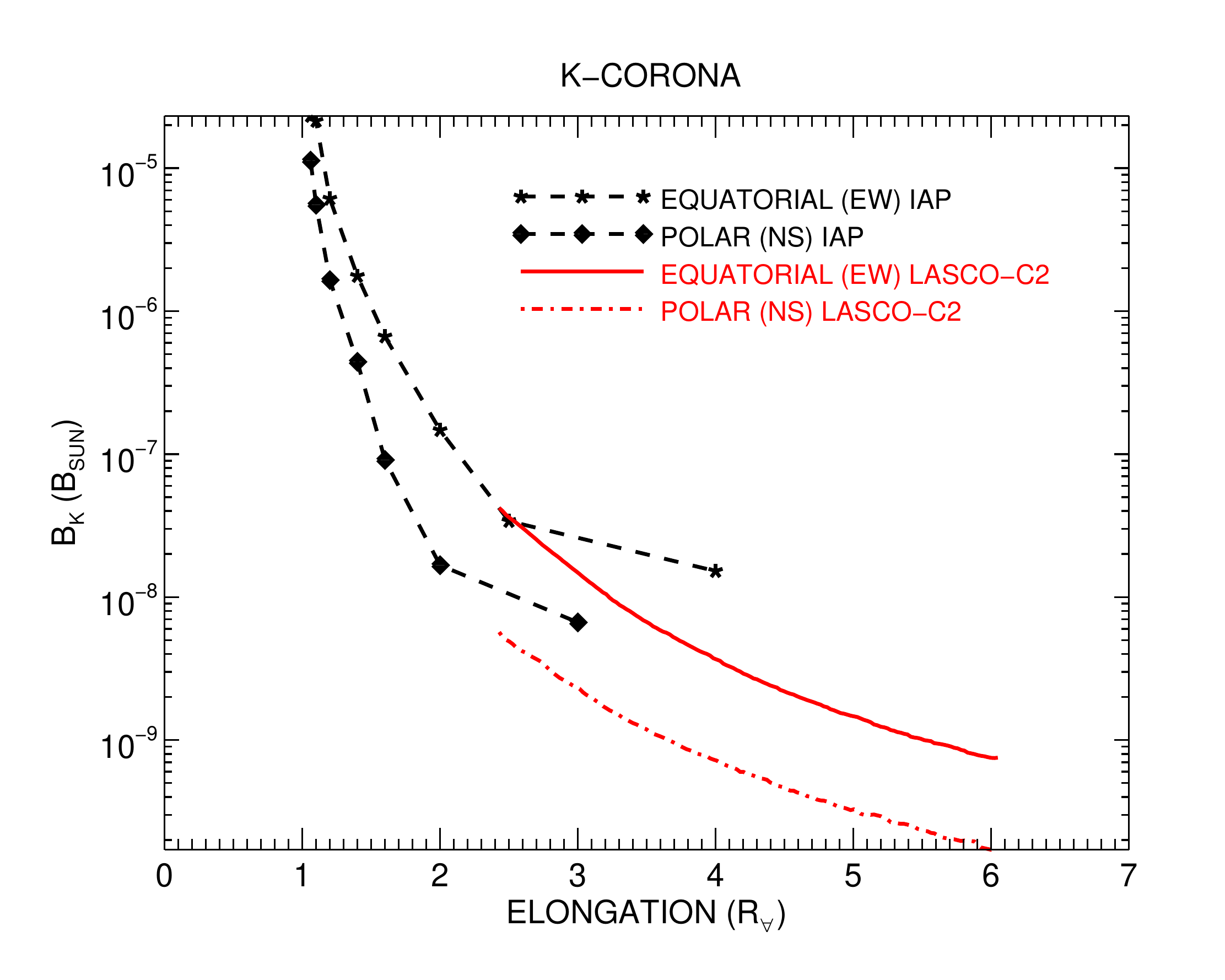} 
\caption{Radial profiles of the radiance of the K-corona derived from ground-based measurements at the eclipse of 1998 (IAP) and quasi-simultaneous images obtained by LASCO-C2.
The east and west profiles have been averaged to yield a single equatorial profile.
Likewise, the north and south profiles have been averaged to yield a single polar profile.}
\label{FigRadprofIAPC2}
  \end{center}
\end{figure}

We report on an additional observation of the 1998 eclipse carried out by a team of Institut d'Astrophysique de Paris (IAP) based at Gury, Guadeloupe.
It did not involve polarization but it gives further insight to the eclipse-C2 intercomparison. 
Photographic images of the corona were obtained with a 624 mm, f/8 Takahashi fluorite refractor and {\it VELVIA} 24$\times$36 mm color film.
The photometric analysis was performed with the IAP Bruckner microdensitometer equipped with a green filter centered at 546 nm by S. Hamot under the supervision of S. Koutchmy and consisted in i) scanning the film along equatorial and polar directions with a long narrow slit oriented tangentially, and ii) recording five stars in the field of view for absolute calibration. 
The sky background measured at large distances was subtracted from the profiles which were then combined to produce two average profiles along the equatorial and polar directions. 
Finally, the F-corona model along these directions of \citet{koutchmy1985f} was subtracted to produce two profiles of the K-corona which were made available to us by S. Koutchmy.
We focus the comparison on these profiles because, in the case of C2, the $B_K$ images are a product of the polarization analysis.
Similar profiles were therefore constructed from the average of the two C2 $B_K$ images that bracketed the eclipse and they are displayed together with the IAP profiles in Figure~\ref{FigRadprofIAPC2}.
If we exclude the two extreme data points at 3 and 4 R${}_\odot$ of respectively the polar and equatorial profiles which are probably affected by residual sky background, the match between the IAP and C2 profiles is quite remarkable.

\vspace*{\stretch{1}}

\subsection{Eclipse on 11 August 1999}

\begin{figure}[htpb!]
\begin{center}
\includegraphics[width=0.6\textwidth]{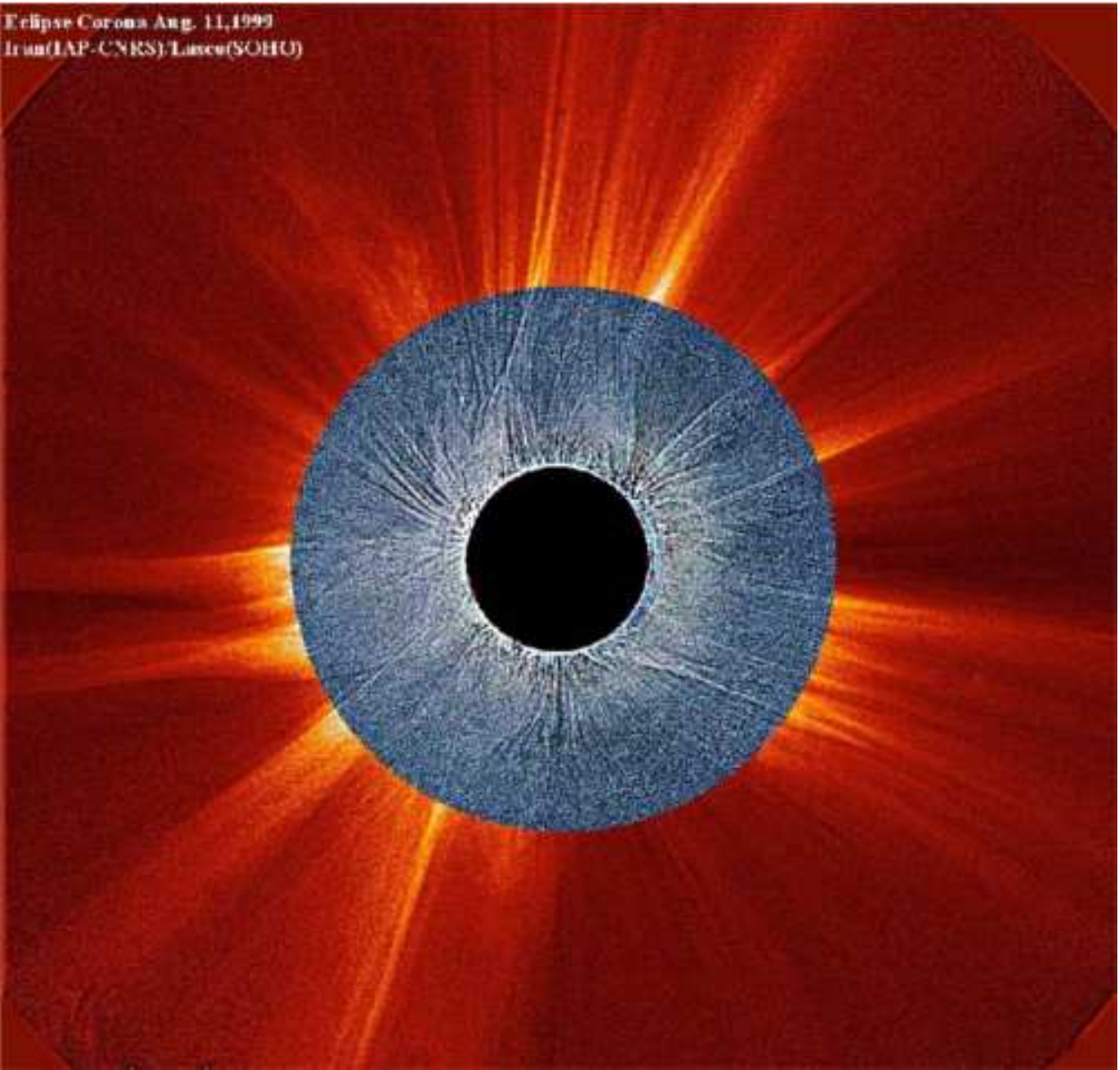} 
\caption{Composite of two images of the eclipse on 11 August 1999, a highly processed image obtained by a team of Institut d'Astrophysique de Paris of the inner corona and a LASCO-C2 image of the outer corona (courtesy S. Koutchmy).
Solar north is up.}
\label{FigEcl1999}
\end{center}
\end{figure}

Figure~\ref{FigEcl1999} displays a qualitative composite constructed by S. Koutchmy of two images, a highly processed one obtained by a team of Institut d'Astrophysique de Paris of the inner corona and a LASCO-C2 image of the outer corona.
This eclipse was observed by the first Author at Chadegan, Iran and he obtained thirty seven photographic images with a 500 mm, f/8 Nikkor objective and {\it Extachrome} (200 ASA) 24$\times$36 mm color film resulting in a \fov of 10$\times$14 R${}_\odot$.
These images covered totality as well as the partial phases before and after the eclipse for calibration purpose based on attenuated images of the solar disk.
Polarization analysis was achieved by a rotating linear polarizer placed in front of the objective and oriented along seven preselected directions separated by $30\deg$. 
The polarization sequence took place at the mid-point of totality (12:03 UT) to offer the best conditions and was preceded and followed by identical sequences of unpolarized images taken with different exposure times.
All photographic images were digitized on 12 bits with a Nikon LS2000 scanner simultaneously in two colors ($R$ and $G$) and at the maximum spatial resolution thus yielding a format of 3894$\times$2592 pixels.
The processing of the $G$ images was performed by M. Bout under the supervision of the first Author and consisted in i) determining the characteristic curve of the film and converting photographic densities to intensities, ii) constructing the vignetting function and correcting the images, iii) determining the geometric parameters (pixel coordinates of the center of the Sun and the spatial scale), iv) deriving the absolute calibration from images of the solar disk after correcting for the differential atmospheric transmission, v) determining and subtracting the background from each image, and vi) performing the polarimetric analysis.

\begin{figure}[htpb!]
\begin{minipage}{1.\textwidth}
\begin{center}
  \includegraphics[width=0.49\textwidth]{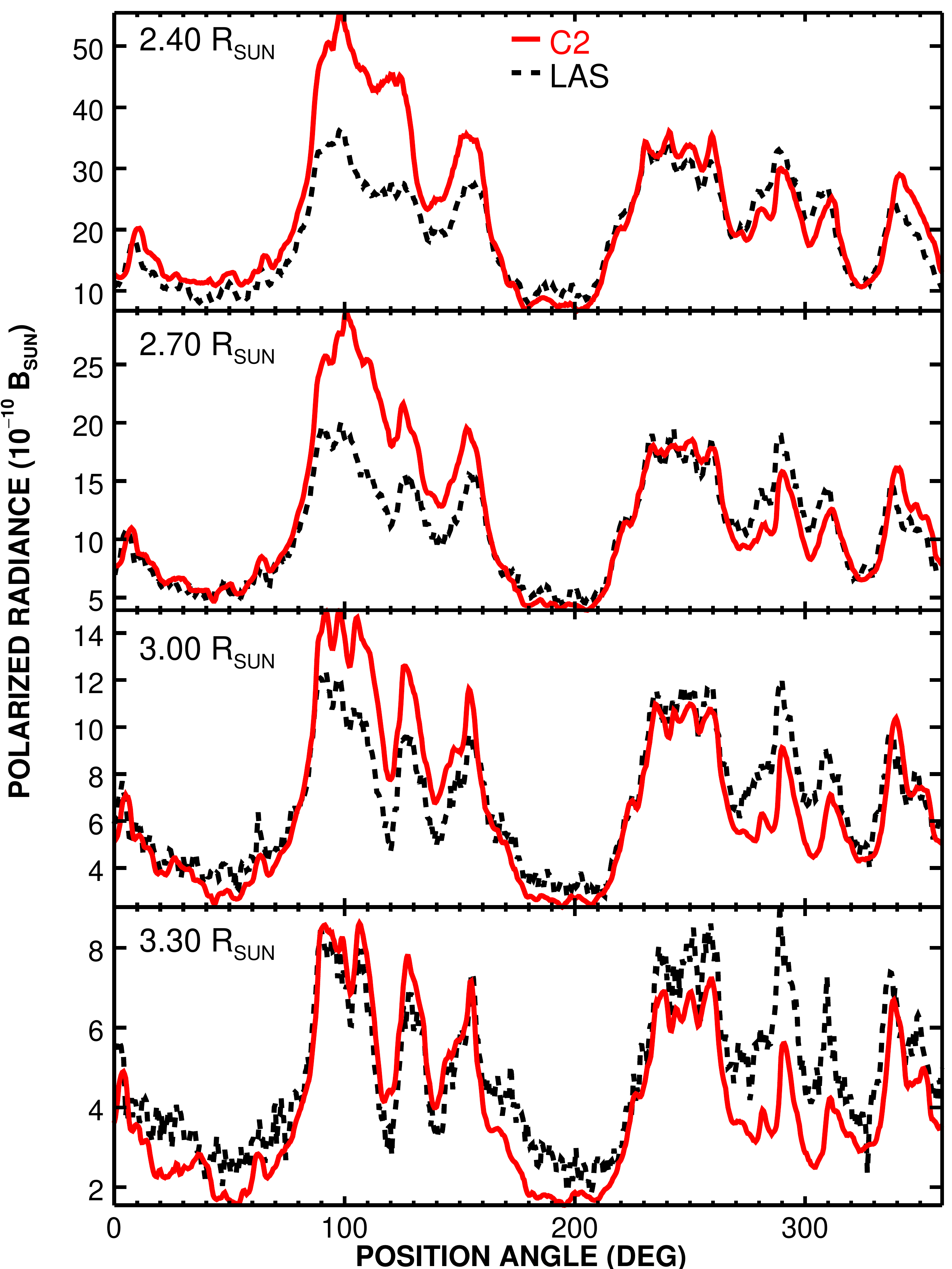} 
  \includegraphics[width=0.49\textwidth]{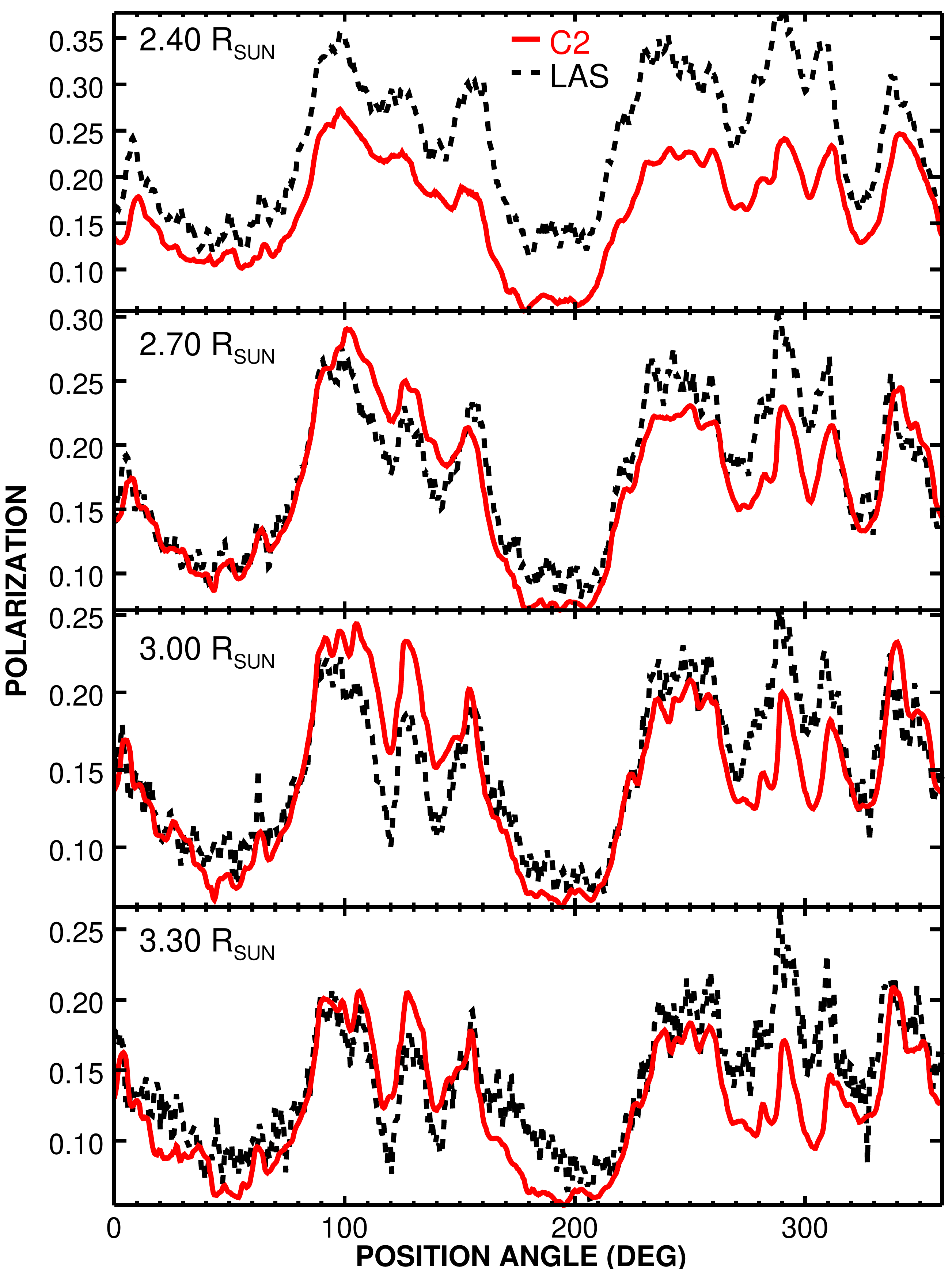}
\caption{Circular profiles extracted at four elongations from LAS and C2 images of the polarized radiance $pB$ (left column) and from images of the polarization $p$ (right column) obtained at the eclipse on 11 August 1999.}
\label{FigCircProfLASC2}
  \end{center}
\end{minipage}
\end{figure}

A LASCO-C2 polarization sequence was taken half an hour before the above observation, precisely at 11:29 UT. 
The comparison with the ground-based observations follows the same procedure as developed for the HAO-C2 case.
Circular profiles were extracted at four heliocentric distances in the overlap region and Figure~\ref{FigCircProfLASC2} displays the polarized radiance and the polarization as functions of position angle. 
The ground-based observations are labeled ``LAS'' for Laboratoire d'Astronomie Spatiale, the former name of Laboratoire d'Astrophysique de Marseille.
The agreement of the $pB$ and $p$ data is globally extremely satisfactory with however minute differences and two main discrepancies.
Concerning $pB$, the broad streamer system in the east-south quadrant is fainter in the case of C2 at elongations $\leq$3 R${}_\odot$.
Concerning the polarization, in the inner corona at 2.4 R${}_\odot$, the C2 data are systematically fainter than the LAS data by a factor of $\approx$0.7.

\begin{figure}[htpb!]
\begin{minipage}{1.\textwidth}
\begin{center}
  \includegraphics[width=0.47\textwidth]{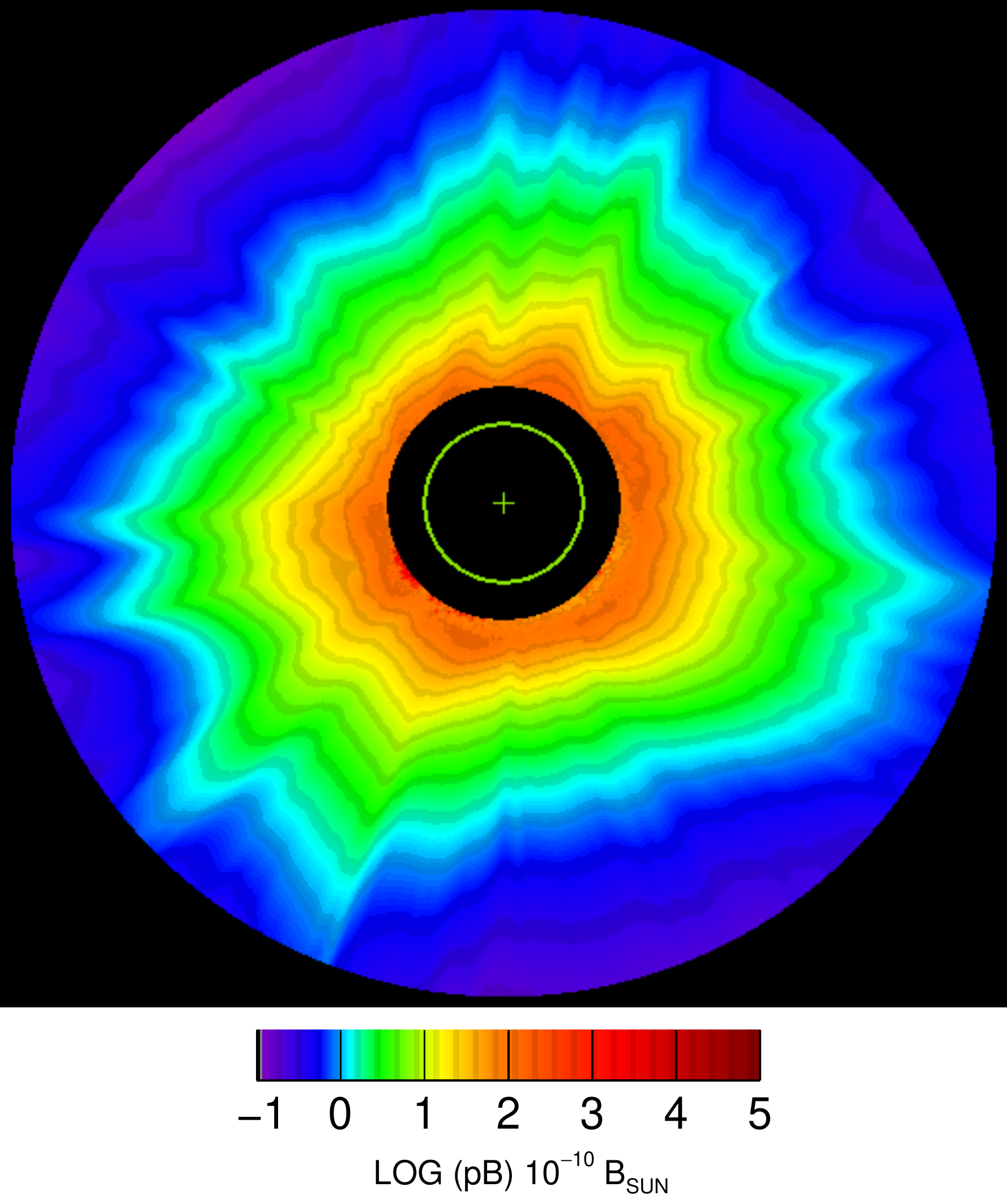} 
\hspace{0.01\textwidth}
  \includegraphics[width=0.47\textwidth]{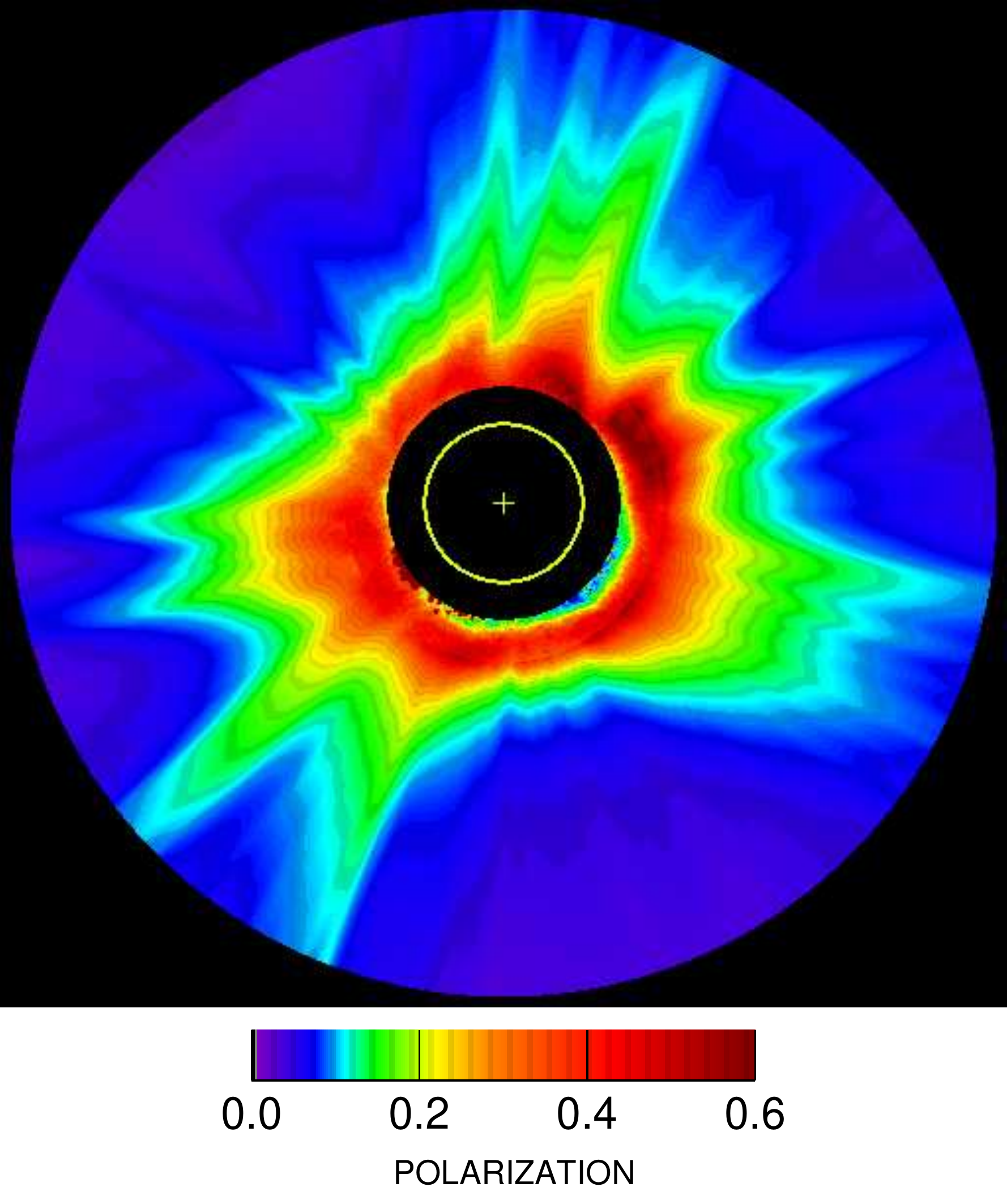}
\hspace{0.01\textwidth}  
\caption{Composites of the LAS and C2 images of the polarized radiance $pB$ (left panel) and the polarization $p$ (right panel) obtained at the eclipse on 11 August 1999.
The yellow circles represent the solar disk with a cross at its center.
Solar north is up.}
\label{FigCompositeLASC2}
  \end{center}
\end{minipage}
\end{figure}

\begin{figure}[htpb!]
\begin{minipage}{1.\textwidth}
\begin{center}
  \includegraphics[width=0.47\textwidth]{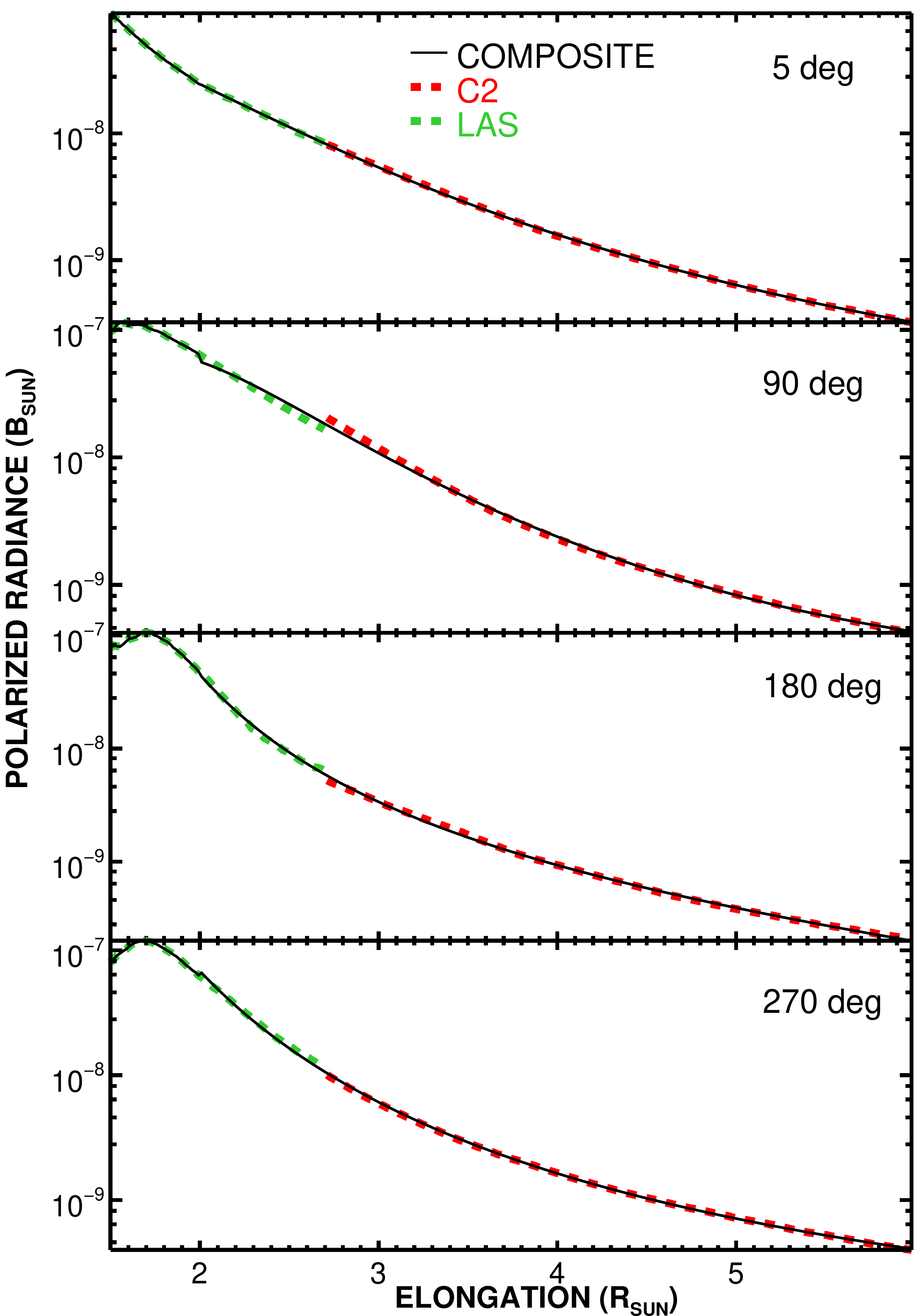} 
\hspace{0.01\textwidth}
  \includegraphics[width=0.47\textwidth]{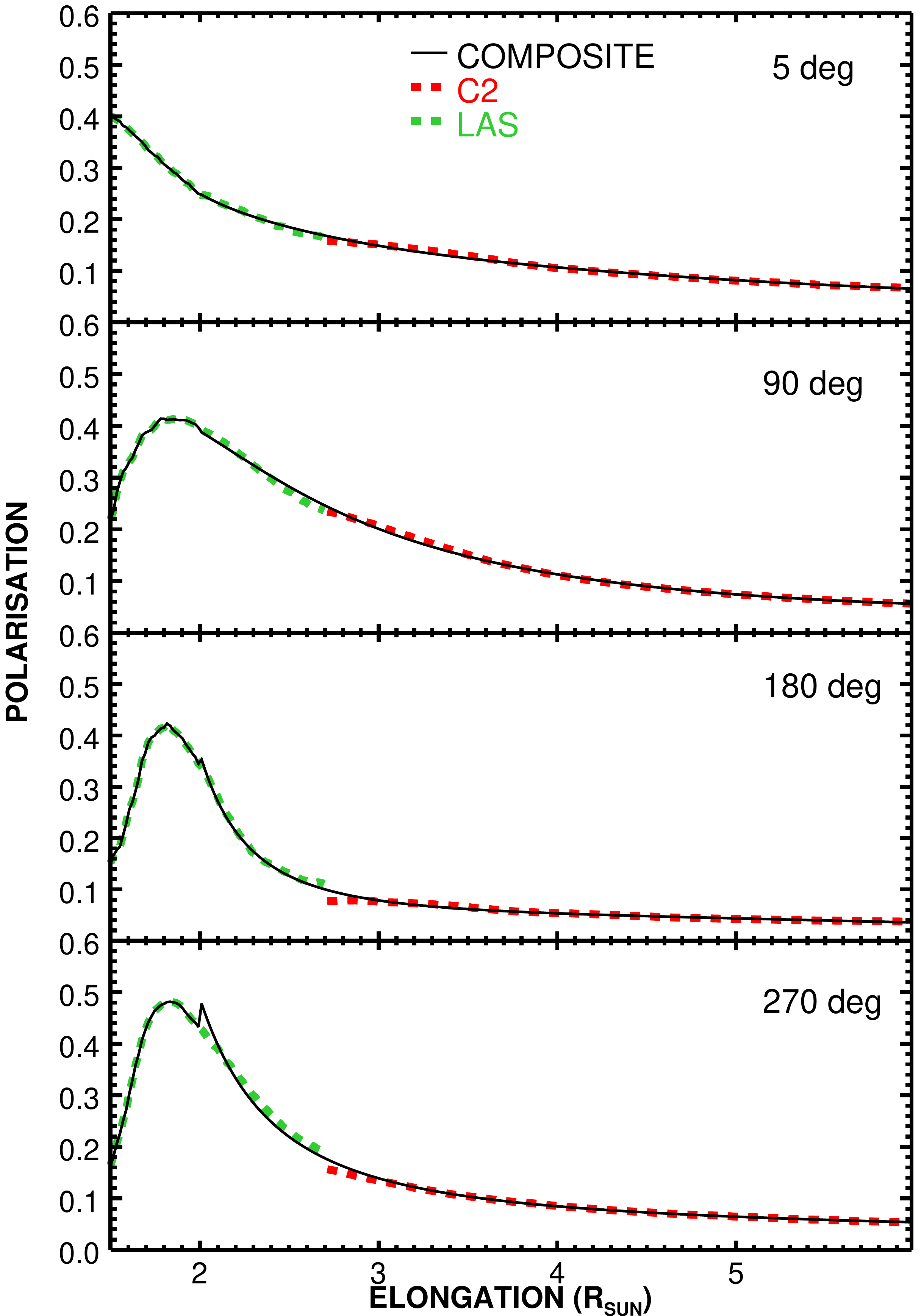}
\caption{Radial profiles along four directions of the composites of the LAS and C2 images of the polarized radiance $pB$ (left column) and the polarization $p$ (right column) obtained at the eclipse on 11 August 1999.
The profiles are along the north ($PA$ = $5\deg$), east ($PA$ = $90\deg$), south ($PA$ = $180\deg$), and west ($PA$ = $270\deg$) directions.
The separate profiles of the LAS and C2 images are over-plotted for comparison.}
\label{FigRadprofCompositeLASC2}
  \end{center}
\end{minipage}
\end{figure}

Likewise the HAO-C2 case, we built two composites merging the LAM data up to 2.7 R${}_\odot$ and the C2 data beyond as displayed in Figure~\ref{FigCompositeLASC2}. 
The $pB$ composite was slightly smoothed to iron out minute imperfections at the junction of the two images but with no incidence at all on the photometry as clearly demonstrated by the four radial profiles shown in Figure~\ref{FigRadprofCompositeLASC2}.
Although less perfect, the polarization composite confirms the excellent continuity between the LAS and C2 data also illustrated by their radial profiles (Figure~\ref{FigRadprofCompositeLASC2}).

\vspace*{\stretch{1}}

\subsection{Eclipse on 29 March 2006}

\begin{figure}[htpb!]
\label{}
\begin{center}
  \includegraphics[width=0.6\textwidth]{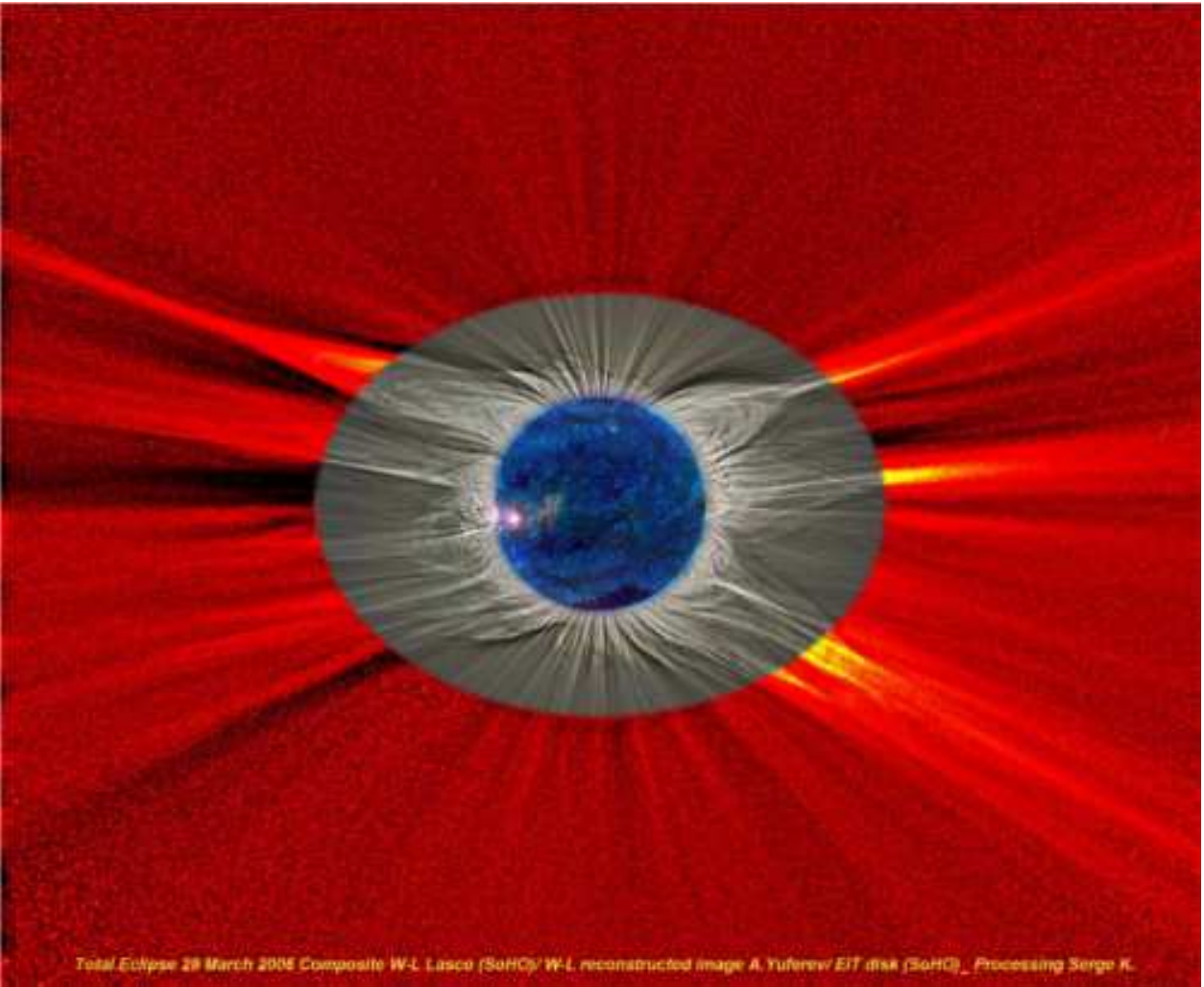}
\caption{Composite of two images of the eclipse on 29 March 2006, a highly processed image of the inner corona obtained by A. Yuferev and a LASCO-C2 image of the outer corona (courtesy S. Koutchmy).
Solar north is up.}
\label{FigEcl2006}
\end{center}
\end{figure}

Figure~\ref{FigEcl2006} displays a qualitative composite constructed by S. Koutchmy combining an EIT image of the solar disk, a highly processed image obtained by A. Yuferev, and a LASCO-C2 image of the outer corona.
Polarization observations were performed by two teams, an italian team from the Osservatorio Astronomico di Torino located at Waw an Namus, Libya, and a french team from Institut d'Astrophysique de Paris located at As Sallum, Egypt.

The instrument setup of the first team named E-KPol is described in \cite{zangrilli2007EKPOL} and implemented a 600 mm, f/12 objective and a cooled CCD camera of 1024$\times$1204 pixels offering a 16-bit digitalization.
The pixel \fov was 8.6 arcsec and the total \fov was 8 R${}_\odot$.
Polarization analysis was achieved by a liquid crystal variable retarder operating at 620 nm with a bandpass of 80 nm and polarized images were obtained at four retardance settings and with three different exposure times.
Absolute calibration was performed with an opal.
The results are presented in \cite{Capobianco2012electro}. 
The $p$ and $pB$ images used in the present analysis were provided to us by G. Capobianco and they combine two full observational sequences, each performed with the four settings and three exposure times.

The instrument setup of the second team implemented a 180 mm, f/5.6 teleobjective and a Canon EOS 350D camera with a CMOS detector of 2080$\times$1854 pixels offering a 12-bit digitalization.
The pixel \fov was 19.84 arcsec and the total \fov was 14 R${}_\odot$.
Polarization analysis was achieved by a rotating Nikon linear polarizer placed in front of the teleobjective and oriented at four preselected directions separated by $45\deg$. 
The final image of the polarization in the green channel of the CMOS detector (effective wavelength of 550 nm and bandpass of 80 nm) combines images obtained with four different exposure times after careful re-centering by correlation; in addition, the linearly polarized sky background was subtracted. 
This image was provided to us by F. S\`evres and S. Koutchmy.

\begin{figure}[htpb!]
\begin{center}
  \includegraphics[width=0.9\textwidth]{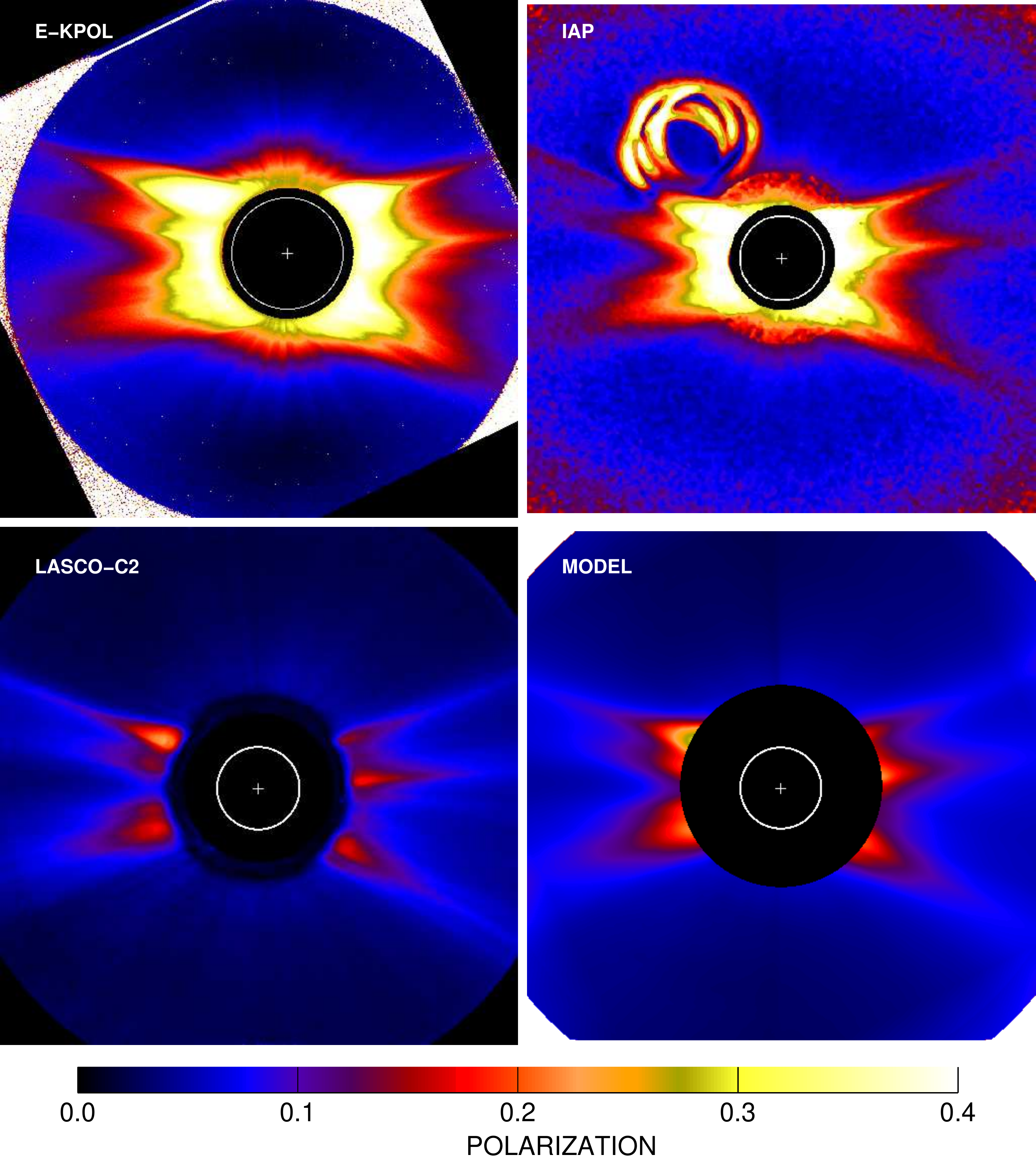} 
\caption{Polarization maps of the corona constructed from E-KPol (upper left panel) and IAP (upper right panel) observations at the eclipse on 29 March 2006, and from a quasi-simultaneous polarization sequence of LASCO-C2 (lower left panel). 
The lower right panel displays a map calculated from models of the K- and F-coronae (see text for detail).
The white circles represent the solar disk with a cross at its center.
Solar north is up.}
\label{FigPolarMaps}
  \end{center}
\end{figure}

A LASCO-C2 polarization sequence was taken approximately 25 minutes after the above observations allowing to obtain $p$ and $pB$ images.
The original polarization maps of the corona from these three observations, E-Kpol, IAP, and C2 are displayed in Figure~\ref{FigPolarMaps}.
Note the ghosts of the bright inner corona resulting from reflections off the polarizer and which preclude comparison in the North-East quadrant of the IAP images.

\begin{figure}[htpb!]
\begin{minipage}{1.\textwidth}
\begin{center}
  \includegraphics[width=0.49\textwidth]{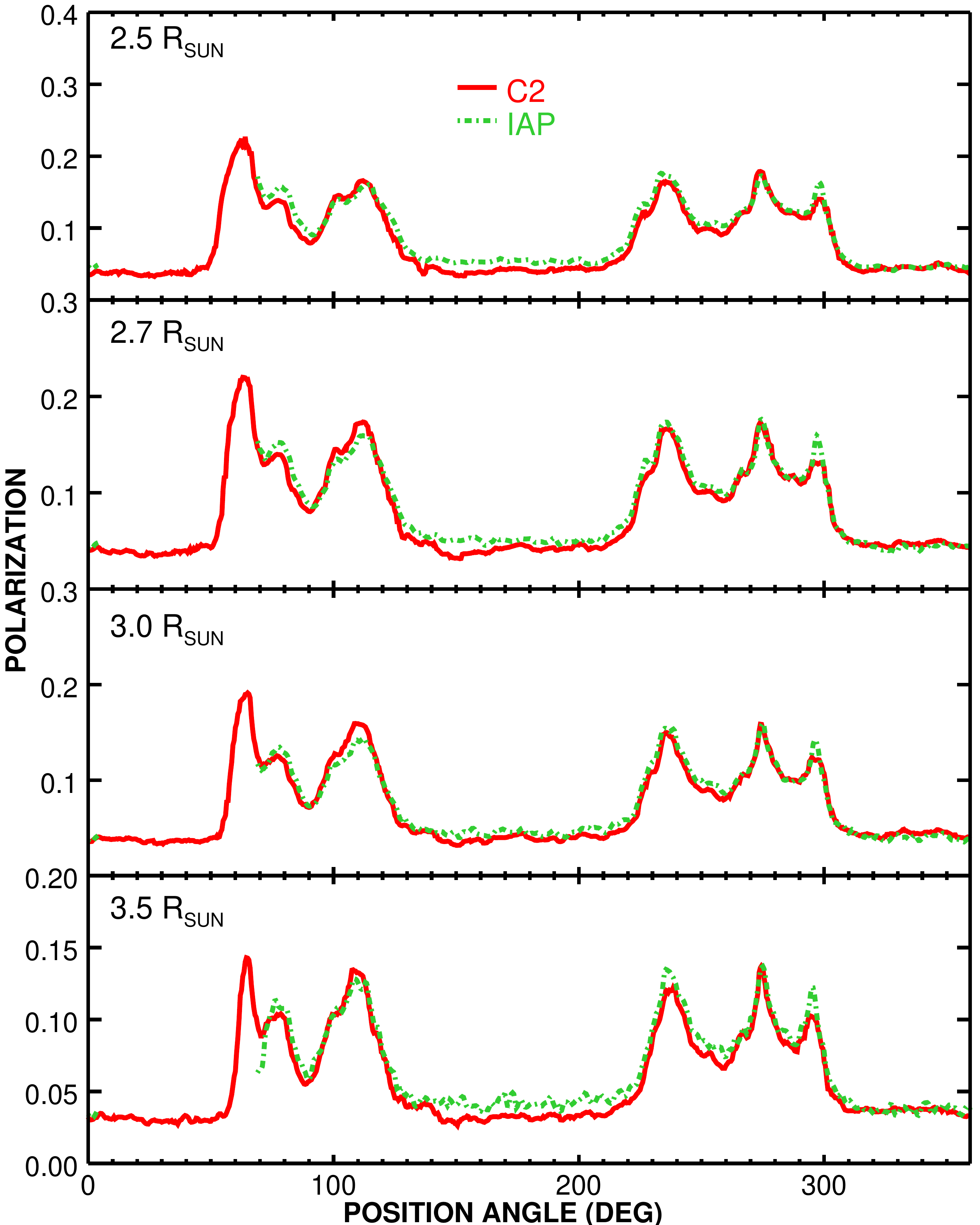} 
\caption{Circular profiles extracted at four solar elongations from IAP and C2 images of the polarization after optimally scaling the former to match the latter.}
\label{FigCircProfIAPC2}
  \end{center}
\end{minipage}
\end{figure}

Likewise the HAO images of 1998, the ground-based images were processed so as to match the spatial scale and the orientation of the C2 images. 
Circular profiles were extracted at four heliocentric distances in the overlap region and Figure~\ref{FigCircProfHAOC2} displays the polarized radiance and the polarization as functions of position angle $PA$.
All polarization profiles follow the same pattern with marked peaks corresponding to the bright streamers of the equatorial belt, but with systematically lower values for C2 with a couple of exceptions.
The E-KPol and IAP results are in agreement for the streamer belt but they diverge for the coronal holes where in fact, the E-KPol and C2 profiles are consistent whereas the IAP data are larger by approximately a factor of 2.
The similarity of the IAP and C2 profiles led us to investigate by linear regression whether there is a simple scaling factor between the two.
This is indeed the case as illustrated in Figure~\ref{FigCircProfIAPC2}, the optimal scaling factor slightly increasing with heliocentric distance, from 0.60 at 2.5 R${}_\odot$ to 0.64 at 3.5 R${}_\odot$.
The radial profiles displayed in Figure~\ref{FigRadProfHAOIAPC2} confirm the above trend with a rather good agreement along the equatorial directions but a clear disagreement along the polar directions; in the latter case, the turnover of the profiles at approximately 3.5 R${}_\odot$ with $p$ increasing beyond is clearly an artifact, probably resulting from imperfect cancellation of the sky contribution.
Figures~\ref{FigCircProfHAOC2} and \ref{FigRadProfHAOIAPC2} display circular and radial profiles extracted from the E-KPol and C2 $pB$ images.
The general agreement is rather good with however a few discrepancies notably in the coronal holes (where the E-KPol values are lower than the C2 values by as much as a factor of approximately 2) and for the peak values in the streamer belt which are almost systematically larger in the case of C2.

\begin{figure}[htpb!]
\begin{center}
  \includegraphics[width=1.\textwidth]{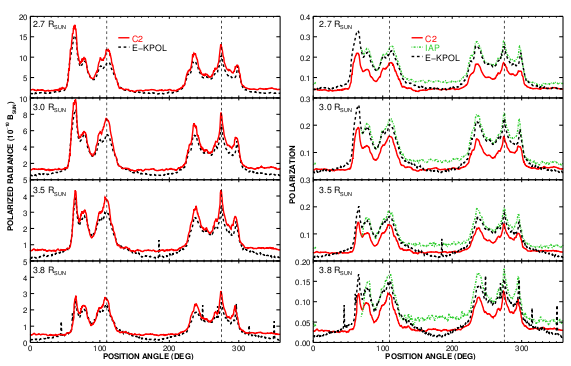} 
\caption{Circular profiles extracted at four solar elongations from E-KPol and C2 images of the polarized radiance $pB$ (left column) and from E-KPOL, IAP, and C2 images of the polarization $p$ (right column) at the time of the eclipse on 29 March 2006.
The vertical dashed lines indicate the locations of the selected radial profiles through two major streamers in the east and west directions.}
\label{FigCircProfHAOSIAPC2}
  \end{center}
\end{figure}

\begin{figure}[htpb!]
\begin{minipage}{1.\textwidth}
\begin{center}
  \includegraphics[width=0.49\textwidth]{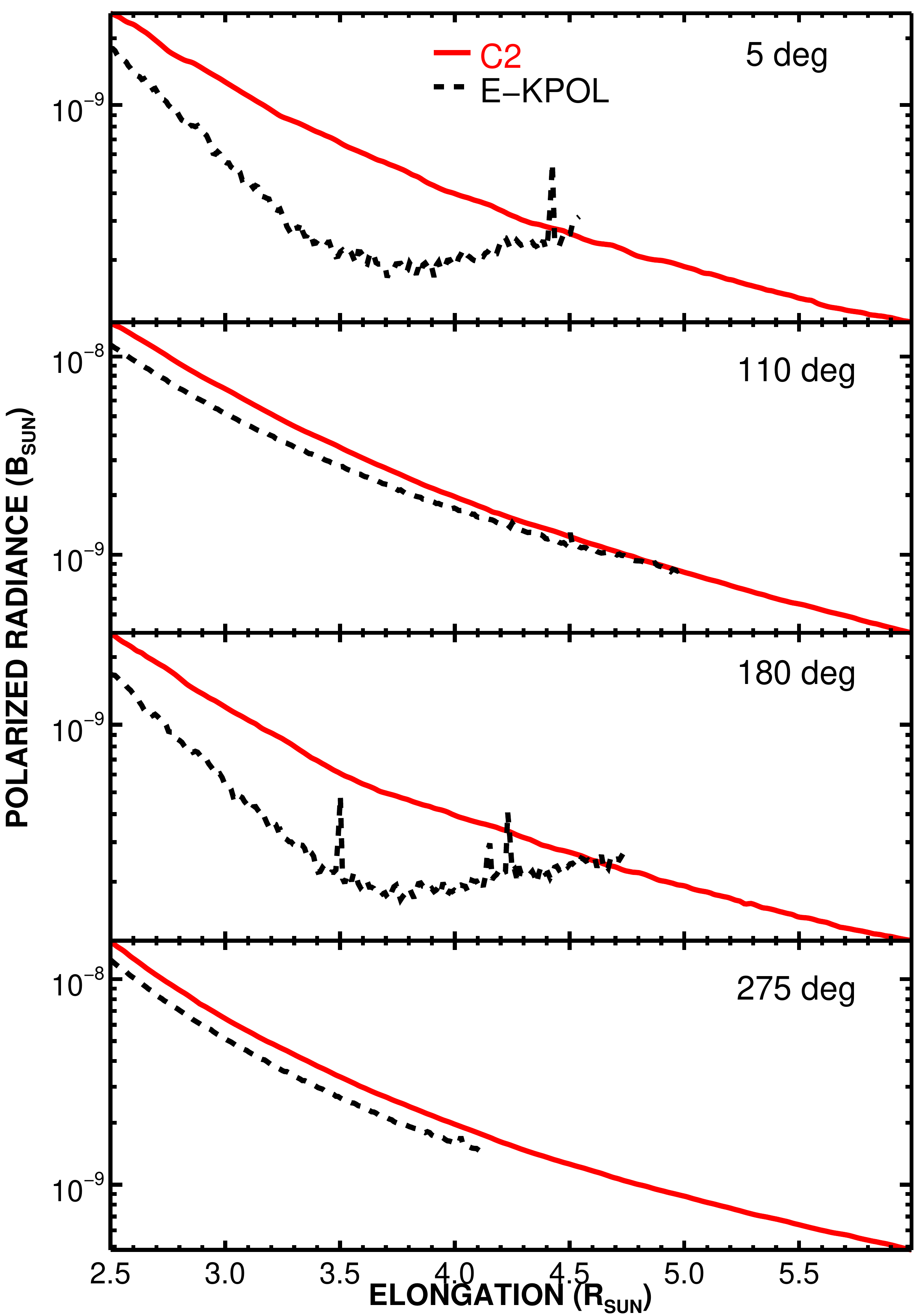} 
  \includegraphics[width=0.49\textwidth]{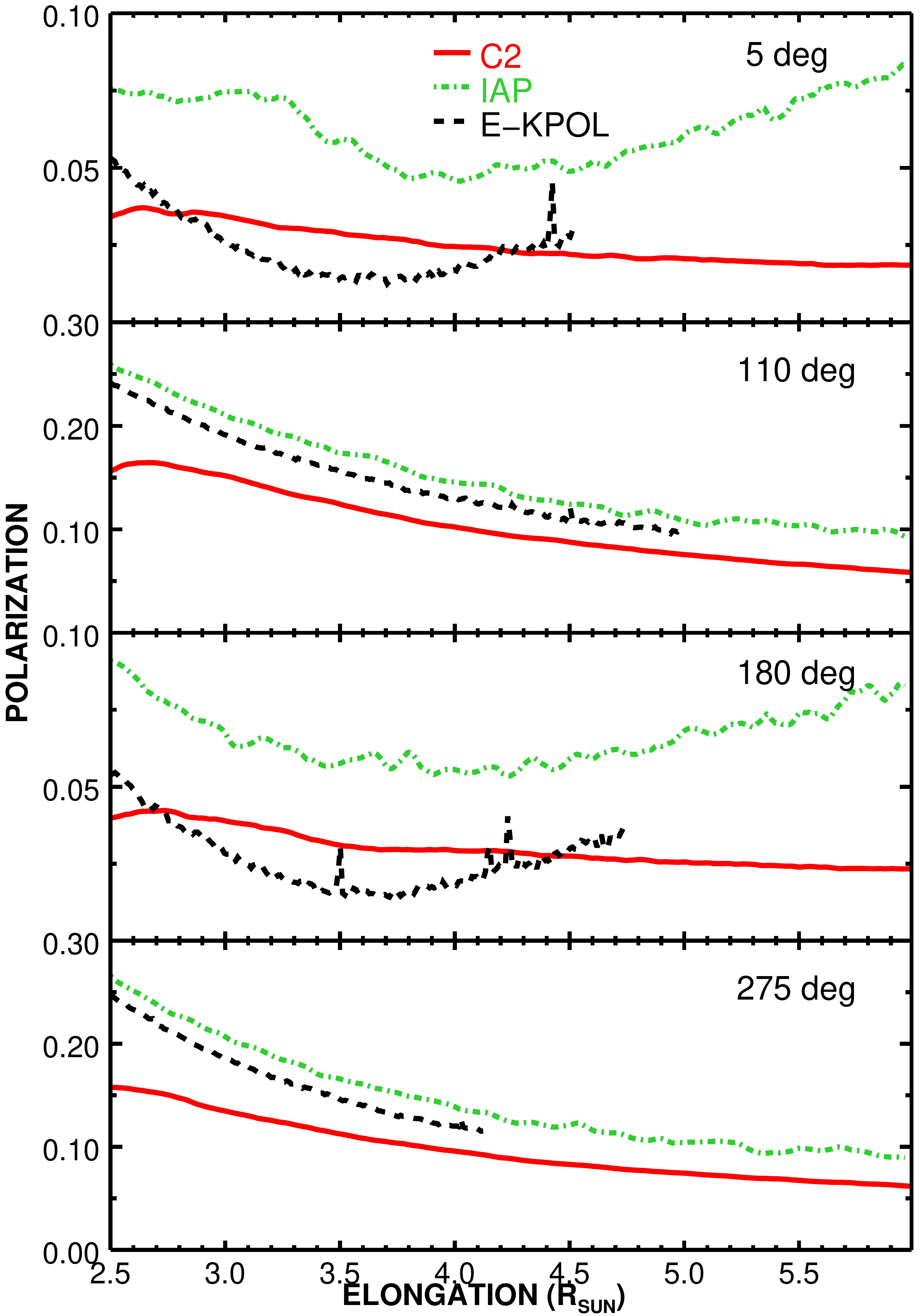}
\caption{Radial profiles along four directions extracted from E-KPol and C2 images of the polarized radiance $pB$ (left column) and from E-KPOL, IAP, and C2 images of the polarization $p$ (right column) at the time of the eclipse on 29 March 2006.
Two profiles are along the north ($PA$ = $5\deg$) and south ($PA$ = $180\deg$) directions and two profiles are along the axis of two major streamers in the east ($PA$ = $80\deg$) and west ($PA$ = $275\deg$) directions as indicated by the dashed lines in Figure~\ref{FigCircProfHAOSIAPC2}.}
\label{FigRadProfHAOIAPC2}
  \end{center}
\end{minipage}
\end{figure}

\subsection{Additional analysis of the results of the Eclipse on 29 March 2006}

The above results for the two eclipses of 1998 and 2006 show that the agreement between C2 and the ground-based observations is excellent for the polarized radiance but less so for the polarization.
We therefore carried out a detailed investigation for the case of the 29 March 2006 eclipse in an attempt to understand the cause(s) of the discrepancies.
We considered that the polarized radiance is robustly determined and proceeded to invert it to derive the 3D distribution of $Ne$ (thereafter ``$Ne$ cube'') using two different methods.
The first one implements the 2D inversion scheme of \cite{quemerais2002two}, already introduced in Subsection~\ref{Ne}, of a C2 $pB$ image obtained on the day of the eclipse at 9:52 UT.
The 3D cube was subsequently constructed by considering a quasi-axial symmetric corona where the eastern hemisphere (resp. western) was generated by rotating the eastern (resp. western) meridional planes of the 2D $Ne$ map.
The inevitable discontinuity along the N-S axis was ironed out by a smoothing operation.
The second one implements the full 3D time-dependent tomographic reconstruction technique developed by \cite{vibert2016time} on a time-series of C2 $pB$ images obtained over a time interval of 15 days centered on the date of the eclipse.
These two solutions match quite well with a trend of the latter one to produce slightly smaller densities then the former.
This is understood on basis of the assumption of quasi-axial symmetry of the first method which forces the bright jets seen on the 2D maps to extend as a belt around the corona which is not necessarily the case.
Incidentally, we considered using the 3D distribution generated by Predictive Science Inc. (available online at http://www.predsci.com/mhdweb/home.php) for the eclipse date but their two polytropic solutions ``std\_101'' and ``std\_201'' were well off our $Ne$ cubes in both geometry and level. 
Finally, Thompson scattering was calculated along lines of sight throughout the two $Ne$ cubes to generate images of the radiance and of the polarized radiance of the K-corona at the date of the eclipse. 

The image of the F-corona resulted from the separation process described in Section~\ref{Sec:Separation} from which an image of the instrumental stray light was subtracted \citep{llebaria2004lessons}.
We further considered the influence of the weak polarization of the F-corona by building a map of $p_F$ assuming elliptical isopleths defined by the north and south profiles given by \cite{koutchmy1985f} thus producing an image of the polarized radiance.
Finally, combining the above results, we generated maps of the polarization of the global K+F corona for three different cases, each case comprising four configurations, 2D/3D inversions and unpolarized/polarized F-corona for a total of 12 combinations.

\begin{itemize}
	\item 
	Case $PS_0$ corresponds to the nominal situation described above.
	\item 
	Case $PS_1$ considers a slightly fainter F-corona by a factor of 1.2.
	This factor results from a comparison between the F-corona obtained from the C2 observations and the equatorial and polar model profiles of \cite{koutchmy1985f}.
	\item 
	Case $PS_2$ considers, in addition to case $PS_1$, a higher K-corona by a factor 1.5 that propels it to a corona of a maximum type.
	This factor results from a comparison of profiles of the electron density between the brightest streamers observed during the eclipse and the spherical model of \citet{baumbach1937strahlung}.
	\end{itemize}
	
\begin{figure}[htpb!]
\label{}
\begin{center}
  \includegraphics[width=0.9\textwidth]{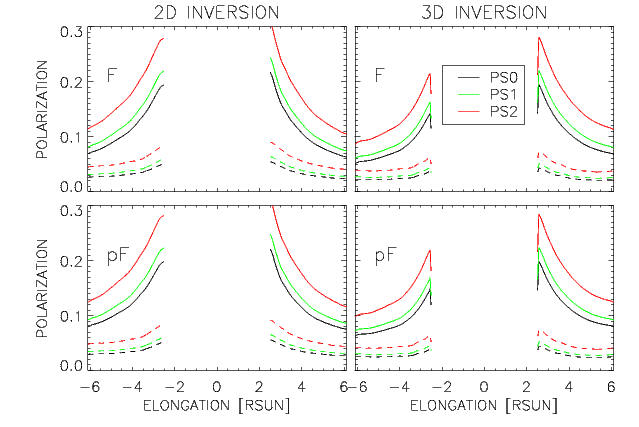} 
\caption{Radial profiles along four directions extracted from different models of the K+F corona as described in the text.
The left (resp. right) column corresponds to the case of the $Ne$ cube obtained from 2D (resp. 3D) inversions.
The upper (resp. lower) row corresponds to the case of an unpolarized (resp. polarized) F-corona.
The solid curves correspond to the two profiles along the axis of two major streamers in the east ($PA$ = $80\deg$) and west ($PA$ = $275\deg$) directions.
The dashed curves correspond to the two profiles along the north and south directions (negative elongations).}
\label{FigPScomparison}
\end{center}
\end{figure}

Figure~\ref{FigPScomparison} displays equatorial and polar profiles of the coronal polarization for the above cases and conspicuously reveals that they are extremely sensitive to even slightly different assumptions.
As noted above, the 2D inversion tends to produce electron densities larger than the 3D inversion thus resulting in larger polarizations, an effect especially marked on the eastern (left) profiles. 
Introducing the polarization of the F-corona has only a minor impact on the profiles, an increase of at most $\approx$ 1\%.
A fainter F-corona by 20\% systematically increases the polarization by typically 1\%.
Finally and as expected, a brighter K-corona has a more drastic effect, systematically increasing the polarization by 2\% and up to 4\% for the polar and equatorial profiles, respectively. 
Ultimately, this exercise reveals that the range of the calculated models of polarization is so large so that they are compatible with the observed E-KPol, IAP, and C2 values. 
However, we point out  that the large values reported by the E-KPol and IAP observations are only compatible with the extreme $PS_2$ case requiring a K-corona of the maximum type which certainly was not the case in March 2006 when solar activity was close to the end of its declining phase, in fact 2.75 year from the solar cycle 23/24 minimum .

\subsection{Eclipse on 21 August 2017}

\begin{figure}[htpb!]
\label{}
\begin{center}
  \includegraphics[width=0.6\textwidth]{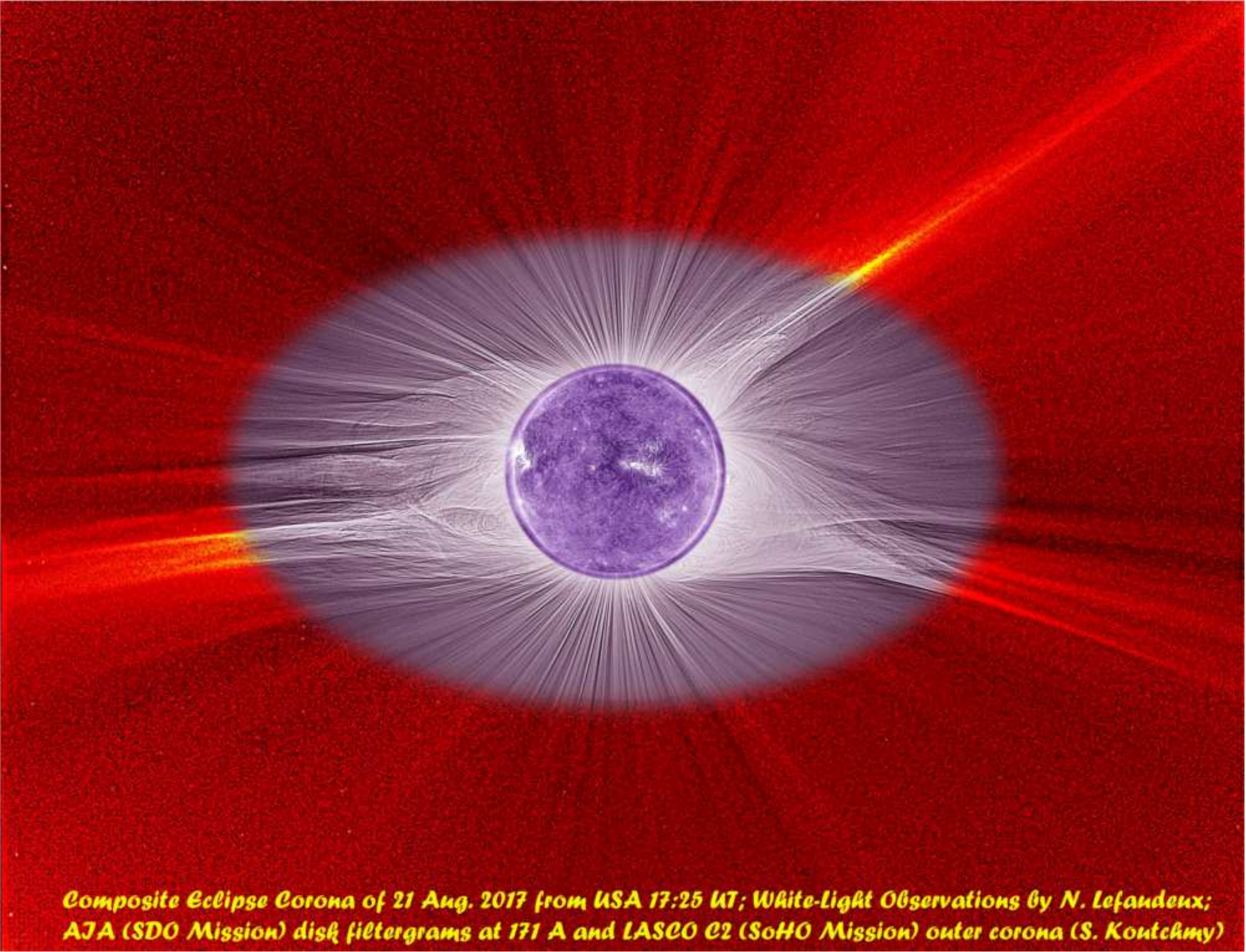}
\caption{Composite of a SDO/AIA disk filtergram at 171~\angstrom \ and of two images of the 21 August 2017 eclipse, a highly processed image of the inner corona obtained by N. Lefaudeux and a LASCO-C2 image of the outer corona (courtesy S. Koutchmy).
Solar north is up.}
\label{FigCompositeLASC22017}
\end{center}
\end{figure}

Figure~\ref{FigCompositeLASC22017} displays a qualitative composite constructed by S. Koutchmy combining a SDO/AIA image of the solar disk, a highly processed image obtained by N. Lefaudeux, and a LASCO-C2 image of the outer corona.
Two teams reported polarization observations of this eclipse.

\begin{itemize}
	\item 
The team of the Rochester Institute of Technology implemented a new device combining a thermoelectrically cooled KAI-04070 CCD detector and a MOXTEK, Inc. micropolarizer array \citep{vorobiev2017imaging}.
This array is aligned and affixed to the CCD such that each set of four CCD pixels samples the electric field intensity along the $0\deg$, $45\deg$, $90\deg$, and $135\deg$ directions.
This is analogous to the use of a color filter array (\eg a Bayer filter mosaic) to create color-sensitive imaging arrays and this results in an imaging polarimeter giving the three Stockes parameters $I$, $Q$, and $U$ of the incident light in a single frame. 
For the eclipse observation, the {\it Rochester Institute of Technology Polarization Imaging Camera} (RITPIC) was fitted with a Bessel R filter and attached to a 530mm f/5 Takahashi FSQ 4-element refractive telescope. 
This resulted in a plate scale of 2.88 arcsec per CCD pixel, that is 5.76 arcsec per image element/superpixel and a total \fov of 4$\times$4 R${}_\odot$.
The observations were obtained from Madras, Oregon.
The limited dynamic range of the sensor required the use of four different exposure times, each exposure being taken sequentially seven times.
The resulting composite image of the polarization of the corona was made available to us by D. Vorobiev.	
	\item
The {\it Rosetta Stone PolarCam} experiment of the NCAR/High Altitude Observatory team was based on an innovative 4-D technology polarization measuring system {\it PolarCam}; it combines a 1.8 Mpixel CCD detector and a micro-polarizer array placed over the sensor with four alternating orientations of linear polarizers \citep{Burkepile2017}.
The optical setup produced a rectangular \fov of 5.6$\times$3.6 R${}_\odot$ and the observations were performed on Casper Mountain, Wyoming at an elevation of 2000~m.
At time of writing, only limited results have been published by \cite{Burkepile2017}, namely two $pB$ circular profiles at 1.12 and 1.35 R${}_\odot$, three $pB$ radial profiles up to 1.8 R${}_\odot$ through three remarkable helmet streamers, and finally a $pB$ radial profile up to 2.5 R${}_\odot$ at a position angle of $268\deg$.
\end{itemize}

\begin{figure}[htpb!]
\begin{minipage}{1.\textwidth}
\begin{center}
  \includegraphics[width=0.47\textwidth]{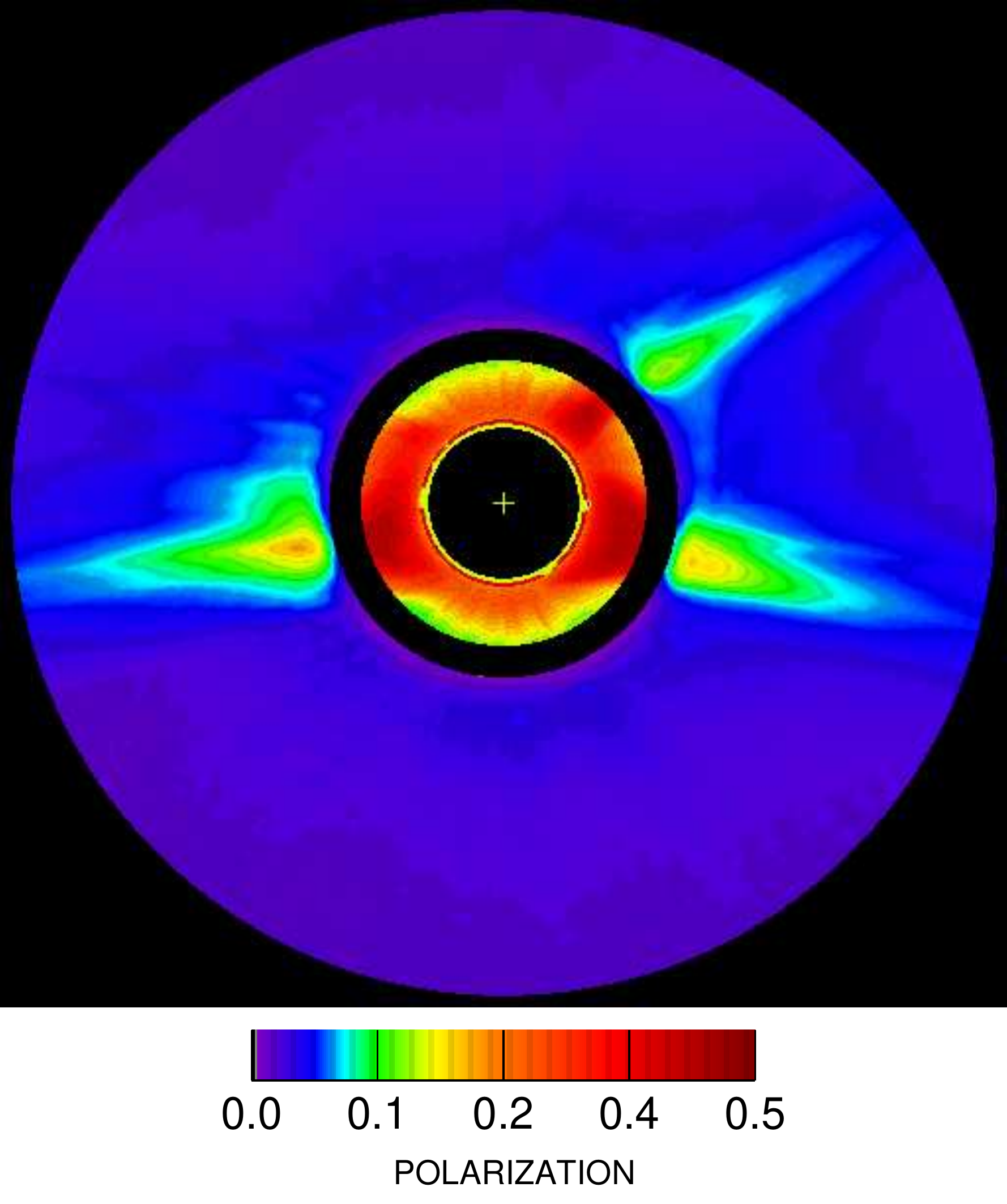} 
\hspace{0.01\textwidth}
  \includegraphics[width=0.47\textwidth]{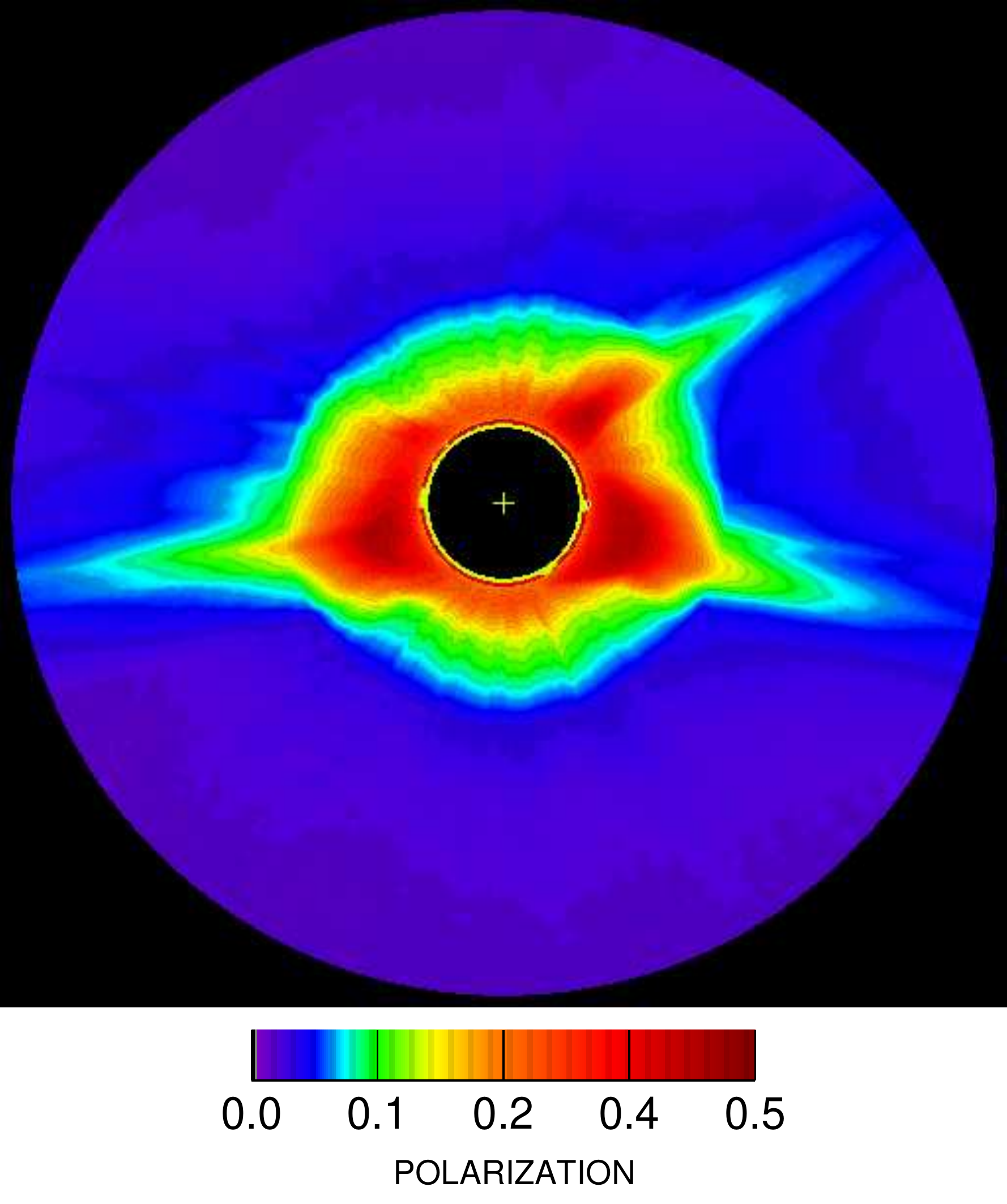}
\hspace{0.01\textwidth}  
\caption{Composites of the {\it Rochester Institute of Technology Polarization Imaging Camera} (RITPIC) and C2 images of the polarization $p$ obtained at the eclipse on 21 August 2017. 
The left panel represents a first version of this composite : a gap appears between the C2 data and the RITPIC data since the fields of view do not overlap. 
The right panel shows an improved version of this composite where the gap is filled with an interpolation between the two images. 
The yellow circles represent the solar disk with a cross at its center.
Solar north is up.}
\label{FigCompositeRITPICC22017}
  \end{center}
\end{minipage}
\end{figure}

\begin{figure}[htpb!]
\begin{minipage}{1.\textwidth}
\begin{center}
  \includegraphics[width=0.47\textwidth]{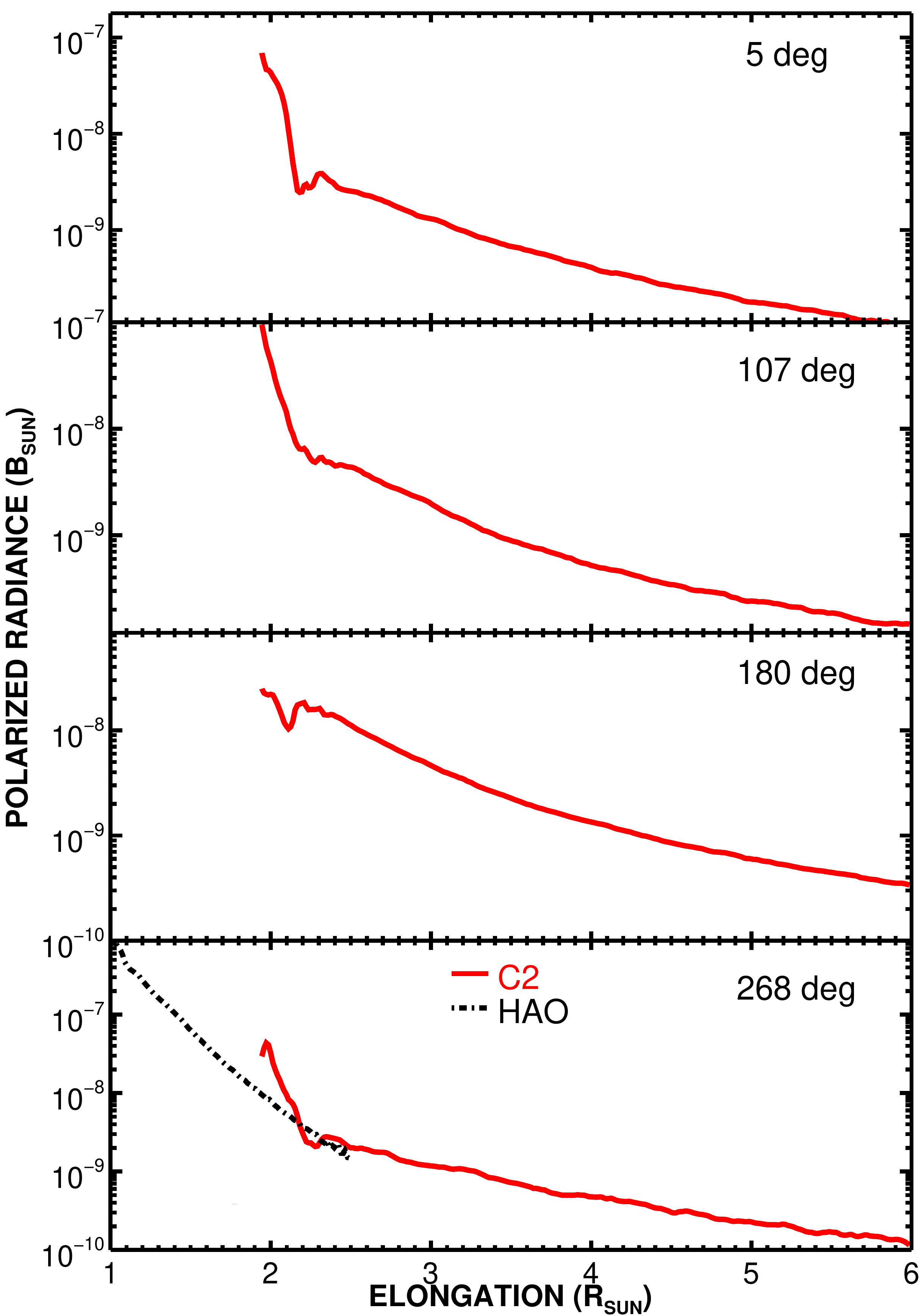} 
\hspace{0.01\textwidth}
  \includegraphics[width=0.47\textwidth]{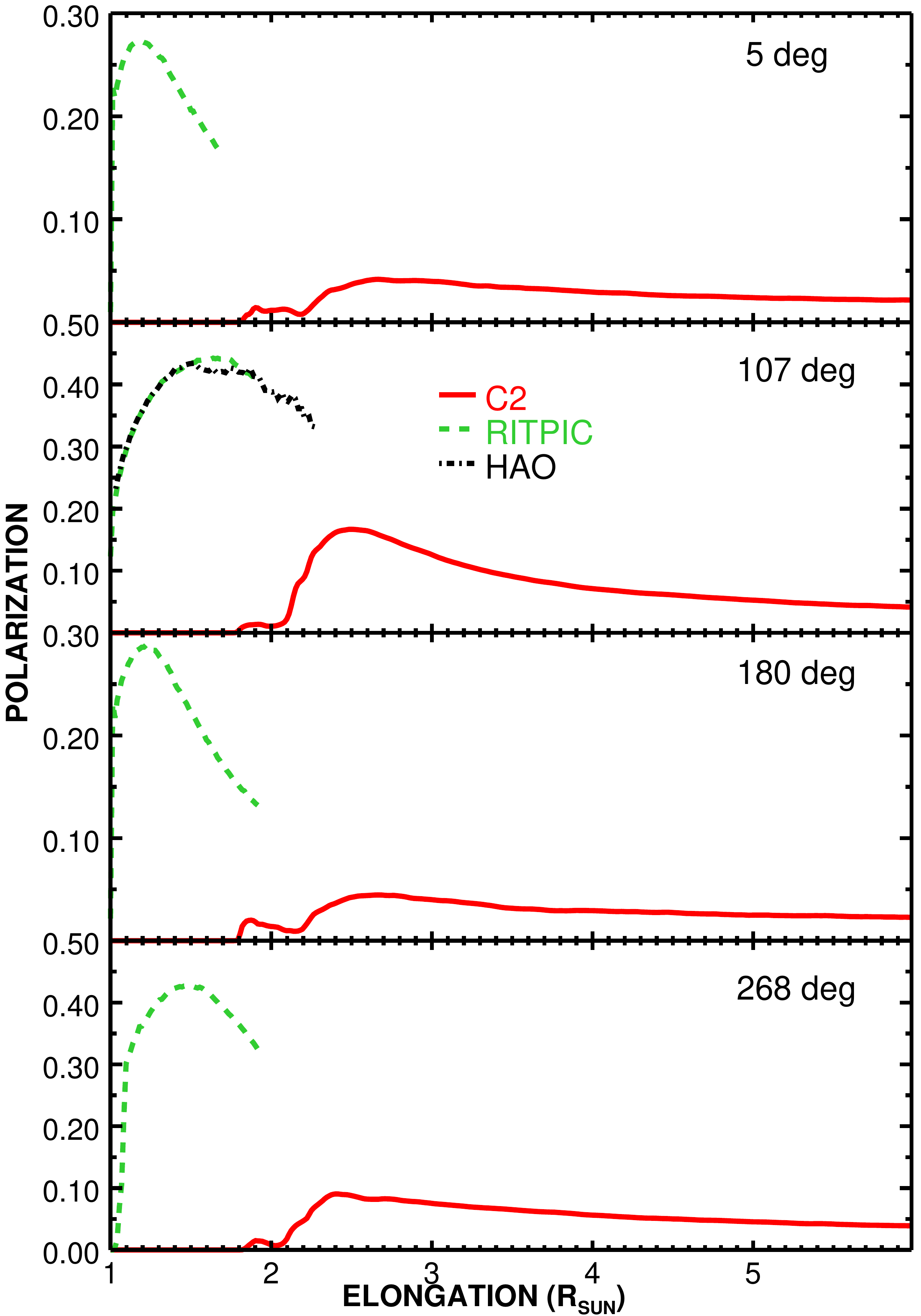}
\caption{Radial profiles along four directions extracted from {\it PolarCam} (HAO) and C2 images of the polarized radiance $pB$ (left column) and from RITPIC, {\it PolarCam}, and C2 images of the polarization $p$ (right column) at the time of the eclipse on 21 August 2017.
Two profiles are along the north ($PA$ = $5\deg$) and south ($PA$ = $180\deg$) directions and two profiles are along the axis of two major streamers in the east ($PA$ = $107\deg$) and west ($PA$ = $268\deg$) directions.}
\label{FigRadprofCompositeRITPICC22017}
  \end{center}
\end{minipage}
\end{figure}

Two LASCO-C2 polarization sequences were taken with the orange filter on the eclipse day at approximately 2:30 and 21 hr that brackets the two ground-based observations, 17:20 at Madras and 17:43 at Casper.
The C2 $p$ and $pB$ images were averaged for the purpose of the comparison with the ground-based data and the procedure follows that developed for the eclipse on 26 February 1998 in the case of the RITPIC polarization image.
Figure~\ref{FigCompositeRITPICC22017} presents two composites of the RITPIC and C2 $p$ images.
The left composite is affected by the gap between the two fields of view (black annulus) which is filled in the right composite by using an interpolation of the radial profiles on polar-transformed images.
The match between the RITPIC and C2 images is excellent and demonstrates the good continuity between the two as confirmed by the four radial profiles displayed in the right panel of Figure~\ref{FigRadprofCompositeRITPICC22017}.
The profile at $PA$ = $107\deg$ includes the HAO result of \cite{Burkepile2017} which is in excellent agreement with the RITPIC profile.
The left panel of Figure~\ref{FigRadprofCompositeRITPICC22017} displays the four profiles of the C2 $pB$ image as well as the HAO profile at $PA$ = $268\deg$ \citep{Burkepile2017}.
Although the overlap is quite narrow, the agreement is clearly quite remarkable.

\section{Conclusion}

The capability of LASCO-C2 to perform accurate polarimetric measurements is affected by several adverse conditions, prominently the very basic polarization analyzer system (three Polaroid foils), the optical design with two folding mirrors, and the fluctuations in exposure times.
Thanks to extensive ground and in-flight calibrations and to our in-depth analysis of these effects, we have been able to fully correct these effects and produce valid results of the polarization of the corona during 24 years.
In addition to our tests of consistency to validate our polarization results, the comparison with eclipses results provided an independent check and we are quite impressed by the quality of several composites that we could produce.
The availability of the resulting unsurpassed data sets of images of the polarized radiance $pB$ and of the radiance of the K-corona open the way to a systematic reconstruction of the three-dimensional distribution of the electron density over two solar cycles by implementing the time-dependent tomography method developed by \cite{vibert2016time}; this will be the topic of a forthcoming paper.

We warn the interested readers that all results published so far which involve the polarization results of LASCO-C2 such as the 3D reconstruction of CMEs based on the polarimetric technique, relied on the standard processing from the ``Solarsoft'' library which does not incorporate the extensive corrections described in this article.
In fact, this standard processing has never been proved to produce the correct polarization of the corona and it most likely corresponds to the first step of our procedure as illustrated in the upper row of Figure~\ref{FigBkC2}.
The only exception is the work performed by \cite{moran2006solar} who attempted to correct for the effects of the folding mirrors but here again, they never verified that their procedure led to a correct polarization state of the corona, particularly its ``tangential'' direction.
We further warn against using the polarization products obtained with the blue and red channels of C2 as the corresponding Mueller matrices have not been determined to the required level of accuracy.
Figure~\ref{Transmissions} indicates furthermore that the principal transmittance $k_{2}$ rapidly increases with increasing wavelength beyond 650~nm so that the rejection in the ``red'' channel is probably insufficient to guaranty a reliable determination of the polarization.

Although a large part of this article is devoted to the characterization of the polarimetric channel of LASCO-C2, the illustrative results cast valuable light on the temporal variation of the radiometric and polarimetric properties of the solar corona. 
We confirm and extend the finding of \cite{barlyaeva2015mid} that they are highly correlated with the variation of the total photospheric magnetic field.
This undoubtedly ascertains the role of the magnetic field in the evolution of the corona.
Our results further highlight the very peculiar behaviour of the corona during SC 24 and the surprising asymmetry between the two polar regions.

All images generated by our procedure are part of the LASCO-C2 Legacy Archive\footnote{\url{http://idoc-lasco-c2-archive.ias.u-psud.fr}} hosted at the Integrated Data and Operation Center (formerly MEDOC) of Institut d'Astrophysique Spatiale.


\begin{acknowledgments}
We thank our former colleagues at the Laboratoire d'Astrophysique de Marseille, M. Bout, F. Ernandez, G. Faury, B. Gard\`es, and C. Peillon  for their contributions to the early phases of this study.
We are grateful to D. Elmore, S. Koutchmy, S. Fineschi, G. Capobianco, E. Balboni, and D. Vorobiev for providing the results of their eclipse observations, to P. Raynal  and J.-L. Reynaud for digitizing the photographic images of the eclipse of 11 August 1999, to J. Wojak for providing the $Ne$ cube, and to Y.-M. Wang for providing the Total Magnetic Field (TMF) data.
We are indebted to S. Koutchmy for constructing the remarkable composites of the four eclipses.
The LASCO-C2 project at the Laboratoire d'Astrophysique de Marseille and the Laboratoire Atmosph\`eres, Milieux et Observations Spatiales is funded by the Centre National d'Etudes Spatiales (CNES).
LASCO was built by a consortium of the Naval Research Laboratory, USA, the Laboratoire d'Astrophysique de Marseille (formerly Laboratoire d'Astronomie Spatiale), France, the Max-Planck-Institut f\"ur Sonnensystemforschung (formerly Max Planck Institute f\"ur Aeronomie), Germany, and the School of Physics and Astronomy, University of Birmingham, UK.
SOHO is a project of international cooperation between ESA and NASA.
\end{acknowledgments}

\vspace{\baselineskip} 
\noindent
\textbf{Disclosure of Potential Conflicts of Interest} The authors declare that they have no conflicts of interest.


\bibliographystyle{spr-mp-sola}
\bibliography{biblio_polar_2019-08-22}
\nocite{*}

\appendix

\section*{Appendix I: Determination of the correction function $S(x,y)$ for the polarized brightness $pB$}
\label{AppendixCross}

The correction function $S(x,y)$ was derived from the polarization sequences performed on 2 and 3 September 1997 when the SoHO spacecraft, starting from its nominal position  $\theta_{\rm r}=0^\circ$, was successively rolled to $\theta_{\rm r}=45^\circ$ and $90^\circ$ (Table \ref{table:rollsequence}).
Each sequence yielded a triplet of polarized images I$_{1}$, I$_{2}$, I$_{3}$ leading to a polarized radiance 
$\mathrm p_{\rm c}I_{\rm c} = p_{\rm cp}I_{\rm cp}$
according to our notations introduced in Section 3.
For simplification in the present discussion, we adopt the standard notation $pB$.
The three sequences therefore produced three $\rm pB$ images denoted as follows:

\begin{description}
\item $\theta_{\rm r}=0^\circ \rightarrow {\rm pB}_{\rm 0} $
\item $\theta_{\rm r}=45^\circ \rightarrow {\rm pB}_{\rm 45} $
\item $\theta_{\rm r}=90^\circ \rightarrow {\rm pB}_{\rm 90} $
\end{description}

\noindent 
Comparing those images requires a rotation by $-\theta_{\rm r}$ and this operation will be denoted by $[\Im ]_{-\theta_{\rm r}}$ where $[\Im ]$ is any image.
$S(x,y)$ is determined by imposing that the three $\rm pB$ images, once corrected by $S(x,y)$ and properly rotated, are identical.
This procedure requires that the corona remained unchanged during the 20 hours of the roll sequence.
This is certainly true for the F-corona as well as for the large-scale K-corona, but small scale variations may present a difficulty.
They will be ignored by considering that $S(x,y)$ is intended to correct for large scale variations only. 
The very validity of our procedure, \ie the existence of an ad-hoc function $S(x,y)$ can only be established by finding a relevant solution as demonstrated in section 7.

Let us form the following ratios of the polarized radiances:
\begin{equation}
 R_{0/45} = {\rm pB}_0/[{\rm pB}_{45}]_{-45}
\end{equation}
\begin{equation}
R_{45/90} = {\rm pB}_{45}/[{\rm pB}_{90}]_{-45}
\end{equation}

\noindent 
The condition that the corrected polarized radiances be identical translates into the two following constraints for $S(x,y)$:

\begin{equation}
 R_{0/45} = \frac{S}{[S]_{-45}}(1+\epsilon_1)
\end{equation}
\begin{equation}
R_{45/90} = \frac{S}{[S]_{-45}}(1+\epsilon_2)
\end{equation}

\noindent 
where $\epsilon_1$ and $\epsilon_2$ account for some uncertainty as the radiances cannot obviously be identical because of small changes in the corona during the time interval elapsed between successive observations and because of the intrinsec noise in the images.
In order to combine the above two equations in a single constraint, we introduce a mask $M=M(x,y)\in\{0,...1\}$ which alleviates the effect of the variations of the inner streamer belt and obtain:
\begin{equation}
  R = \frac{M.R_{0/45}+[M]_{45}.R_{45/90}}{M+[M]_{45}}
\end{equation}

\noindent 
$S(x,y)$ is now determined by imposing that it is positive, smooth and that it satisfies the relationship:
\begin{equation}
S = R[S]_{-45}  
\end{equation} 

\noindent 
To solve the above equation, we switch from cartesian $(x,y)$ to polar $(r,\phi)$ coordinates and we change the scale from linear to logarithmic (homomorphic transform).
We then obtain:

\begin{eqnarray}
\log S(r,\phi) & = & \log R(r,\phi) + \log S(r,\phi-\frac{\pi}{4}) .
\end{eqnarray}

\noindent 
As all these functions are periodic in $\phi$, we proceed by looking for a periodic function $\psi(r,\phi)$ satisfying:
\begin{eqnarray}
\psi(r,\phi)-\psi(r,\phi-\frac{\pi}{4}) & = & \log R(r,\phi) .
\end{eqnarray}

\noindent 
In the Fourier space, this equation becomes:

\begin{eqnarray}
  \Psi^*(r,\omega)(1-e^{-i\pi \omega/4}) & = & \widehat{\log R}^*(r,\omega)  
\end{eqnarray}

\noindent  
$\psi(r,\phi)$ and $R(r,\phi)$ being periodic, both sides may be expressed as Fourier series:
\begin{eqnarray}
  \Psi_k(r)(1-e^{-i\pi k/4}) & = & \widehat{\log R}_k(r) \qquad \forall k \in \{0,...N-1\},  
\end{eqnarray} 
where $N$ is the number of angular samples. 
This differential problem is only partially solvable: indeed, it can be noted that the average value remains undefined, as well as the harmonic terms corresponding to  $k \in \{0,8,16,\dots,8n\}$. 
In practise, and because of the noise, the determination can well be limited to the first five harmonic terms:
\begin{eqnarray*}
\Psi_n(r) & = & \frac{\widehat{\log R}_k(r)}{1-e^{-ik\pi/4}}  \qquad k = \{1,2,\dots,5\} \\ .   
\end{eqnarray*}

\noindent 
The radial variations of the $\Psi_k(r)$ coefficients are strongly correlated (thus offering some extra robustness), and a third order polynomial of $r$ is ample. 

In practice the missing values in the $(r,\phi)$ frame due to the square format of the initial images poses an additional problem alleviated by limiting the expansion series to the first terms in the outer part of image where data are missing.  

The polar form of the function $\log S = \psi(r,\phi)$ is obtained with the limited Fourier series :
\begin{eqnarray*}
\psi(r,\phi) & \simeq & \sum_1^5 \Psi_k(r)\;e^{-i2\pi k\phi} 
\end{eqnarray*}
From $\psi(r,\phi)$ the derivation of $S(x,y)$ is straightforward.

\begin{figure}[htpb!]
	\centering
	\includegraphics[width=0.7\textwidth]{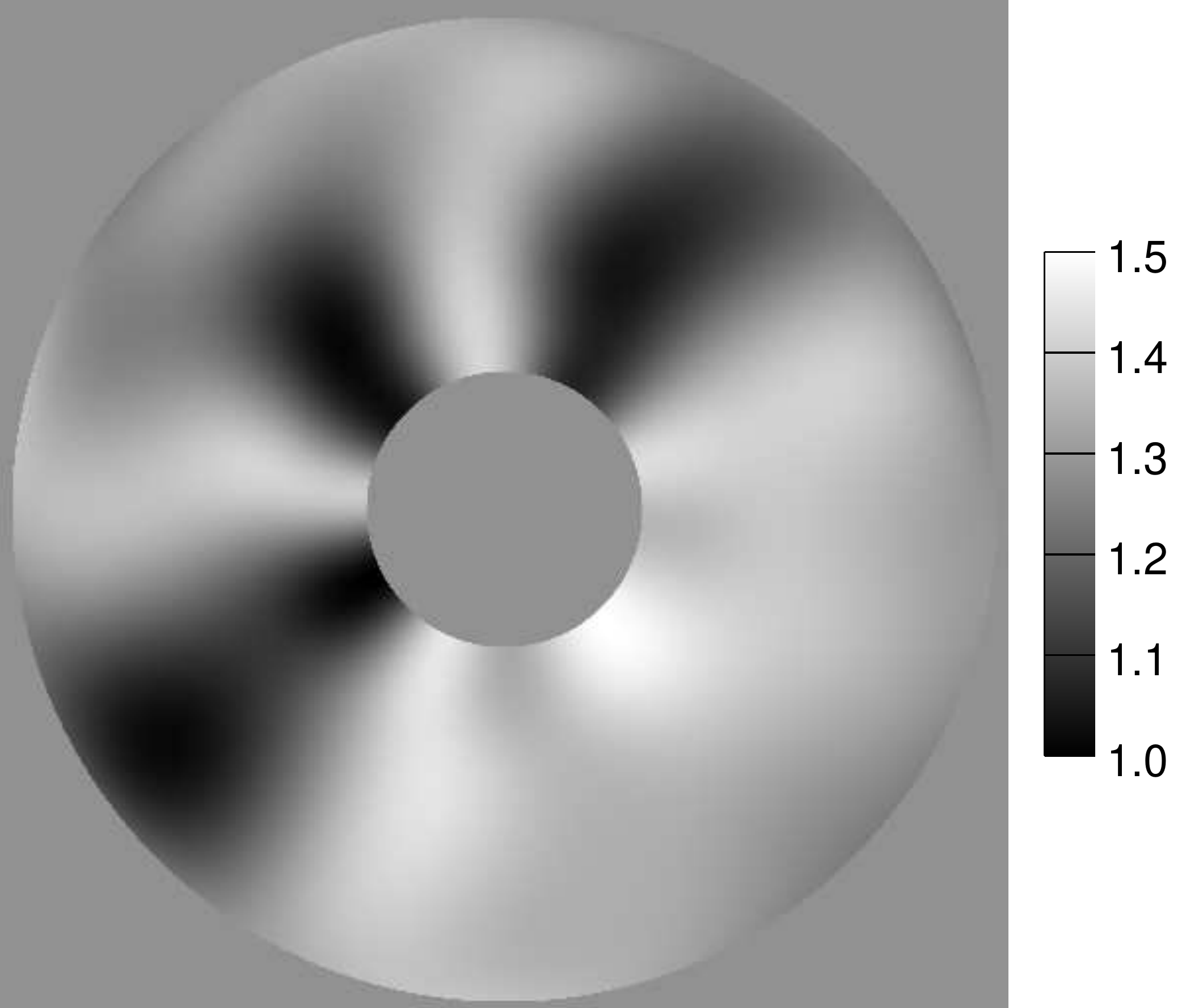}
	\caption{The LASCO-C2 correction pattern $\psi_{(r,\theta)}$ for the polarized brightness image, built from the three images during September, 1997 rotation sequence.}
	\label{FigCrossC2}
\end{figure}



\end{article} 

\end{document}